\documentclass[a4paper,oneside]{book}
\usepackage[latin2]{inputenc}
\usepackage{t1enc}
\usepackage{graphicx}
\usepackage{amsmath}
\usepackage{stmaryrd}
\usepackage{cite}

\def\lsi{\raise0.3ex\hbox{$<$\kern-0.75em\raise-1.1ex\hbox{$\sim$}}}
\def\gsi{\raise0.3ex\hbox{$>$\kern-0.75em\raise-1.1ex\hbox{$\sim$}}}
\def\Tr{{\rm Tr}}
\def\Re{{\rm Re}}
\def\Im{{\rm Im }}
\newcommand{\be}{\begin{equation}}
\newcommand{\ee}{\end{equation}}
\newcommand{\bea}{\begin{eqnarray}}
\newcommand{\eea}{\end{eqnarray}}
\date{}

\begin{document}
\title{{\LARGE The phase diagram of quantum chromodynamics}\\[1cm]
}
\author{Z. Fodor, S.D. Katz \\[5mm]
{\em University of Wuppertal, Germany}\\ {\em Eotvos University, Budapest, Hungary} \\[8cm]}
\maketitle

\tableofcontents

\chapter{Introduction}
Quantum chromodynamics (QCD) is the theory of strong interactions.
The elementary particles of QCD --contrary to the other particles
described by the Standard Model (SM) of particle physics-- can 
not be observed directly.
The Lagrangian of QCD is given by quarks and gluons. Instead of
free quarks and gluons we observe bound state hadrons.

One of the most important features of QCD is asymptotic freedom.  
At small energies the coupling is strong, the value of the coupling constant
is large. For large energies the coupling constant decreases and 
approaches zero.
Since the coupling constant is large at small energies, we can not use 
one of the most powerful methods of particle physics, the perturbative 
approach. For large enough energies the coupling gets smaller, thus asymptotic 
freedom opens the possibility to use perturbative techniques. In this regime
scattering processes can be treated perturbatively. The results are in 
good agreement with the experiments.

At small energies (below about 1~GeV) the bound states and their 
interactions can be described only by non-perturbative methods. The most
systematic non-perturbative technique today is lattice field theory.
The field variables of the Lagrangian are defined on a discrete space-time
lattice. The continuum results are obtained by taking smaller and 
smaller lattice spacings ($a$) and extrapolating the results to vanishing 
$a$.
Though lattice field theory has been an active field for 30 years,
the first continuum extrapolated full results appeared only recently.

Another consequence of asymptotic freedom that the coupling decreases 
for high temperatures (they are also characterized by large energies).
According to the expectations at very high temperatures (Stefan-Boltzmann 
limit) the typical degrees of freedom are no longer bound state hadrons      
but freely moving quarks and gluons. Since there are obvious qualitative  
differences between these two forms of matter, we expect a phase transition
between them at a given temperature $T_c$. The value of $T_c$ can be 
estimated to be the typical QCD scale ($\approx 200$~MeV).

At large baryonic densities the Fermi surface is at large energy, thus 
we observe a similar phenomenon, the typical 
energies are large, the coupling is small. Also in this case we
expect a phase transition between the phases characterized by small and 
large energies. 
In QCD the thermodynamic observables are related to the grand canonical 
partition function. Therefore, the baryonic density can be tuned 
by tuning the baryonic chemical potential ($\mu$). If we increase the chemical
potential the corresponding $T_c$ values decrease. Thus, one obtains
a non-trivial phase diagram on the $T$--$\mu$ plane.

Understanding the T>0 and $\mu$>0 behaviour of QCD is not only a theoretical
question. In the early Universe (about $10^{-5}$ after the Big Bang)
the T>0 QCD transition resulted in hadrons, which we observe today, 
and even more, which we are made of. The nature of the transition
(first order phase transition, second order phase transition or 
an analytic crossover) and its characteristic scale ($T_c$) tell a 
lot about the history of the early Universe and imply
important cosmological consequences. Since in the early Universe the number of 
baryons and antibaryons were almost equal we can use $\mu$=0 to a 
very good approximation.

One of the most important goal of heavy ion experiments is to understand and    
experimentally determine the phase diagram of QCD.
The determination of the temperatures and/or chemical potentials in
a heavy ion collision is far from being trivial. The larger the energy
the closer the trajectory (the $\mu$-T pairs, which characterise
the time development of the system) to the $\mu$=0 axis.
Earlier heavy ion experiments (e.g. that of the CERN SPS accelerator)
mapped relatively large $\mu$ regions (approximately 150-200~MeV), whereas
present experiments of the RHIC accelerator runs around 40~MeV.
The heavy ion program of the LHC accelerator at CERN 
will study QCD thermodynamics
essentially along the $\mu=0$ axis.
The most important physics goal of the CBM experiment of the FAIR
accelerator at GSI in Darmstadt is to understand the QCD phase diagram
at large baryonic chemical potentials.

Knowledge on the large density region of the phase diagram can    
guide us to understand the physics of neutron stars (e.g. the existence 
of quark matter in the core of the neutron stars). 

In this summary we will study the QCD transition at non-vanishing
temperatures and/or densities. We will use lattice gauge theory
to give non-perturbative predictions. As a first step, we determine the nature
of the transition (first order, second order or analytic crossover)
and the characteristic scale of the transition (we call it 
transition temperature) at vanishing baryonic densities.
According to the detailed analyses there is no singular 
phase transition in the system, instead one is faced with an
analytic --crossover like-- transition between the phases 
dominated by quarks/gluons and that with hadrons (from now on
we call these two different forms of matter phases).
As a consequence, there is no unique transition temperature. Different 
quantities give different $T_c$ values (which are then defined as the 
most singular point of their temperature dependence).
We will determine the equation of state as a function of the 
temperature and baryonic chemical potential.

\begin{figure}
\centerline{\includegraphics*[width=7cm]{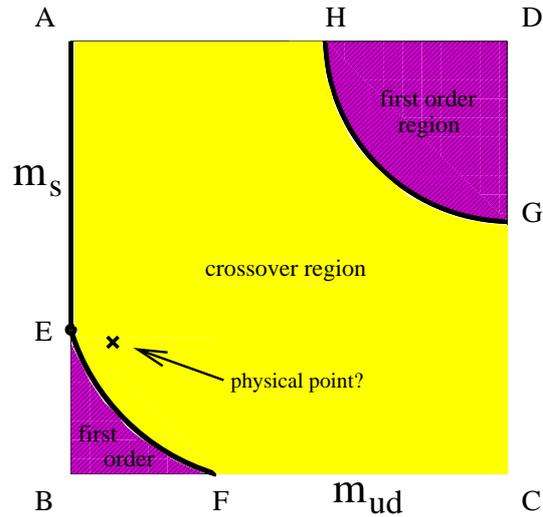}}
\caption{\small\label{fig_phase_ud}
The conjectured phase diagram of QCD on the hypothetical $m_s$--$m_{ud}$ plane 
(strange quark mass versus light --up and down-- quark mass, from
now on we use degenerate light quark masses). The middle region
corresponds to analytic crossover. In the purple (lower left and upper
right) regions one expects a first order phase transition. On the boundaries
between the first order phase transition regions and the crossover region
and along the AE line the transition is of second order.}
\end{figure}

As a second step we leave the $\mu=0$ axis and study phenomena at 
non-vanishing baryonic densities. As we will see this is quite difficult,
any method is spoiled by the sign problem, which we will discuss in detail.
In the last 25 years several results were obtained for $\mu=0$ (though
they were not extrapolated to the continuum limit). Until quite
recently there were no methods, which were able to give any information
on the non-vanishing chemical potential part of the phase diagram. In 2001
a method was suggested, with which the first informations were obtained
and several questions could be answered.
Using this --and other-- methods we determine the phase diagram for
small values of the chemical potential, we locate the critical
point of QCD and similarly to the  $\mu=0$ case we calculate the
equation of state (note, that these results are not yet in the continuum 
limit, they are obtained at relatively large lattice spacings; for the 
continuum extrapolated results larger computational resources are necessary
than available today).

\section{The phase diagram of QCD}
Before we discuss the results let us summarize the qualitative picture
of the QCD phase diagram. Figure \ref{fig_phase_ud} shows
the conjectured phase diagram of QCD as a hypothetical function of the $m_{ud}$ 
light quark mass and $m_{s}$ strange quark mass. 
In nature these quark masses are fixed and they correspond to a single
point on this phase diagram. The figure shows our expectations for the
nature of the transition. QCD is a gauge theory, which has two limiting
cases with additional symmetries. One of these limiting cases correspond
to the infinitely heavy quark masses (point D of the diagram).
This is the pure SU(3) Yang-Mills theory, which has not only the SU(3)
gauge symmetry but an additional Z(3) center symmetry, too. At high
temperatures this Z(3) symmetry is spontaneously broken. The order
parameter which belongs to the symmetry is the Polyakov loop.
The physical phenomenon, which is related to the spontaneous 
symmetry breaking is the deconfining phase transition. At high 
temperatures the confining feature of the static potential disappears.
The first lattice studies were carried out in the pure SU(2) gauge 
theory~\cite{Kuti:1980gh, McLerran:1980pk}. The transition turned out to be
a second order phase transition. Later on the increase of the computational
resources allowed to study the SU(3) Yang-Mills theory. It was
realized that in this case we are faced with a first order phase transition,
which happens around 270~MeV in physical units~\cite{Celik:1983wz,Kogut:1982rt,Gottlieb:1985ug,Brown:1988qe,Fukugita:1989yb}.

The other important limiting case corresponds to vanishing
quark masses (points A and B).
In this case the Lagrangian has an additional global symmetry, namely
chiral symmetry. Left and right handed quarks are transformed 
independently. Point A corresponds to a two flavour theory ($N_f=2$),
whereas the three flavour theory ($N_f=3)$ is represented by point B.
The chiral symmetry can be described by $U(N_f)_L \times U(N_f)_R$.
At vanishing temperature the chiral symmetry is spontaneously broken,
the corresponding Goldstone bosons are the pseudoscalar mesons
(in the $N_f=2$ case we have three pions). Since in nature the quark masses
are small but non-vanishing the chiral symmetry is only an approximative
symmetry of the theory. Thus, the masses of the pions are small but 
non-zero (though they are much smaller than the masses of other hadrons).
At high temperatures the chiral symmetry is restored. There is a phase transition between the low temperature chirally broken and the high temperature chirally symmetric phases. 
The corresponding order parameter is the chiral condensate.
For this limiting case we do not have reliable lattice results (lattice
simulations are prohibitively expensive for small quark masses, thus
to reach the zero mass limit is extremely difficult). There are model studies,
which start from the underlying symmetries of the theory. These studies
predict a second order phase transition for the $N_f=2$ case, which belongs
to the O(4) universality class. For the $N_f=3$ theory these studies
predict a first order phase transition~\cite{Pisarski:1983ms}.
For intermediate quark masses we expect an analytic 
crossover (see Figure~\ref{fig_phase_ud}). One of the most important questions
is to locate the physical point on this phase diagram, thus to determine
the nature of the T>0 QCD transition for physical quark masses.

\begin{figure}
\centerline{\includegraphics*[width=8cm]{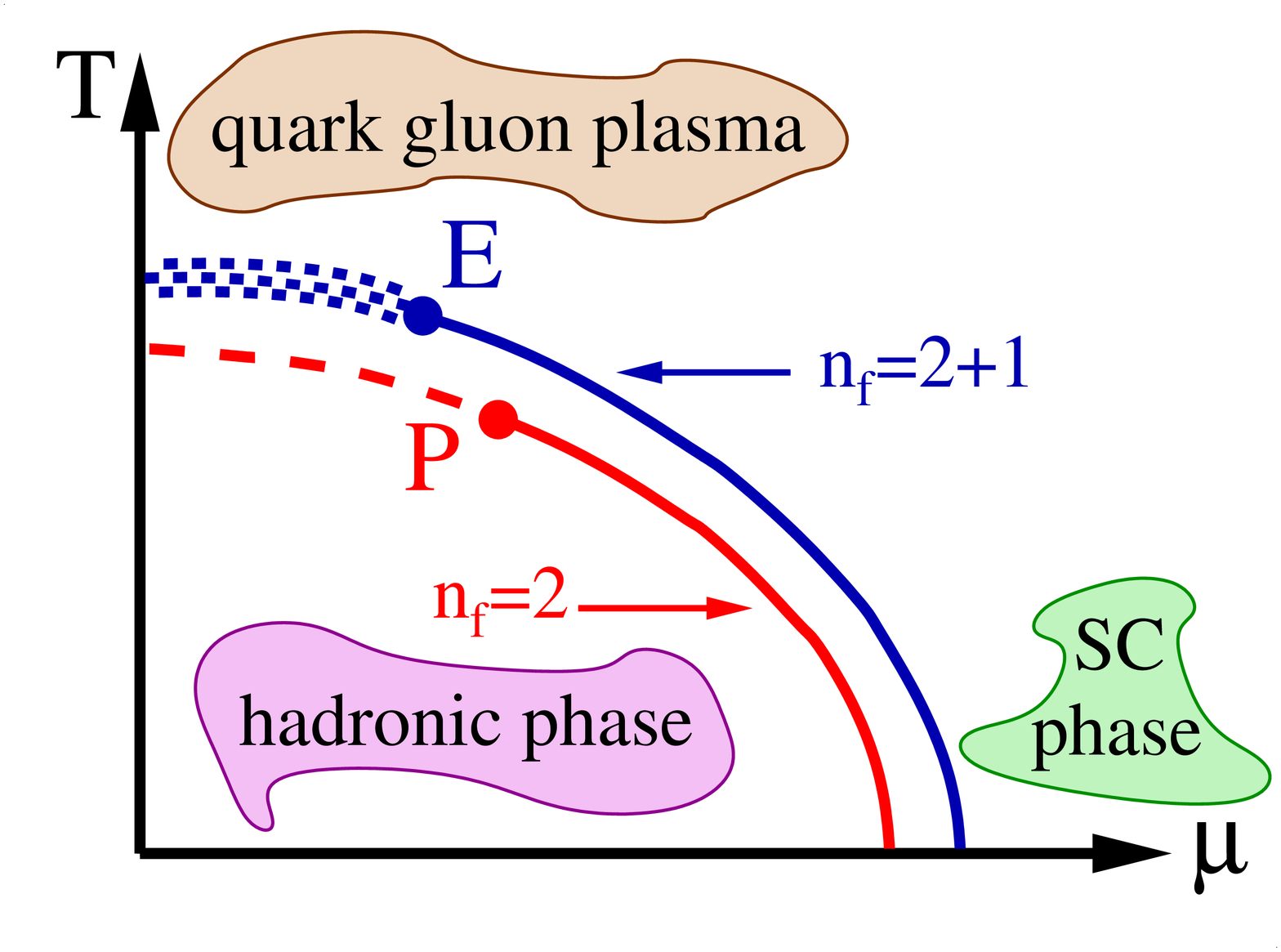}}
\caption{\small\label{fig_phase}
The most popular scenario for the $\mu$--$T$ phase diagram of QCD. For
the massless $N_f=2$ case (red curve) we find a $P$ tricritical point
between the second order (dashed line) and first order (solid lime)
regions. For physical quark masses (two light quarks and another
somewhat heavier strange quark: $N_f=2+1$, represented by the blue curves)
the crossover (dotted region) and first order phase transition (solid line)
regions are separated by a critical point $E$.
}
\end{figure}

The most popular scenario for the $\mu$--$T$ phase diagram of QCD can be seen 
on Figure \ref{fig_phase}. At T=0 and at large chemical potentials
model calculations predict a first order phase 
transition~\cite{Klevansky:1992qe}. In two flavour massless QCD there is a 
tricritical point between the second order phase transition region 
(which starts at the second order point at $\mu=0$) and the first
order phase transition region at large chemical potentials.
As we will see QCD with physical quark messes is in the crossover region, 
thus in this case we expect a critical (end)point $E$ between the crossover 
and first order phase transition regions.

A particularly interesting picture is emerging at large chemical
potentials. Due to asymptotic freedom at large densities we obtain
a system with almost non-interactive fermions. Since quarks 
attract each other, it is easy to form Cooper-pairs, which results
in a colour superconducting phase. The discussion of this interesting
phenomenon is beyond the scope of the present summary. 

The structure of the present work can be summarized as follows.
In chapter \ref{chap_lat} we summarize the necessary
techniques of lattice gauge theory. Chapter
\ref{chap_mu0} discusses the $\mu=0$ results. The nature
of the transition is determined, its characteristic scale is calculated
($T_c$) and the equation of state is given.
We discuss the $\mu\neq 0$ case in chapter \ref{chap_mu}. 
The source of the sign problem is presented and the multi-parameter
reweighting is introduced. We determine the phase diagram, the critical
point and the equation of state. 
Chapter \ref{chap_sum} summarizes the results and provides a 
detailed outlook. Based on the available techniques and computer
resources we estimate the time scales needed to reach the various milestones
of lattice QCD thermodynamics.

\chapter{QCD thermodynamics on the lattice}\label{chap_lat}
We summarize the most important ingredients of lattice QCD. Instead of 
providing a complete introduction we focus on those elements of the
theory and techniques, which are essential to lattice thermodynamics.
A detailed introduction to other fields of lattice QCD can be found in 
Ref.~\cite{Montvay:1994cy}.

Thermodynamic observables are derived from the grand canonical partition
function. The Euclidean partition function can be given by the following
functional integral:               
\be\label{part_func}
Z=\int {\cal D}U {\cal D}\bar{\psi} {\cal D}\psi e^{-S_E(U,\bar{\psi},\psi)},
\ee
here $U$ represent the gauge fields (gluons), whereas $\psi$ and $\bar{\psi}$ 
are the fermionic fields (quarks). QCD is an SU(3) gauge theory
with fermions in the fundamental representation. Thus, at various space-time points the four components of the $U$ gauge
field can be given by SU(3) matrices for all four directions. 
The fermions are represented by non-commuting Grassmann 
variables.

The Boltzmann factor is given by the Euclidean action, which is a functional
of the gauge and fermionic fields.
Equation (\ref{part_func}) contains additional parameters (though
they are not shown in the formula explicitely).
These parameters are the $\beta$ gauge coupling (it is related to
the continuum gauge coupling as $\beta=6/g^2$), the quark masses
($m_i$) and the chemical potentials ($\mu_i$). 
For simplicity equation (\ref{part_func}) describes only one
flavour. More than one flavour can be described by introducing
several $\psi_i$ fields. In nature there are six quark flavours. 
The three heaviest flavours ($c,b,t$) are much heavier than the typical
energy scales in our problem. They do not appear as initial
or final states and they can not be produced at the typical energy scales.
Their effects can be included by
a simple redefinition of the other bare parameters (for some quantities
they should be included explicitly as dynamical degrees of freedom, however,
we will not discuss such processes). 
The three other quarks are the $u,d$ and $s$ quarks. The masses of the 
$u,d$ quarks are much smaller than the typical hadronic scale, therefore
one can treat them as degenerate degrees of freedom (exact SU(2)
symmetry is assumed). This approximation is satisfactory, since 
the mass difference between the $u$ and $d$ quarks can explain
only $\approx$50\% of the mass difference between different pions.
For the remaining $\approx$50\% the electromagnetic interaction is
responsible (the up and down quarks have different electric charges). 
Including the mass differences would mean that one should include
an equally important feature of nature, namely the electromagnetic
interactions, too. This is usually far beyond the precision lattice
calculations can reach today. Assuming $m_u=m_d$ is a very good 
approximation, the obtained results are quite precise, 
uncertainty related to this choice is clearly subdominant. For the 
degenerate up and down quark mass we use the shorthand notation $m_{ud}$.
The $s$ quark is 
somewhat heavier, its mass is around the scale of the $\Lambda$ parameter
of QCD.
In typical lattice applications one uses the $m_u=m_d<m_s$ setup, which
is usually called as $N_f=2+1$ flavour QCD.

In order to give the integration measure 
(${\cal D}U {\cal D}\bar{\psi} {\cal D}\psi$)
one has to regularize the theory.
Instead of using the continuum formulation we introduce a hypercubic 
space-time lattice $\Lambda$. The fields are defined on the sites (fermions)
and on the links (gauge fields) of this lattice. It is easy to show
that this choice automatically respects gauge invariance. For a 
given site $x \in \Lambda$ four $(x;\mu)$ links can be defined
 (here $\mu$ denotes the direction of the link, $\mu=1\dots 4$).
Using this choice the integration measure is given by 
\be                                           
{\cal D}U {\cal D}\bar{\psi} {\cal D}\psi=\prod_{x\in\Lambda,\mu=1\dots 4}dU_{x;\mu}
\prod_{x \in\Lambda}d\psi_x \prod_{x\in\Lambda}d\bar{\psi}_x
\ee

With this regularization one can imagine the functional integral as a sum
of the Boltzmann factors $\exp(-E/kT)$ over all possible
$\left\{U,\psi,\bar{\psi}\right\}$ configurations (here we use the
$k=1$ convention). Thus, our system corresponds to a four-dimensional
classical statistical system. The energy functional is simply replaced by 
the Euclidean action. An important difference is that in statistical
physics the temperature is included in the Boltzmann factor, whereas
in our case it is related to the temporal extent of the lattice (it is
the inverse of it). It is easy to show that using periodic boundary
conditions for the bosonic fields and antiperiodic boundary conditions
for the fermionic fields our equation (\ref{part_func})
reproduces the statistical physics partition function.

\section{The action in lattice QCD}
The lattice regularization means that one should discretize the Euclidean 
action $S_E$. This step is not unambiguous. There are several lattice
actions, which all lead to the same continuum action. The difference 
between them is important, since these differences tell us how fast they approach the continuum result as we decrease the lattice spacing.
Calculating a given $A$ observable on the lattice of a lattice spacing
$a$, the result differs from the continuum one
\be\label{kont_korr}
\left<A\right>_a=\left<A\right>+{\cal O}(a^\eta).
\ee
The power $\eta$ depends on the way we discretized the action. The 
larger the power $\eta$ the better the action (for large $\eta$
we can obtain a result, which is quite close to the continuum one,
already at large lattice spacing). 

The most straightforward discretization is obtained by simply taking
differences at neighbouring sites to approximate derivatives.
Actions, which have better scaling behaviour (larger $\eta$ or smaller
prefactor) are called improved actions.

In the following paragraphs we summarize the most important actions.

The action $S_E$ usually can be written as a sum $S_E=S_g+S_f$,
where $S_g$ is the gauge action (it depends only on the gauge fields)
and $S_f$ is the fermionic action (it depends both on the gauge and
fermionic fields).

The simplest gauge action is the Wilson gauge action
which is the sum of the 
\be
U_P(x;\mu\nu)=U_{x;\mu}U_{x+a\hat{\mu};\nu}U^{\dagger}_{x+a\hat{\nu};\mu}U^{\dagger}_{x;\nu}
\ee
plaquettes. Here $\hat{\mu}$ denotes the unit vector in
the $\mu$ direction. The Wilson action reads:
\be\label{hatas_wilson_gauge}
S_{g,\rm{ Wilson}}=-\beta\left(\frac{1}{3}\sum_{x,\mu<\nu}\Re\Tr U_P(x;\mu\nu)-1\right)
\ee
This action is the simplest real, gauge invariant expression, which can
be constructed using the gauge fields. One can show, that in the continuum
limit the above expression leads to the usual Yang-Mills gauge action.

One can improve the action by adding other gauge invariant terms.
The simplest such improvement term is provided by the $2\times 1$
rectangles, for which  --analogously to the plaquette term-- we multiply 
the SU(3) link matrices around the rectangle. Denoting this term
by $U_{2\times 1}(x;\mu\nu)$ one obtains the following action 
\be\label{hatas_sym}
S_{g}=-\frac{\beta}{3}\left(c_0\sum_{x,\mu<\nu}\Re\Tr U_P(x;\mu\nu)+
c_1\sum_{x,\mu\neq\nu}\Re\Tr U_{2\times 1}(x;\mu\nu)\right)
\ee
It can be shown that this choice improves the scaling. On the 
tree level the condition $c_0+8c_1=1$ should be fulfilled and 
$c_1=-1/12$. This is the (tree level improved) Symanzik gauge action.
Other improvements use also chair-like closed paths and non-perturbative
coefficients. Note, however that the tree level improvement is usually
enough for thermodynamic studies, the main source of difficulties is
in the fermionic part (further improvements in the gauge sector can be 
considered as a sort of ``over-killing''). 

Discretizing the fermionic fields is more difficult than discretizing
gauge fields.  
The naive discretization leads to the following action
\be
S_{f,\mbox{\small naive}}=\sum_x\left[am\bar{\psi}\psi+
\frac{1}{2}\sum_{\mu=1\dots 4}
\left(\bar{\psi}_xU_{x;\mu}\gamma_\mu\psi_{x+a\hat{\mu}}-
\bar{\psi}_xU^\dagger_{x-a\hat{\mu};\mu}\gamma_\mu\psi_{x-a\hat{\mu}}
\right)\right].
\ee
in the free case ($U=1$) the propagator has 16 poles in the Brillouin 
zone (we expected only one). Thus, contrary to the continuum case our 
lattice action describes 16 degenerate fermions instead of 1 fermion.

There are several ways to resolve this problem. The two most
popular solutions are the Wilson and the Kogut-Susskind regularizations.
The problem is related to the fact that the continuum fermion action
contains only first derivatives. The basic idea of the Wilson fix
is to add a second derivative term --Wilson term-- to the action: 
$a\bar{\psi}\partial_\mu\partial_\mu\psi$.
This term vanishes in the continuum limit. For non-vanishing lattice
spacings the Wilson term increases the masses of the 15 non-physical 
modes so that they are at the cutoff scale ($1/a$).
As we approach the continuum limit these 15 particles decouple.
Generally, one can use a Wilson term with an arbitrary coefficient $r$.
The usual choice is $r=1$. In this case the action reads
\be\label{hatas_wil}
S_{f,\rm{Wilson}}=\sum_x\left[\bar{\psi}\psi+
\kappa\sum_{\mu=1\dots 4}
\left(\bar{\psi}_xU_{x;\mu}(1+\gamma_\mu)\psi_{x+a\hat{\mu}}+
\bar{\psi}_xU^\dagger_{x-a\hat{\mu};\mu}(1-\gamma_\mu)\psi_{x-a\hat{\mu}}
\right)\right].
\ee
Here the fields are rescaled appropriately. The disadvantage of  
Wilson fermions is the loss of chiral symmetry for vanishing
quark masses. This symmetry is restored only in the continuum limit.
The quark mass receives an additive renormalization
and the asymptotic scaling (c.f. equation (\ref{kont_korr})) is linear
in $a$.

Kogut and Susskind introduced another formalism, namely the {\it staggered} 
fermions. The spinor components of the fermionic field are distributed
among the corners of a $2^4$ hypercube. This leads to a diagonal
expression in the spinor index. By using only 1 out these 4 diagonal 
components one can reduce the number of degrees of freedom by a factor of 4.
This action describes 16/4=4 fermions of the same mass.
The action can be written as
\be\label{hatas_stag}
S_{f,\rm{staggered}}=\sum_x\left[am\bar{\chi}\chi+
\frac{1}{2}\sum_{\mu=1\dots 4}\alpha_{x;\mu}
\left(\bar{\chi}_xU_{x;\mu}\chi_{x+a\hat{\mu}}-
\bar{\chi}_xU^\dagger_{x-a\hat{\mu};\mu}\chi_{x-a\hat{\mu}}
\right)\right],
\ee
where $\alpha_{x;\mu}=(-1)^{x_1+\dots+x_{\mu-1}}$. 
Contrary to the naive or Wilson fermion formulations the $\chi$ field has
only one spin component. For simplicity we use the Greek letter $\psi$
also for staggered fermions. The most important advantage of the 
staggered formalism is, that the action has a $U(1)_L\times U(1)_R$ symmetry 
(which is a remnant of the original chiral symmetry). Due to this symmetry
there is no additive mass renormalization. The asymptotic scaling
is better than for Wilson fermions, it is proportional to $a^2$. An
additional advantage is of computational nature. Since we do not have
Dirac indices the computations are faster. The most important
disadvantage of the staggered fermions is the fourfold degeneracy
of the fermions. Later we discuss the technique, which allows one to use less than four fermions.

In principle, there are several other fermion formulations. Note, however,
that the Nielsen-Ninomiya no-go theorem excludes any continuum-like 
fermion formulations~\cite{Nielsen:1980rz,Nielsen:1981xu}.
According to this theorem one can not have a local fermion formulation
with a proper continuum limit for one flavour, which respects chiral 
symmetry. Recently, it was possible to construct a fermion formulation, which
fulfills the above conditions and respects a modified (lattice-like) 
chiral symmetry~\cite{Neuberger:1997fp,hep-lat/9802011}.
These fermions are called chiral lattice fermions. They represent
a mathematically elegant formulation with many important features, which
make lattice calculations unambiguous and straightforward. Unfortunately, 
they are extremely CPU demanding, they require approximately two orders
of magnitude more computer time than the more traditional Wilson 
or staggered fermions. The first steps in order to develop reliable     
algorithms have been made and exploratory studies have been carried out
on relatively small lattices~\cite{Fodor:2003bh,Cundy:2004pz,DeGrand:2004nq,Egri:2005cx,Lang:2005jz,Fukaya:2006vs}. 
We expect that in the near future important results will be obtained by
using chiral lattice fermions.                   

Both eq. (\ref{hatas_wil}) and (\ref{hatas_stag}) are bilinear in 
the fermionic fields (it is true for other actions, too): 
\be\label{hatas_bilin}
S_f=\sum_{x,y}\bar{\psi}_x M_{xy}(U) \psi_y,
\ee
here the specific form of the matrix $M$ can be derived from eq. 
(\ref{hatas_wil}) and (\ref{hatas_stag}). 
Since the fermionic fields are represented by Grassmann variables
it is difficult to treat them numerically. We do not know
any technique, which can be used as effectively as the bosonic 
importance sampling methods. 
Fortunately, the fermionic integrals can be evaluated exactly. Using    
the known Grassmann integration rules one obtains:                         
\be
\int {\cal D}\bar{\psi} {\cal D}\psi e^{-S_f}=\det M(U),
\ee
Thus the partition function (\ref{part_func}) can be written as follows:
\be\label{part_func_det}
Z=\int{\cal D}U \det M(U) e^{-S_g(U)}=
\int{\cal D}U e^{-\left\{S_g(U)-\ln\det M\right\}}.
\ee
This simple step resulted in an effective theory, which contains only  
bosonic fields. The action reads: $S_{\rm{eff.}}=S_g-\ln\det M$.
Unfortunately this action is non-local. Due to the fermionic
determinant fields at arbitrary distances interact with each other
(the original action $S_E=S_g+S_f$ is local in the field variables).
This non-locality is the most important source of difficulties. It is 
much more demanding to study full QCD (with dynamical fermions) than
pure SU(3) gauge theory.

For the 2+1 flavour theory we need different fermionic fields. Each
fermionic integration results in a fermionic determinant. These determinants
depend explicitely on the quark masses: 
\be
Z(m_1,m_2,\dots m_{N_f})=\int{\cal D}U 
\det M(m_1;U)\det M(m_2;U)\dots \det M(m_{N_f};U)
e^{-S_g(U)}.
\ee
For Wilson fermions the above formula can be used directly for 2+1 flavours.
For staggered fermions another trick is needed. Since one $\psi$ field
describes four fermions in the staggered formalism one uses the fourth
root trick. The reason for that is quite simple. For more than one field 
one uses powers of the determinant. Analogously for one flavour
one uses the fourth root of the determinant.
We expect that the partition function
\be\label{frac_pow}
Z(N_f)=\int{\cal D}U 
\left[\det M(U)\right]^{N_f/4}
e^{-S_g(U)}
\ee
describes $N_f$ flavours in the staggered theory. Note, however, that the    
locality of such a model is not obvious. As we saw the partition function
\ref{part_func_det}) was the result of a local theory, 
which is not necessarily the case for (\ref{frac_pow}). This question
is still debated in the literature (see e.g. ~\cite{Bernard:2006ee}).       
Though the theoretical picture is not clear, all numerical results show    
that the fourth root trick most probably leads to a proper description
of one flavour of QCD. 

In the rest of this work we will deal with staggered theory, only. Since
staggered fermions are computationally less demanding than other fermion
formulations, the vast majority of the results in the literature are 
obtained by using staggered fermions. Another reason why the staggered 
fermions are so popular for thermodynamic studies is related to the fact 
that staggered fermions are invariant under (reduced) chiral symmetry,
which might play an important role for questions such as chiral 
symmetry restoration (at the finite temperature QCD transition).

In numerical simulations we use finite size lattices of $N_s^3N_t$. The
three spatial sizes are usually the same ($N_s$), they give the spacial 
volume of the system, whereas the temporal        
extension in Euclidean space-time is directly related to the temperature:  
\begin{align}\label{VT_def}
V=(N_sa)^3, && T=\frac{1}{N_ta}.
\end{align}  
Lattices with $N_t \ge N_s$ are called ``zero temperature'' lattices, and  
lattices with $N_t \ll N_s$ are called ``non-zero temperature'' lattices. 
In thermodynamic studies a small temperature region around the transition 
temperature
is the main focus of the analyses (an exception is the determination of the   
equation of state, which can be studied at much higher temperatures, too).
According to $T=1/(N_t a)$ one can fix the temperature by using smaller and
smaller lattice spacings and larger and larger $N_t$ temporal extensions.
Thus, the resolution of an analysis is usually characterized by the temporal
extension. In the literature one finds typically $N_t$ values of 4, 6, 8 
and 10, which correspond to lattice spacings (at and around $T_c$) 
of approximately $a=$0.3, 0.2, 0.15 and
0.12~fm, respectively. We give here only approximative values and it 
is impossible to give precise values for the lattice spacings, particularly
for these relatively coarse lattices. The reason for this ``no-go''
observation can be summarized as follows. QCD predicts only dimensionless
combinations of observables. These combinations are only approximated on the
lattices, they have $a^\eta$ scaling corrections, which vanish as we 
approach the continuum limit. Since different combinations have different
scaling corrections, the lattice spacing can not be given unambiguously.

The lattice spacing defines a cutoff $\Lambda \sim 1/a$. One of the most
important source of difficulties is related to the fact that we want to  
ensure that all masses we study are smaller than the cutoff, whereas
all Compton wave-lengths (which are proportional to the inverse masses) 
are much smaller than the size of the lattice
(otherwise one has large finite volume effects). 
Since in QCD we have masses, which are quite different (the mass ratio of 
the nucleon and pion is about seven) we are faced with a multi-scale problem.
This results in a quite severe lower bound on $N_t$. In earlier works the   
only way to deal with such a multi-scale problem was to ignore that in nature we have such a phenomenon. People
used a much smaller nucleon to pion mass ratio than the physical one, 
thus they used quite heavy      
quark masses, which resulted in heavy pion masses. Since the transition
is related to the chiral features of the theory (we speak about chiral
transition) this approximation is clearly non-physical. Another reason to
use larger than physical pion masses is of algorithmic nature. 

\section{Correlators}
The expectation value of an observable $O$ can be given as a functional
integral over the $U$ and $\psi_i, \bar{\psi}_i$ fields:
\be
\left< O\right>=\frac{1}{Z}
\int {\cal D}U {\cal D}\bar{\psi} {\cal D}\psi O\left[ U,\bar{\psi},\psi\right]
e^{-S_E(U,\bar{\psi},\psi)}.
\ee
Quantum field theories can be defined by operators. Formally, defining
the theory by $\hat{O}$ operators or defining it by the above functional
integral are identical. The results in the two formalisms are the same 
$\langle\hat{O}\rangle=\langle O\rangle$.

At zero temperature typical choices of operators are n-point functions
of the fields. Particularly important n-point functions are the two-point
functions (propagators). E.g. for pions the interpolating operator $\hat{O}$
can be given as $\hat{O}=\hat{\bar{\psi}}_u\gamma_5\hat{\psi}_d$. The u,d
indices denote up and down quarks. The (Euclidean) time evolution of the 
$\hat{O}$ operator is given by the Hamiltonian $\hat{H}$ by the usual way:
$\hat{O}(t)=e^{t \hat{H}}\hat{O}(0)e^{-t \hat{H}}$. Thus, inserting a complete
set of energy eigenstates $\left|n\right>$, the two-point
function can be written as
\be
\left<0\right|\hat{O}(t)\hat{\bar{O}}(0)\left|0\right>=
\sum_n\left|\left<0\right|\hat{O}\left|n\right>\right|^2e^{-(E_n-E_0)t}.
\ee
For large $t$ values the above function is dominated by the term with
the smallest $E_n$ (assuming that its prefactor is non-vanishing). Thus,
for a given chanel the exponential decay of the two-point function of
the operator $\hat{O}$ (with the proper quantum numbers) provides us
with the smallest energy (mass). The correlation length $\xi$ is proportional
to the inverse of the mass. In lattice units $\xi=1/(ma)$, where $\xi$
denotes this dimensionless correlation length (in lattice units).
Using the appropriate operators one can determine the masses of hadrons 
on the lattice.

\section{Continuum limit}
The final goal of lattice QCD is to give physical answers in the continuum
limit. Results at various lattice spacings `$a$' are considered as 
intermediate steps. Since the regularization (lattice) is inherently
related to the non-vanishing lattice spacing it is not possible
to carry out calculations already in the continuum limit in our   
lattice framework. The continuum physics appears as a limiting result.
Obviously, the $a\rightarrow 0$ limit should be carried out according 
to eq. (\ref{kont_korr}). During this procedure the physical
observables, more precisely their dimensionless combinations should
converge to finite values. On the way to the continuum limit one should  
tune the parameters of the Lagrangian as a function of the lattice spacing.  
The renormalization group equations tell us how the parameters
of the Lagrangian depend on the lattice spacing. For small gauge coupling
(thus, for large cutoff or close to the continuum limit) the 
perturbative form of the renormalization group equations can be 
used. For somewhat larger gauge couplings one should use non-perturbative 
relationships.

As we have seen, the correlation length $\xi$ of a hadronic interpolating 
operator is proportional to the inverse mass of the hadron $\xi=1/(ma)$.
In order to reach the continuum limit the lattice spacing in
physical unit should approach zero: $a\rightarrow 0$. Since the hadron  
mass is a finite value in the same physical units, the correlation length
$\xi$ should diverge. Thus the continuum limit of lattice QCD is 
analogous to the critical point of a statistical physics system (which is
also characterized by a diverging correlation length). The Kadanoff-Wilson   
renormalization group of statistical physics can be used for lattice QCD,
too. The renormalization group transformation tells us how to change 
the parameters of the lattice action (Lagrangian) in order to obtain the
same large distance behaviour (the small distance behaviour is not 
important for us, it merely reflects our discretization process).
This was the original idea of Wilson: one has to carry out a few 
renormalization group transformation with increasing `$a$' and describe QCD 
by an action which is ``good enough'' at these large lattice spacings. After
these steps the action can be used for a numerical solution. Unfortunately,
the renormalization group transformation procedure results in an action,
which is far too complicated to be used~\footnote{Note, that actions
with very good scaling properties can be constructed by using the  
renormalization group transformations and reducing the number of 
terms in the action~\cite{Hasenfratz:1993sp,DeGrand:1995ji}}.
Usually, when one changes the lattice spacing (e.g. all the way to the
continuum limit) the form of the action remains the same, only its
parameters are changed. The way the parameters change is called 
renormalization group flow or line of constant physics (LCP). 
It can be obtained by choosing a few dimensionless
combinations of observables and demanding that their values remain
the same ``predefined'' value 
as we change the lattice spacing. Using different sets
of observables result in different LCPs; however, these different
LCPs merge when we approach the continuum limit. The LCPs are  
usually determined by non-perturbative techniques. The simplest 
procedure is to measure the necessary dimensionless combinations
at various parameters of the action (bare parameters) and interpolate 
to those bare parameters, at which the dimensionless combinations take their
predefined value. A few iterative steps are usually enough to 
reach the necessary accuracy.

\section{Algorithms}\label{sect_alg}
The determination of expectation values of various observables is the
most important goal of lattice gauge theory. To that end one has to   
evaluate multi-dimensional integrals. Since the dimension can be as high
as $10^9$, which is the state of the art these days, it is a non-trivial
numerical work. The systematic mapping of such a multi-dimensional 
function is clearly impractical. The only known method to handle the problem
is based on Monte-Carlo techniques. We chose some configurations randomly
and calculate the observables on these fields. The seemingly simplest method
is to generate configurations with a uniform distribution (uniform in the
integration measure). This choice is quite inefficient, since the Boltzmann
factor exponentially suppresses most of these terms. Only a few     
configurations would give a sizable contribution, and using a uniform
distribution the probability of finding these configurations is extremely
small. The most efficient technique, which is available today, is based
on importance sampling. The configurations are not uniformly generated, 
instead one uses a distribution $p\propto e^{-S_E}$ for the generation.
Thus, those configurations, which contribute with a large Boltzmann weight    
are chosen more probably ($e^{-S_E}$ is large) than those, which contribute
with a small Boltzmann factor ($e^{-S_E}$ is small). For the case of QCD
the Euclidean action $S_E$ contains the bosonic and the fermionic fields,
which are represented by Grassmann variables. There is no known importance
sampling based procedure for Grassmann variables. As we discussed already one has to evaluate
the fermionic integral explicitely. This integration leads to               
(\ref{part_func_det}). Thus, a procedure based on importance sampling
uses the distribution $p(U)\propto \det M(U) e^{-S_g(U)}$ for generating the 
configurations. Let us assume that we have an infinitely large ensemble of 
configurations, given by the above distribution. The expectation
value of an observable $O$ can be calculated as 
\be
\left<O\right>=\lim_{N\to\infty} \frac{1}{N}\sum_{i=1}^{N}O(U_i).
\ee
In practice our ensemble is always finite, thus the $N\to\infty$ limit 
can not be achieved, $N$ remains finite. The lack of this infinite limit       
results in statistical uncertainties. One standard technique to   
determine the statistical errors is the {\it jackknife} 
method~\cite{Montvay:1994cy}.

A crucial ingredient of any method based on importance sampling is
the positivity of $\det M(U)$ (it should be a positive real number) for
all possible $U$ gauge configurations. Otherwise the expression       
$\det M(U) e^{-S_g(U)}$ can not be interpreted as a probability. In order to 
illustrate the importance of this condition we shortly summarize the simplest
importance sampling based Monte-Carlo method, the Metropolis algorithm. 
All known techniques  
represent a Markov chain, in which the individual configurations of the ensemble 
are obtained from the previous configurations. The Metropolis 
algorithm consists of two steps. In the first step we change     
the configuration $U$ randomly and obtain a new configuration $U'$.
Obviously, this configuration has a different Boltzmann factor. In the
second step we take account for this difference and accept this new $U'$    
configuration, as a member of our ensemble, with the probability 
\be
P(U'\leftarrow U)=\min\left[
1,e^{-\Delta S_g}\frac{\det M(U')}{\det M(U)}\right],
\ee
where $\Delta S_g=S_g(U')-S_g(U)$. If the configuration $U'$ was not 
accepted (the probability of this case is $1-P(U'\leftarrow U)$), we
keep the original configuration $U$. It can be easily seen that       
$0\le P(U'\leftarrow U) \le 1$ is fulfilled only if $\det M$ is 
positive and real.

Interestingly, this non-trivial condition is fulfilled, it is a 
consequence of the $\gamma_5$ hermiticity of the $M$ fermion matrix
(or in other words Dirac operator)
\be
M^\dagger=\gamma_5 M \gamma_5.
\ee
This equality can be easily checked both in the continuum formulation 
(\ref{hatas_wil}) or for the lattice formulations (\ref{hatas_stag}).

If $v$ is an eigenvector of $M^\dagger$ with eigenvalue $\lambda$ then
$\lambda v=M^\dagger v=\gamma_5 M \gamma_5 v$. This gives
$\lambda \gamma_5 v=M \gamma_5 v$. Thus, $\lambda$ is an eigenvalue
of $M$. It can be similarly shown that an eigenvalue of $M$ is 
also an eigenvalue of $M^\dagger$. Thus, $M$ and $M^\dagger$ have the same
eigenvalues. As a result, these eigenvalues are either real or appear
in conjugate pairs. As a consequence $\det M$ is always real. 
In the continuum theory
and for staggered fermions the massless Dirac operator has only purely
imaginary eigenvalues, thus the real eigenvalues of the massive Dirac 
operator are always positive (they are equal to the quark mass).
In these cases the condition $\det M\ge 0$ is fulfilled and the equality 
appears only for vanishing quark masses. For Wilson fermions negative
eigenvalues might appear. Note; however, that these eigenvalues 
disappear as we approach the continuum limit. For Wilson fermions the        
most straightforward method is to use two degenerate quarks,
thus $(\det M)^2$ can be used. Alternatively one can take one flavour
with $\left| \det M \right|$. Since in the continuum limit 
$\det M$ is a positive real number, taking the absolute value does not    
influence the continuum limit.

It is important to note already at this stage that for non-vanishing
baryonic chemical potentials the $\gamma_5$ hermiticity of the Dirac
operator is not fulfilled, thus the determinant is not necessarily 
a real positive number. The partition function itself will be always
real, thus one can use the real part of the integrand. Note, however,
that the real part of the determinant can be positive or negative. In the 
sum large cancellations appear between the terms with different signs. 
This is the sign problem, which makes studies at non-vanishing chemical
potential extremely difficult. As a consequence importance sampling such as the Metropolis algorithm does not work.

The Metropolis algorithm generates the configurations according to the
proper distribution. Unfortunately, it is a quite inefficient algorithm.   
There are two reasons for that. First of all, one has to calculate the       
determinant of the Dirac operator in each step, which is quite CPU demanding,
it scales with the third power of $N_s^3 N_t$. Secondly, the subsequently
generated 
configurations in the Markov chain are not independent of each other 
(for rejected configurations they are even
identical). As it turns out the autocorrelation is huge.

There are several algorithms on the market, which are much more
efficient. The most widely used method is the so-called Hybrid Monte-Carlo 
(HMC) algorithm~\cite{Polonyi:1983tm,Scalettar:1986uy,Duane:1987de}.
We shortly summarize the basic ideas to this technique. The determinant 
of any positive definite, hermitian $H$ can be written as a functional 
integral over bosonic fields
\be\label{pseudoferm}
\det H=\frac{\int{\cal D}\Phi^\dagger{\cal D}\Phi 
e^{-\Phi^\dagger H^{-1} \Phi}}{\int{\cal D}\Phi^\dagger{\cal D}\Phi e^{-\Phi^\dagger \Phi}}.
\ee
(On a finite lattice the integral is a large --but finite--  dimensional 
integral.)

Since the fermion matrix $M$ is usually not hermitian we usually
use $H=M^\dagger M$. This choice describes two fermions/flavours in the 
continuum or in the Wilson formalism (or 8 fermions in the staggered 
formalism, for staggered fermions one uses the word taste instead 
of flavour). In order to describe $N_f$ fermion flavors (tastes)  
$H=(M^\dagger M)^{N_f/2}$ is used (in the staggered case  
$H=(M^\dagger M)^{N_f/8}$ is the proper choice). Note, that these steps
result in several problems, which we will discuss later.

The partition function for two degenerate quarks reeds
\be\label{part_func_pseudoferm}
Z=C\int {\cal D}U {\cal D}\Phi^\dagger{\cal D}\Phi 
e^{-S_g(U)-\Phi^\dagger \left(M^\dagger M \right)^{-1} \Phi},
\ee
where the denominator of  (\ref{pseudoferm}), which gives only an  
irrelevant prefactor to $Z$, is denoted by $1/C$.
Let us assign to each lattice link a traceless anti-hermitian matrix 
$\Pi_{x\mu}$, for which $\Pi^2/2=\sum_{x;\mu}\left|\Pi_{x\mu}\right|^2/2$.
Multiplying $Z$ by the constant 
$1/C'=\int {\cal D} \Pi \exp \left(-\Pi^2/2\right)$
we obtain
\be\label{HMC}
Z=C'C\int {\cal D}\Pi{\cal D}U {\cal D}\Phi^\dagger{\cal D}\Phi 
e^{-\Pi^2/2-S_g(U)-\Phi^\dagger \left(M^\dagger M \right)^{-1} \Phi}.
\ee
For fixed $\Phi^\dagger, \Phi$ one can define a function
\be
{\cal H}(U,\Pi)=\Pi^2/2+S_g(U)+\Phi^\dagger \left(M(U)^\dagger M(U) \right)^{-1} \Phi,
\ee
which depends on the $U_{x\mu}$ and $\Pi_{x\mu}$ matrices (here $x\mu$
parameterizes the links). We can consider $H$ as the Hamiltonian of  
a classical many-particle system with general coordinates of $U_{x\mu}$
and general momenta of $\Pi_{x\mu}$. It is possible to solve the 
canonical equations of motions in a fictious time $t$. Along such
trajectories the Hamiltonian ${\cal H}$ is constant. Thus, for the     
$U$ and $P$ fields we can introduce a Metropolis step (thus a 
new $U'$ and $P'$) for which the integrand of (\ref{HMC}) does not
change and the acceptance probability is 1. The update of the 
$\Phi^\dagger, \Phi$ fields is done by a global heatbath. 
The calculation of the trajectories are done numerically, thus the 
Hamiltonian is not conserved exactly. It can be shown that for the  
{\it leapfrog} integration (which is the most common choice in the
literature) the change in the Hamiltonian is proportional to 
the integration step-size squared:
$\Delta {\cal H} \propto \varepsilon^2$. In order to have an exact 
algorithm one has to carry out at the end of each trajectory 
an additional Metropolis accept/reject step. 
This concludes the necessary steps of a Hybrid Monte-Carlo algorithm,  
which we summarize here. 
\begin{enumerate}
\item For a fixed gauge field $U$ we generate $\Pi$, $\Phi^\dagger$ and 
$\Phi$ configurations. The generation is done via a global heatbath 
according to the integrand. 
\item The canonical equation of motions are integrated numerically
from $t=0$ to $T$ using a step-size of $\varepsilon$. The usual choice is
$T=1$.
\item The configuration $U'$ at the end of the trajectory is accepted with
probability of 
\be
P(U'\leftarrow U)=\min\left[
1,e^{-\Delta {\cal H}}\right].
\ee
$\Pi'$, $\Phi^\dagger$ and 
$\Phi$ are not needed any more, for the next trajectory they will 
be regenerated as discussed in our first step. 
\end{enumerate}
It can be proven, that repeating this procedure gives the proper 
distribution for the gauge configurations. The most demanding
calculation numerically is the integration of the second step and 
the calculation of the term $\left(M^\dagger M \right)^{-1} \Phi$
in the third step. An identical question is to solve the linear equation of
\be
\Phi=\left(M^\dagger M \right)\chi.
\ee
The standard procedure is the conjugate gradient method. Since the 
matrix is sparse this method gives 
the solution, of the necessary precision, in $c\cdot L_s^3L_t$ steps. 
The coefficient $c$ is proportional to the condition number
of the $M$ matrix. The smallest eigenvalue of the matrix $M$ is the quark 
mass (for the Wilson formalism, even smaller eigenvalues can appear).
The largest eigenvalue is a quark mass independent constant. Thus, the
time needed for the computation is inversely proportional to the quark mass.
This is the reason for the increase of the CPU costs for small quark 
masses, which makes calculations in lattice QCD with physical quark masses
quite challenging.

As we discussed, for one flavour or taste fractional powers of the 
expression $\left(M^\dagger M \right)$ should be taken. In this case 
the standard conjugate gradient method can not be applied:
$\left(M^\dagger M \right)^{-N_f/2} \Phi$ (for the staggered  
formalism the power $-N_f/8$ should be used). It can be shown that 
in the second step of the algorithm only integer powers of the fermion
matrix is needed, and the inversions can be carried out. In the third
step, however, the fractional power can not be avoided. Until recently
no efficient method was known to treat fractional powers, the most widely
used method, the $R$ algorithm~\cite{Gottlieb:1987mq} simply omitted the 
third step. For small enough $\varepsilon^2$ the change in the Hamiltonian
was small, too: $\Delta {\cal H} \propto \varepsilon^2$. The method was 
not exact. In order to produce unambiguous results one had to carry out
a $\varepsilon \to 0$ extrapolation, which was usually omitted. 

Recently, a new method appeared in the literature, which solved this
problem. In this rational Hybrid Monte-Carlo (RHMC) algorithm the fractional
powers are approximated by rational functions. Using 10-15 orders one
can~\cite{Clark:2006wp} approximate the fractional powers upto 
machine precision. Using this technique all three steps of the Hybrid
Monte-Carlo method can be done for arbitrary $N_f$ exactly. 
Interestingly enough, the exact rational Hybrid Monte-Carlo algorithm turned 
out to be faster than the non-exact $R$ algorithm.

\chapter{Results at zero chemical potential}\label{chap_mu0}
We start the review of recent results with the $\mu=0$ case. Results
on the order of the transition, the absolute temperature of the transition
and the equation of state will be discussed.

All thermodynamics studies are based on two main steps. The observables
relevant for locating and describing the transition are determined on
high temperature ($N_s \gg N_t$) lattices. Thus, $T>0$ simulations is 
one of the necessary ingredients.

In order to set the parameters of the action and to give temperatures in
physical values (in MeV), some observables (as many of them as many parameters
the action has) have to be compared to their experimental values.
The parameters of the action have to be tuned so that these selected 
observables agree with their experimental values. Since sufficiently
high precision experimental values, such as hadron masses, are currently
only available at zero temperature,
this step can only be completed via $T=0$ simulations. Since the parameters
of the action, which are then used for the $T>0$ simulations, 
are set in this step, it is useful to start with the $T=0$ simulations 
and then proceed with the $T>0$ ones.

\section{Choice of the action}
In chapter \ref{chap_lat} we have seen that the choice of the lattice
action has a significant impact on the continuum extrapolation. On the 
one hand, an improved
action can make it possible to do a reliable extrapolation from larger
lattice spacings than with an unimproved action. 
On the other hand the computational needs of improved actions are often 
much higher than in the unimproved case. In the following we review the 
actions used by different collaborations in large scale lattice 
thermodynamics calculations. 

In the gauge sector typically 
the (\ref{hatas_sym}) Symanzik improved action is
used either with tree level coefficients or with tadpole improvement.
This improves the scaling of the action significantly compared to the
unimproved Wilson action at an acceptable cost.

In the fermionic sector upto now 
all large scale thermodynamics studies were 
carried out with staggered fermions. The main reasons why most collaborations
take this choice is the computational efficiency and the remnant 
chiral symmetry of staggered fermions. The MILC collaboration uses ASQTAD fermions,
the RBC-Bielefeld collaboration uses p4 improved 
fermions and the Wuppertal-Budapest group stout improved fermions.
The two former are described in detail in Ref.~\cite{Heller:1999xz}
while the latter was originally introduced in~\cite{Morningstar:2003gk}
and the used parameters can be found in Ref.~\cite{Aoki:2005vt}.

\begin{figure}
\centerline{\includegraphics*[width=8cm]{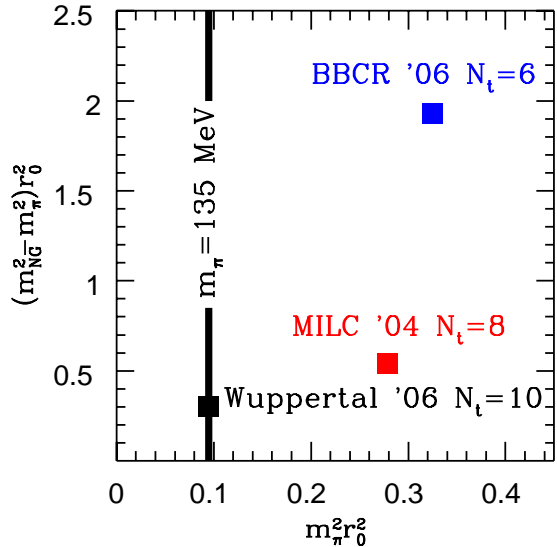}}
\caption{\small\label{fig_stout}
Taste symmetry violation of three different lattice actions:
ASQTAD improved action used by the MILC collaboration~\cite{Bernard:2005mf},
p4 action used by the RBC-Bielefeld collaboration~\cite{Cheng:2006qk} 
and the stout improved action used by the Wuppertal-Budapest 
group~\cite{Aoki:2006we}.
The taste symmetry violation is characterized by the difference
of the squares of the two lightest pions. All quantities are normalized
by the $r_0$ Sommer scale. The vertical line indicates the physical pion mass.
}
\end{figure}

Free staggered fermions describe four degenerate quark flavors. In the
interacting case, however, due to taste symmetry violation the quark
masses and the corresponding pseudoscalar masses will only become
degenerate in the continuum limit. This feature is also present in the 
2+1 flavor theory obtained via the rooting trick.
At the lattice spacings typically
used for thermodynamics studies, the second lightest pseudoscalar mass
can easily be three-four times heavier than the lightest one. Since the
order of the transition depends on the number of quark flavors, it is 
desirable to use an action where taste symmetry violation is significantly
reduced. Figure~\ref{fig_stout} shows the taste symmetry violation for 
the three actions discussed above. We can see that stout smearing is
the most effective in reducing taste symmetry violation.

\section{T=0 simulations}\label{sect_T0}
\subsection{Determination of the LCP}

In lattice calculations of QCD thermodynamics we usually determine
some observables at several different temperatures. Since the temperature
is inversely proportional to the temporal extent of the lattice:
$T=1/(N_ta)$, there are two ways  to change the temperature. One can either
change $N_t$ or the lattice spacing. Since $N_t$ is an integer, the first
possibility gives reach only for a discrete set of temperatures. Therefore
the temperature is usually tuned by changing the lattice spacing at fixed
$N_t$. This means that, as discussed in chapter \ref{chap_lat}, 
while changing the lattice spacing we have to properly tune the parameters
of the action to stay on the Line of Constant Physics (LCP).

Since the action has three parameters ($\beta$ and the quark masses), we have 
to choose three physical quantities. Usually one of these quantities
is used to set the physical scale while two independent 
dimensionless ratios of the three quantities defines the LCP. 
We have to choose such quantities whose experimental values are well known.
Since according to chiral perturbation theory the pseudoscalar meson masses
($m_{PS}$) are directly connected to quark masses ($m^2_{PS} \propto m_q$), 
they are good candidates to set the quark mass parameters. In case of
2+1 flavors this means the masses of pions ($m_\pi$) and kaons ($m_K$).
For the third quantity there are several possibilities. It is useful to
choose an observable which has a weak quark mass dependence.

Up to very recently the most common way to set the physical scale
was via the static quark-antiquark potential. Both the MILC and RBC-Bielefeld
collaborations are still using this technique. On the lattice
the static potential can be determined with the help of Wilson-loops.
A $W_{x;\mu}(R,T)$ Wilson-loop of size $R\times T$ is an observable
similar to the plaquette where we take the product of the links along a
rectangle of size $R\times T$. The first, spatial direction of the rectangle
is characterized by $\mu=1\dots 3$, whereas the second direction is always 
the Euclidean time. One can define the Wilson-loop average as:
\be
W(R,T)=\Re \Tr \sum_{x;\mu=1\dots 3}W_{x;\mu}(R,T)
\ee
It can be shown that the free energy (at zero temperature the potential energy)
of a system with an infinitely heavy
quark-antiquark pair separated by a distance $R$ is given by
\be
V(R)=-\lim_{T\to\infty} \frac{1}{T}\ln W(R,T).
\ee
There are two useful quantities which can be easily obtained from $V(R)$ and
they are usually used for scale setting.
The first one is the $\sigma$ string tension which is defined as
$\sigma=\lim_{R\to\infty}dV(R)/dR$. While $\sigma$ is a useful quantity
in the pure gauge theory
--where the potential is linear for large $R$--, in QCD it does not
exist in a strict sense. At large distances pair creation will lead
to string breaking and the potential will saturate. Nevertheless,
$\sigma$ is still sometimes used to set the scale in full QCD.

The second quantity obtained from $V(R)$ is the Sommer parameter, $r_0$,
which is defined implicitly by~\cite{Sommer:1993ce}:
\be
\left.R^2\frac{dV(R)}{dR}\right|_{R=r_0}=1.65.
\ee

Both quantities have the great disadvantage that they can not
be measured directly by experiments. Their values can only be 
estimated from e.g. heavy meson spectroscopy. The value of the
string tension is $\sqrt{\sigma}\approx$440~MeV, while for $r_0$ 
the most accurate values are based on lattice calculations (where the
scale was set with some other quantity of course)~\cite{Aubin:2004wf,Gray:2005ur}: 
$r_0=$0.469(7)~fm, other values are 0.444(3) (based on the pion decay 
constant\cite{Alexandrou:2008tn}, 0.467(33) from QCDSF~\cite{Gockeler:2005rv} 
and 
0.492(6)(7) from PACS-CS~\cite{Aoki:2008sm}.
Note, that there are 
several sigma differences between these results. This fact emphasizes the general observation, that the determination of $r_0$ is difficult, and that the systematic errors are underestimated.

It may be desirable to use a quantity which is well known experimentally.
The nucleon mass may seem as an obvious choice, however, on the lattice
spacings typical in thermodynamics studies, an accurate lattice determination
of the nucleon mass is difficult. Another choice, which is often used in
the literature, is the $\rho$ meson mass. Unfortunately as it is a resonance,
its precise mass determination would require a detailed analysis of its
interaction with the decay products.

The quantity used by the Wuppertal-Budapest group is the leptonic
decay constant of the kaon: $f_K=159.8$~MeV, whose experimental 
value is known to about one percent accuracy and it can be precisely
determined on the lattice~\footnote{Note, that very recently the experimental
value of $f_K$ has slightly decreased~\cite{Amsler:2008zz}}. 
Let us now illustrate the determination 
of the LCP and the scale setting with the $m_\pi,m_K,f_K$ choice.

\begin{figure}
\centerline{\includegraphics*[width=10cm, bb=16 170 592 610]{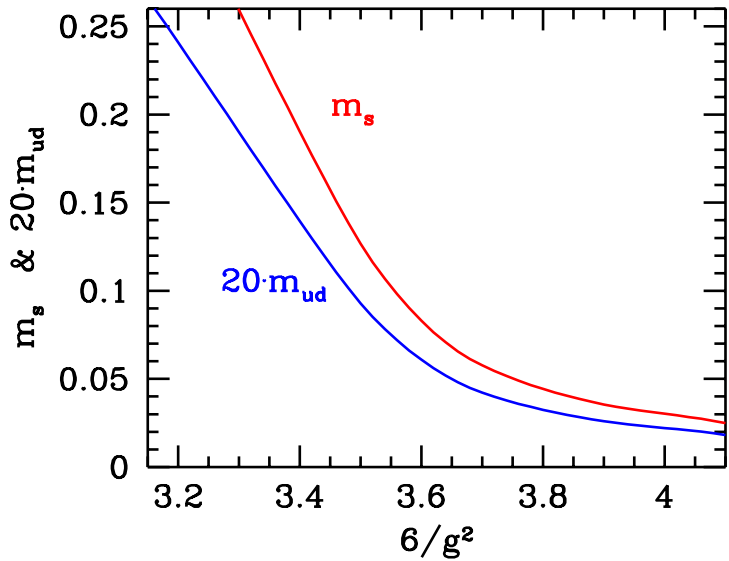}}
\caption{\small\label{fig_lcp2}
Line of constant physics defined via $m_\pi$, $m_K$ and $f_K$.
}
\end{figure}

For any set of the dimensionless bare parameters 
( $\beta$, $am_{ud}$ and $am_s$) we can determine 
$am_\pi$, $am_K$ and $af_K$ on the lattice.
For a fixed $\beta$ we can set $am_{ud}$ and $am_s$ such that
the ratios $(am_\pi)/(af_K)$ and $(am_K)/(af_K)$ agree with the 
physical $m_\pi/f_K$ and $m_K/f_K$ ratios. This way we have an 
$am_{ud}(\beta)$ and an $am_s(\beta)$ function. We call these functions LCP.
The lattice spacing is given by the third quantity: 
$a=(af_K)/(159.8 \rm{MeV})$. Figure~\ref{fig_lcp2} shows the LCP obtained this
way using stout improved staggered fermions.

We have to note here, that the LCP is not unique, it depends on the 
three quantities used for its definition. However, all LCP's should merge
together towards the continuum limit.

\begin{figure}[t!]
\centerline{\includegraphics*[width=10cm]{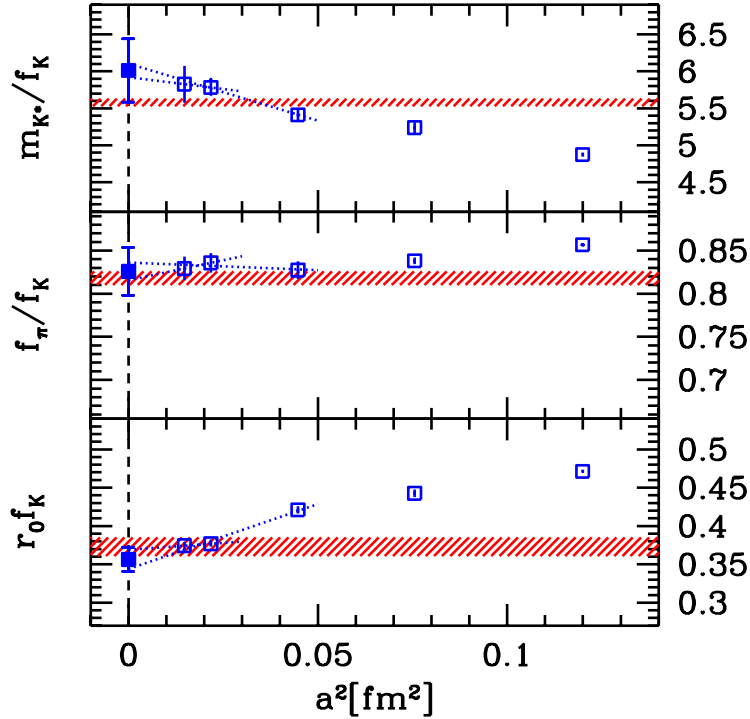}}
\caption{\small\label{fig_T0}
Mass of the $K^{*}$ meson, the pion decay constant and the $r_0$ Sommer parameter
(from top to bottom). All three quantities are normalized by $f_K$ or its inverse.
We present the results obtained at five lattice spacings, the continuum
extrapolated values are also shown. The continuum extrapolations
were carried out using the two or three finest lattice spacings (dashed lines).
The red bands indicate the experimental values with their uncertainties
in the first two cases. For $r_0 f_K$ the MILC lattice result is shown.
}
\end{figure}

Once the LCP is fixed and the scale is set with the help of the three selected
quantities, the expectation values of all other observables are predictions
of QCD. If QCD is the correct theory of the strong interaction, these 
predictions should be in agreement with the corresponding experimental 
values (if there are any) in continuum limit. Figure~\ref{fig_T0} shows
the  $m_{K^*}$ mass of the $K^{*}$ meson, the $f_\pi$ pion decay constant
and the $r_0$ Sommer parameter (all normalized by $f_K$ or its inverse). The
results again were obtained with stout improved staggered fermions along the
LCP shown on Figure~\ref{fig_lcp2}. The continuum extrapolation has been
carried out using the two or three finest lattice spacings. The difference
of these extrapolations account for the systematic uncertainty of the
results. In case of $m_{K^*}$ and $f_\pi$ we compared the results to the
experimental values, while $r_0$ was compared to the results of the MILC
collaboration~\cite{Aubin:2004wf}.

\section{The order of the QCD transition}

The nature of the QCD transition affects our understanding of the
universe's evolution~(see e.g. Ref.~\cite{Schwarz:2003du}). In a strong 
first order phase transition scenario the quark-gluon plasma super-cools before 
bubbles of hadron gas are formed. 
These bubbles grow, collide and merge during which gravitational 
waves could be produced~\cite{Witten:1984rs}. Baryon 
enriched nuggets could remain 
between the bubbles contributing to dark matter.
Since the hadronic phase is the initial condition for nucleosynthesis, the 
above picture with inhomogeneities could have a strong 
effect on it~\cite{Applegate:1985qt}. 
As the first order phase transition weakens, these effects become less 
pronounced. Recent calculations provide strong evidence that 
the QCD transition is an analytic transition (what we call here a crossover), thus the 
above scenarios -and many others-  are ruled out.

There are some QCD results and model calculations to determine the 
order of the
transition at $\mu$=0 and $\mu$$\neq$0 for different fermionic contents 
(c.f.~\cite{Pisarski:1983ms,Celik:1983wz,Kogut:1982rt,Gottlieb:1985ug,Brown:1988qe,Fukugita:1989yb,Halasz:1998qr,Berges:1998rc,Schaefer:2004en,Herpay:2005yr}). 
Unfortunately, none of these
approaches can give an unambiguous answer for the order of the transition
for physical values of the quark masses. The only known systematic 
technique which could give a final
answer is lattice QCD.  

When we analyze the nature and/or the absolute scale of the $T>0$ QCD 
transition for the physically relevant case two ingredients
are quite important. 

First of all, one should use physical quark masses. 
As Figure \ref{fig_phase_ud} 
shows the nature of the transition depends on the quark mass,
thus for small or large quark masses it is a first order phase transition, 
whereas for intermediate quark masses it is an analytic crossover. Though
in the chirally broken phase chiral perturbation theory provides a 
controlled technique to gain information for the quark mass dependence, it
can not be applied for the $T>0$ QCD transition (which deals with the
restoration of the chiral symmetry). In principle, the behavior of 
different quantities in the critical region 
(in the vicinity of the second order phase transition line) might give
some guidance. However, a priori it is not known how large this region is.
Thus, the only consistent way to eliminate uncertainties related to 
non-physical quark masses is to use physical quark masses (which is, of 
course, quite CPU demanding).

Secondly, the nature of the $T>0$ QCD transition is known to suffer from
discretization errors 
\cite{Karsch:2003va,deForcrand:2007rq,Endrodi:2007gc}. Let us mention one example. The three
flavor theory with a large, $a\approx 0.3$~fm lattice spacing and
standard action predicts a critical pseudoscalar mass 
of about 300~MeV. This point separates the first order 
and crossover regions of Figure \ref{fig_phase_ud}. If we
took another discretization, with another discretization error, 
e.g. the p4 action and the same lattice spacing, the
critical pseudoscalar mass turns out to be around 70~MeV (similar effect is
observed if one used stout smearing improvement and/or finer lattices). 
Since the physical
pseudoscalar mass (135~MeV) is just between these two values, the discretization
errors in the first case would lead to a first order transition, whereas 
in the second case to a crossover. 
The only way to resolve this inconclusive situation is to carry out a 
careful continuum limit analysis. 

Since the nature of the transition influences the absolute
scale ($T_c$) of the transition --its value, mass dependence, uniqueness 
etc.-- the above comments are valid for the determination of $T_c$, too.

\begin{figure}\begin{center}
{\includegraphics[width=13cm,bb=45 545 570 690]{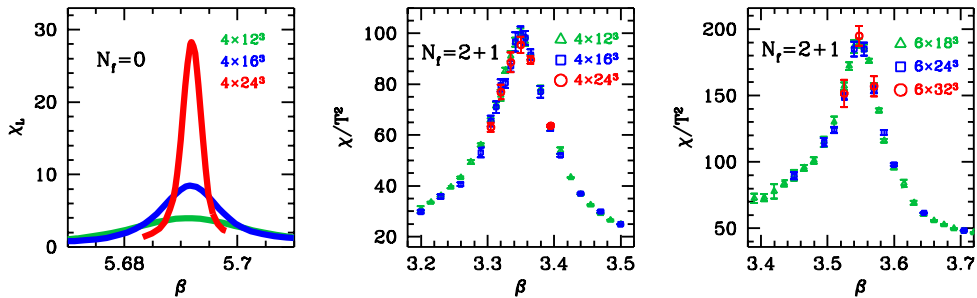}}
\end{center}\caption{\label{susc_446}
The volume dependence of the susceptibility peaks for pure SU(3) gauge 
theory (Polyakov-loop susceptibility, left panel) and for full QCD (chiral
susceptibility on $N_t$=4 and 6 lattices, middle and right panels, 
respectively). 
}
\end{figure}

In order to determine the nature of the transition one should apply finite size
scaling techniques for the chiral susceptibility~\cite{Aoki:2006we}.
$\chi=(T/V)\cdot (\partial^2\log Z/\partial m_{ud}^2)$. 
This quantity shows a pronounced peak
as a function of the temperature. For a first order phase transition, such
as in the pure gauge theory, the peak of the analogous Polyakov
susceptibility gets more and more singular as we increase the volume (V). The
width scales with 1/V the height scales with volume (see left panel of Figure
\ref{susc_446}). 
A second order transition shows a
similar singular behavior with critical indices. 
For an analytic transition 
(crossover) the peak width and height saturates to a constant value. 
That is what we observe in full QCD on $N_t$=4 and 6 lattices (middle and
right panels of Figure \ref{susc_446}). We see an order of magnitude difference
between the volumes,
but a volume independent scaling. It is a clear indication for a crossover. 
These results were obtained with physical quark masses for two sets of 
lattice spacings. Note, however, that for a final conclusion the important
question remains: do we get the same volume independent scaling 
in the continuum; 
or we have the unlucky case what we had for 3 flavor QCD 
(namely the
discretization errors changed the nature of the transition for the physical
pseudoscalar mass case)?

One can carry out a finite size scaling analysis with the continuum
extrapolated height of the renormalized susceptibility. 
The renormalization of the chiral susceptibility
can be done by taking the second derivative of
the free energy density ($f$) with respect to the renormalized mass
($m_r$).
The logarithm of the partition function contains quartic divergences. These
can be removed by subtracting the free energy at $T=0$:
$f/T^4$ =$-N_t^4$$\cdot [\log Z(N_s,N_t)/(N_t N_s^3)-\log Z(N_{s0},N_{t0})/(N_{t0} N_{s0}^3
)]$.
This quantity has a correct
continuum limit. The subtraction term is obtained at $T$=0, for which
simulations are carried out on lattices with $N_{s0}$, $N_{t0}$ spatial
and temporal extensions (otherwise at the same parameters of the action).
The bare light quark mass ($m_{ud}$) is related to $m_r$ by the
mass renormalization constant $m_r$=$Z_m$$\cdot$$m_{ud}$.
Note that $Z_m$
falls out of the combination
$m_r^2$$\partial^2$/$\partial$$m_r^2$=$m_{ud}^2$$\partial^2$/$\partial$$m_{ud}^2
$.
Thus, $m_{ud}^2\left[\chi(N_s,N_t)-\chi(N_{s0},N_{t0})\right]$ also has a
continuum limit
(for its maximum values for different $N_t$, and in the continuum limit
we use the shorthand
notation $m^2$$\Delta \chi$).

\begin{figure}[h!]
\centerline{\includegraphics*[bb=20 500 592 690,width=17cm]{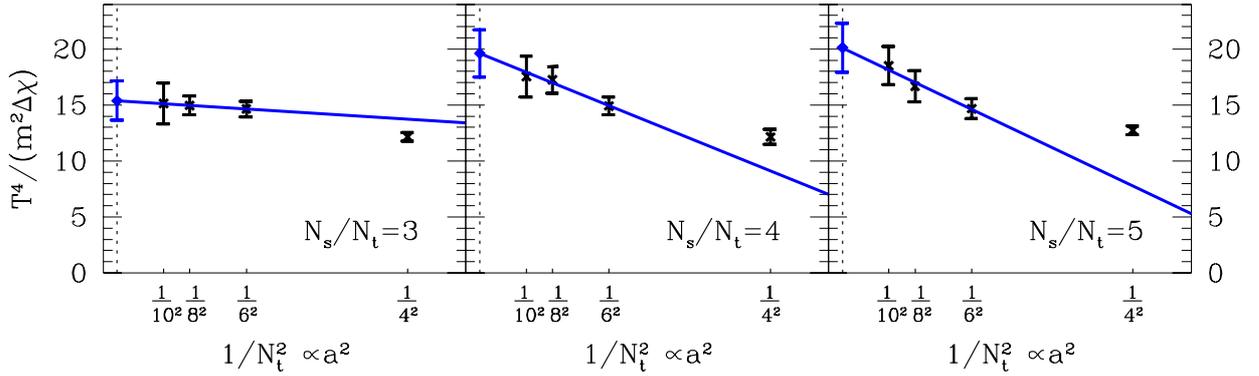}}
\caption{\label{cont_ex}
Normalized susceptibilities $T^4/(m^2\Delta\chi)$
for the light quarks for aspect
ratios r=3 (left panel) r=4 (middle panel) and r=5 (right panel)
as functions of the lattice spacing. Continuum 
extrapolations are carried out for all three physical volumes and the
results are given by the leftmost blue diamonds.
}
\end{figure}

\begin{figure}\begin{center}
\includegraphics[width=6.5cm,angle=0,bb=18 180 570 605]{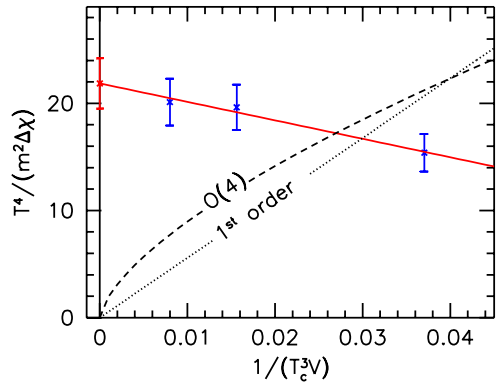}
\end{center}
\caption{\label{cont_scal}
Continuum extrapolated susceptibilities $T^4/(m^2\Delta\chi)$
as a function of 1/$(T_c^3V)$.  For true phase transitions the infinite 
volume extrapolation should be consistent with zero, whereas for an 
analytic crossover the infinite volume extrapolation gives a 
non-vanishing value. The continuum-extrapolated susceptibilities show no 
phase-transition-like volume dependence, though the volume changes by a 
factor of five.
The V$\rightarrow$$\infty$ extrapolated value is 22(2) which is
11$\sigma$ away from zero. For illustration, we fit the expected
asymptotic behaviour for first-order and O(4) (second order)
phase transitions shown by dotted and dashed lines, which results in 
chance probabilities of $10^{-19}$ ($7\times10^{-13}$), respectively. 
}
\end{figure}

In order to carry out the finite volume scaling in the continuum limit
three different physical volumes were taken.
For these volumes the dimensionless combination
$T^4/m^2 \Delta \chi$ was calculated at 4 different
lattice spacings: 0.3~fm was always off, otherwise the continuum
extrapolations could be carried out. Figure~\ref{cont_ex} shows these extrapolations.
The volume dependence of the continuum extrapolated
inverse susceptibilites is shown on Figure~\ref{cont_scal}.

The result is consistent with an approximately constant
behaviour, despite the fact that there was a factor of 5 difference in the
volume. The chance probabilities, that statistical fluctuations
changed the dominant behaviour of the volume dependence 
are negligible. As a conclusion we can say
that the staggered QCD transition at $\mu=0$ is a crossover.

\section{The transition temperature}

\begin{figure}\begin{center}
\includegraphics[width=12cm,bb=0 180 570 620]{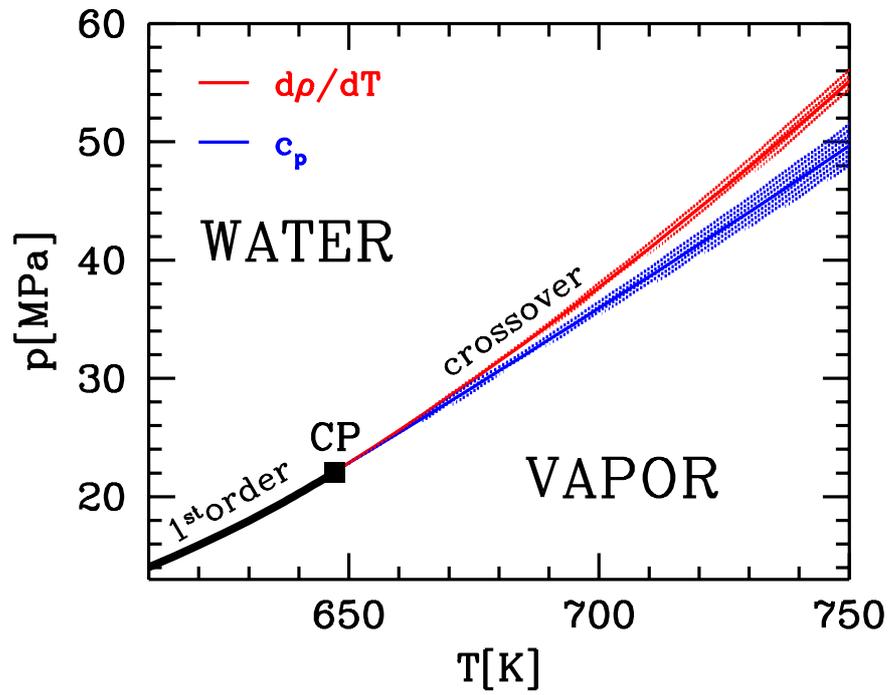}
\end{center}
\caption{\label{steam}
The water-vapor phase diagram. 
}
\end{figure}

An analytic crossover, like the QCD 
transition has no unique $T_c$. A particularly nice 
example for that is the water-vapor transition (c.f. Figure \ref{steam}). 
Up to about 650~K the transition 
is a first order one, which ends at a second order
critical point.  For a first or second order phase transition the 
different observables (such as density or heat capacity)
have their singularity (a jump or an infinitely high peak) at the same
pressure.  However, at even higher temperatures the transition is an
analytic crossover, for which the most singular points are different. The
blue curve shows the peak of the heat capacity 
and the red one the inflection point of
the density. Clearly, these transition temperatures are different, which
is a characteristic feature of an analytic transition (crossover). 

\begin{figure}\begin{center}
\includegraphics*[width=12cm]{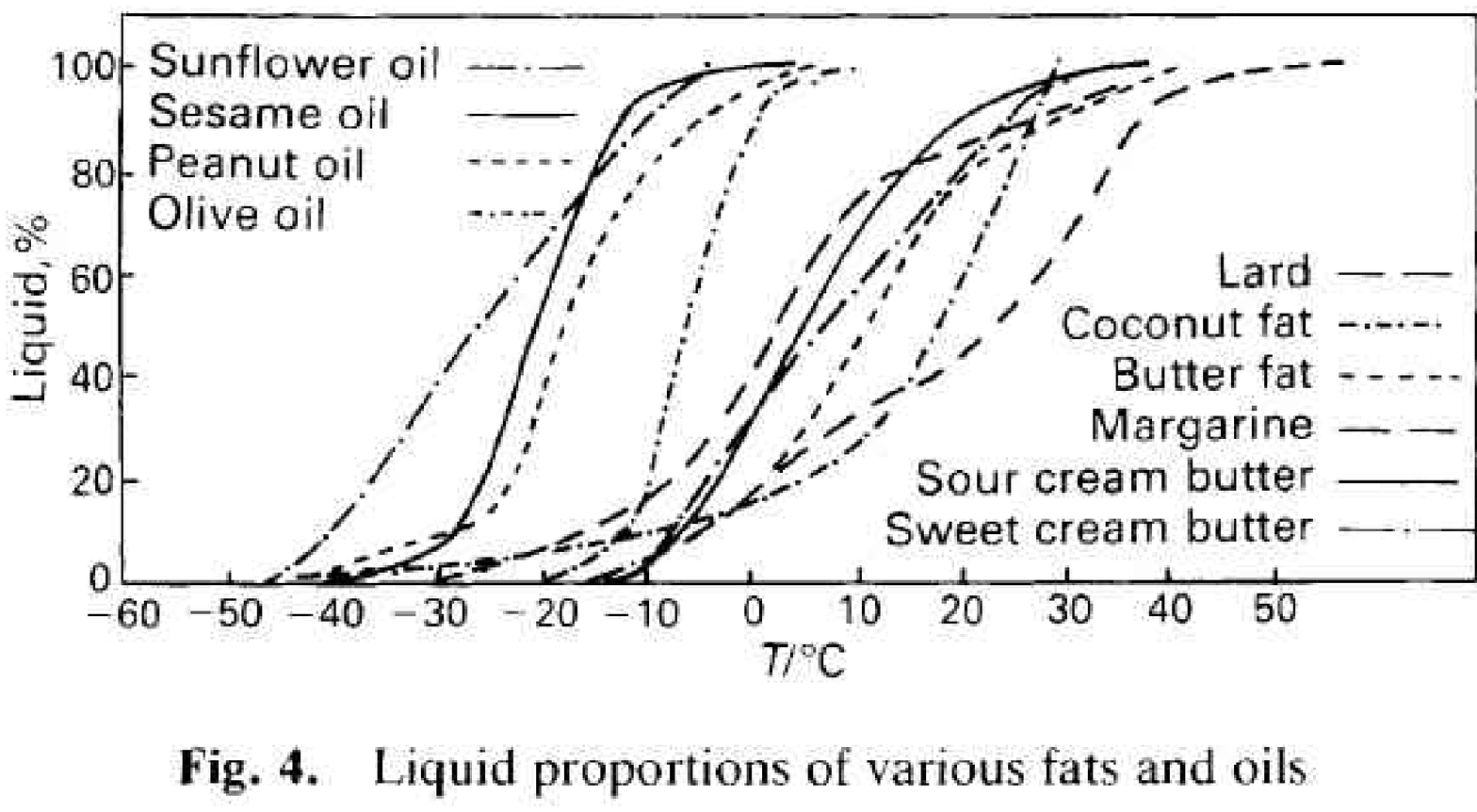}
\end{center}
\caption{\label{butter}
Melting curves of different natural fats.}
\end{figure}

There is another --even more often experienced-- example for broad transitions,
namely the melting of butter. As we know the
melting of ice shows a singular behavior. The transition is of first order,
there is only one value of the temperature at which the whole transition
takes place at 0$^o$C (for 1~atm. pressure). 
Melting of butter\footnote{Natural fats 
are mixed triglycerides of fatty acids from $C_4$ to
$C_{24}$, (saturated or unsaturated of even carbon numbers).} 
shows analytic behaviour. The transition
is a broad one, it is a crossover (c.f. Figure \ref{butter} for the 
melting curves of different natural fats). 

Since we have an analytic crossover also in QCD, we expect very similar 
temperature dependence for the quantities relevant in QCD (e.g. 
chiral condensate, strange quark number susceptibility or Polyakov loop).

There are three lattice results on $T_c$ in the literature based on
large scale calculations. The MILC collaboration studied the unrenormalized
chiral susceptibility~\cite{Bernard:2004je}. The possibility of different quantities leading
to different $T_c$'s was not discussed. They used $N_t$=4,6 and 8 lattices, but
the light quark masses were significantly higher than their physical values.
The lightest ones were set to 0.1$\cdot m_s$. A combined chiral and continuum
extrapolation was used to reach the physical point. Furthermore, they used
the non-exact R algorithm. Their result is $T_c=169(12)(4)$~MeV, where the first
error comes from the finite $T$ runs, whereas the second one from the
scale setting.

\begin{figure}\begin{center}
\includegraphics*[width=10cm]{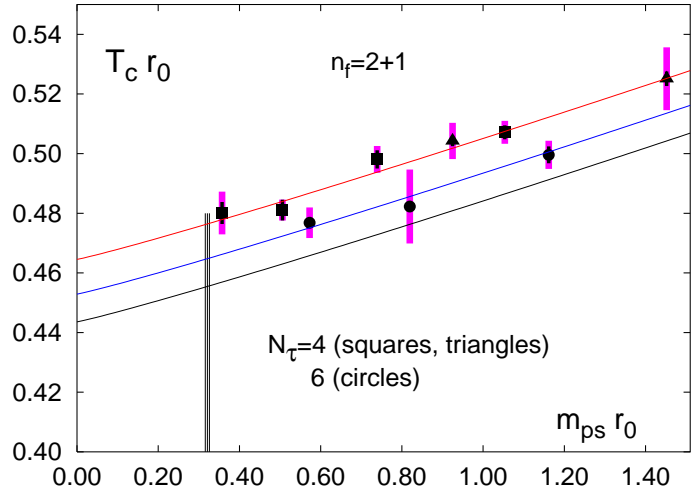}
\end{center}
\caption{\label{tc_bie}
$T_c\cdot r_0$ determined by the RBC-Bielefeld 
collaboration~\cite{Cheng:2006qk}. Squares and triangles correspond to two 
slightly different strange quark masses on $N_t=4$, while circles show 
$N_t=6$ results. The red and blue lines show the chiral extrapolations
at these lattice spacings and the black line is the continuum estimate. The 
vertical line indicates the physical point.
}
\end{figure}

The RBC-Bielefeld collaboration has published results obtained from $N_t=4$ 
and 6 lattices~\cite{Cheng:2006qk}. They have ongoing investigations with 
$N_t=8$. They use almost physical quark masses on $N_t=4$ and somewhat higher
on $N_t=6$. They study the unrenormalized chiral susceptibility and
the Polyakov-loop susceptibility. They claim that both quantities give the 
same $T_c$. Figure~\ref{tc_bie} shows their chiral extrapolation for
their two lattice resolutions. Their result is $T_c=192(7)(4)$~MeV, where
the first error is the statistical one and the second is the systematic
estimate coming from the different extrapolations.

The Wuppertal-Budapest group investigated three different quantities: the
renormalized chiral susceptibility, the renormalized Polyakov-loop and the
quark number susceptibility. The transition temperature obtained from
the chiral susceptibility was found to be significantly smaller than the ones
given by the other two quantities.

\begin{figure}\begin{center}
\centerline{\includegraphics[height=15.3cm, bb=190 160 410 710]{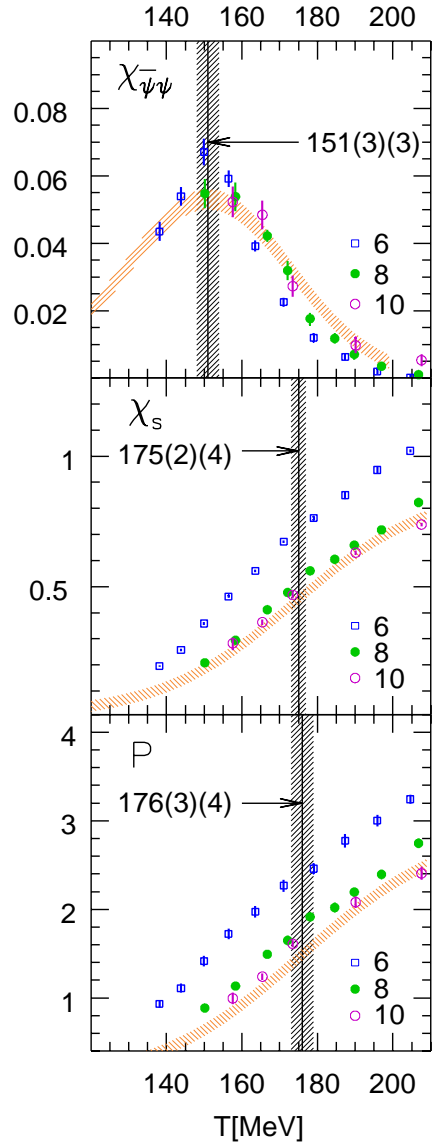}}
\end{center}\caption{\label{fig:susc}
Temperature dependence of the renormalized
chiral susceptibility ($m^2\Delta \chi/T^4$), the strange
quark number susceptibility ($\chi_s/T^2$)
and the renormalized Polyakov-loop ($P_R$) in the transition region. The 
different symbols show the results for $N_t=6,8$ and
$10$ lattice spacings (empty boxes for $N_t=6$, filled and open circles for $N_t=8$
and $10$).
The vertical bands indicate the corresponding transition temperatures and
their uncertainties coming from the T$\neq$0 analyses. This error is
given by the number in the first parenthesis, whereas the error of the
overall scale determination is indicated by the number in the second
parenthesis. The orange bands show the continuum limit estimates for the
three renormalized quantities as
a function of the temperature with their uncertainties.
}
\end{figure}

\begin{figure}[t!]
\centerline{\includegraphics*[width=13cm,bb=18 520 589 704]{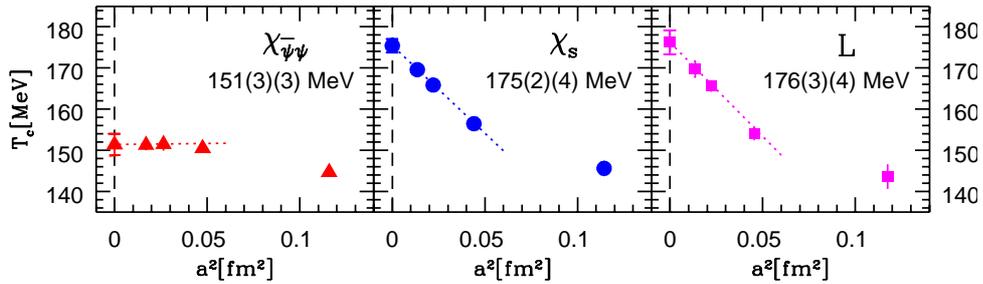}}
\caption{\label{fig:tc}
Continuum limit of the transition temperatures obtained from the renormalized chiral
susceptibility ($m^2\Delta \chi/T^4$), 
strange quark number susceptibility ($\chi_s/T^2$) and renormalized
Polyakov-loop ($P_R$). 
}
\end{figure}

The upper panel of 
Figure \ref{fig:susc} shows the temperature dependence of the 
renormalized chiral susceptibility
for different temporal extensions ($N_t$=6, 8 and 10). The $N_t=4$ results
are not yet in the scaling region, thus they are not plotted. 
For small enough lattice spacings, thus close to the continuum limit,
these curves should coincide.  
The two smallest lattice
spacings ($N_t=8$ and $10$) are already consistent with each other,
suggesting that they are also consistent with the continuum limit 
extrapolation (indicated by the orange band). 
The curves exhibit pronounced peaks. We define the 
transition temperatures by the position of these peaks. 
The left panel of Figure \ref{fig:tc} shows the transition 
temperatures in
physical units for different lattice spacings obtained from the
chiral susceptibility. As it can be seen
$N_t$=6, 8 and 10 are already in the scaling region, thus a safe
continuum extrapolation can be carried out.
The T=0 simulations resulted in a $2\%$ error on the overall scale.
The final result for the transition temperature based on the chiral
susceptibility reads:  
\begin{equation} 
T_c(\chi_{\bar{\psi}\psi})=151(3)(3) {\rm ~MeV},
\end{equation} 
where the first error comes from the T$\neq$0, the second from
the T=0 analyses.  

For heavy-ion experiments the quark number susceptibilities are 
quite useful, since
they could be related to event-by-event fluctuations.
The second transition temperature is obtained from the strange quark number
susceptibility,  which is defined via~\cite{Bernard:2004je}
\begin{equation}
\frac{\chi_{s}}{T^2}=\frac{1}{TV}\left.\frac{\partial^2 \log Z}{\partial \mu_{s} ^2
}\right|_{\mu_{s}=0}, 
\end{equation}
where $\mu_s$ is the strange quark chemical potential (in
lattice units). Quark number susceptibilities have the convenient property,
that they automatically have a proper continuum limit, there is no need for
renormalization.

The middle panel of
Figure \ref{fig:susc} shows the temperature dependence of the
strange quark number susceptibility
for different temporal extensions ($N_t$=6, 8 and 10).
As it can be seen, the two smallest lattice
spacings ($N_t=8$ and $10$) are already consistent with each other,
suggesting that they are also consistent with the continuum limit
extrapolation. This feature indicates, that they are closer to 
the continuum result than our statistical uncertainty.

The transition temperature can be defined as the peak in the temperature
derivative of the strange quark number susceptibility, 
that is the inflection point of the
susceptibility curve. 
The middle panel of Figure \ref{fig:tc} shows the transition
temperatures in
physical units for different lattice spacings obtained from the
strange quark number susceptibility. As it can be seen
$N_t$=6, 8 and 10 are already in the $a^2$ scaling region, thus a safe
continuum extrapolation can be carried out.
The continuum extrapolated value for the transition temperature 
based on the strange quark number susceptibility is
significantly higher than the one from the chiral susceptibility. The 
difference is 24(4)~MeV.  For the transition temperature in the continuum 
limit one gets:
\begin{equation}
T_c(\chi_s)=175(2)(4) {\rm ~MeV},
\end{equation}
where the first (second) error is from the T$\neq$0 (T=0) temperature
analysis.
Similarly to the 
chiral susceptibility analysis, the curvature at the peak can be used
to define a width for the transition. 
\begin{equation}
\Delta T_c(\chi_s)= 42(4)(1) {\rm ~MeV}.
\end{equation}

In pure gauge theory the order parameter of the deconfinement transition is 
the Polyakov-loop: 
\begin{equation} 
P=\frac{1}{N_s^3}\sum_{\bf x} {\rm tr} 
[U_4({\bf x},0) U_4({\bf x},1) \dots U_4({\bf x},N_t-1)]. 
\end{equation} 
P acquires a non-vanishing expectation value in the deconfined phase, 
signaling the spontaneous breakdown of the Z(3) symmetry. When fermions 
are present in the system, the physical interpretation of the 
Polyakov-loop expectation value is more complicated. 
However, its absolute value can be related to the quark-antiquark 
free energy at infinite separation: 
\begin{equation} 
|\langle P \rangle |^2 = 
\exp(-\Delta F_{q\bar{q}}(r\to \infty)/T). 
\end{equation} 
$\Delta F_{q\bar{q}}$ is the difference of the free energies 
of the quark-gluon plasma with and without the quark-antiquark pair.

The absolute value of the Polyakov-loop vanishes in the continuum 
limit. It needs renormalization. This can be 
done by renormalizing the free energy of the quark-antiquark pair
\cite{Kaczmarek:2002mc}. 
Note, that QCD at T$\neq$0 has only the  
ultraviolet divergencies which are already present at T=0. 
In order to remove these divergencies at a given lattice spacing
a simple renormalization condition can be used\cite{Fodor:2004ft}: 
\begin{equation}
V_R(r_0)=0
\end{equation} 
with $r_0=0.46$~fm, where the potential is 
measured at T=0 from Wilson-loops. The above
condition fixes the additive term in the potential at a given lattice
spacing. This additive term can be used at the same lattice spacings
for the potential obtained from Polyakov loops, or equivalently
it can be built into the definition of the renormalized 
Polyakov-loop.
\begin{equation} 
|\langle P_R \rangle | = |\langle 
P \rangle | \exp(V(r_0)/(2T)), 
\end{equation}
where $V(r_0)$ is the unrenormalized potential  
obtained from Wilson-loops.

The lower panel of
Figure \ref{fig:susc} shows the temperature dependence of the
renormalized Polyakov-loops
for different temporal extensions ($N_t$=6, 8 and 10).
The two smallest lattice
spacings ($N_t=8$ and $10$) are approximately in 1-sigma agreement
 (our continuum limit estimate is indicated by the orange band).

Similarly to the strange quark susceptibility case
the transition temperature is defined as the peak in the temperature
derivative of the Polyakov-loop, that is the inflection point of the
Polyakov-loop curve. 
The right panel of Figure \ref{fig:tc} shows the transition
temperatures in
physical units for different lattice spacings obtained from the
Polyakov-loop. As it can be seen
$N_t$=6, 8 and 10 are already in the scaling region, thus a safe
continuum extrapolation can be carried out. 
The continuum extrapolated value for the transition temperature 
based on the renormalized Polyakov-loop is 25(4)~MeV
higher than the one from the chiral susceptibility. 
For the transition temperature in the continuum 
limit one gets:
\begin{equation}
T_c(P)=176(3)(4) {\rm ~MeV},
\end{equation}
where the first (second) error is from the T$\neq$0 (T=0) temperature
analysis. The width of the transition is
\begin{equation}
\Delta T_c(P)=38(5)(1) {\rm ~MeV}.
\end{equation}

\begin{figure}\begin{center}
\includegraphics[width=13cm,angle=0,bb=0 0 765 85]{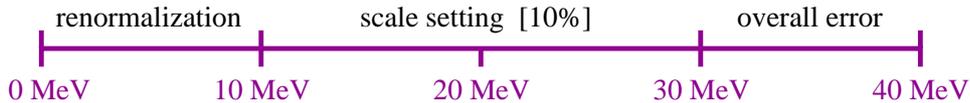}
\end{center}\caption{\label{tc_diff}
Possible contributions to the 40~MeV difference between the results
of Refs. \cite{Aoki:2006br} and \cite{Cheng:2006qk}.
}
\end{figure}

\begin{figure}\begin{center}
\includegraphics*[height=6.0cm,bb=300 165 592 420]{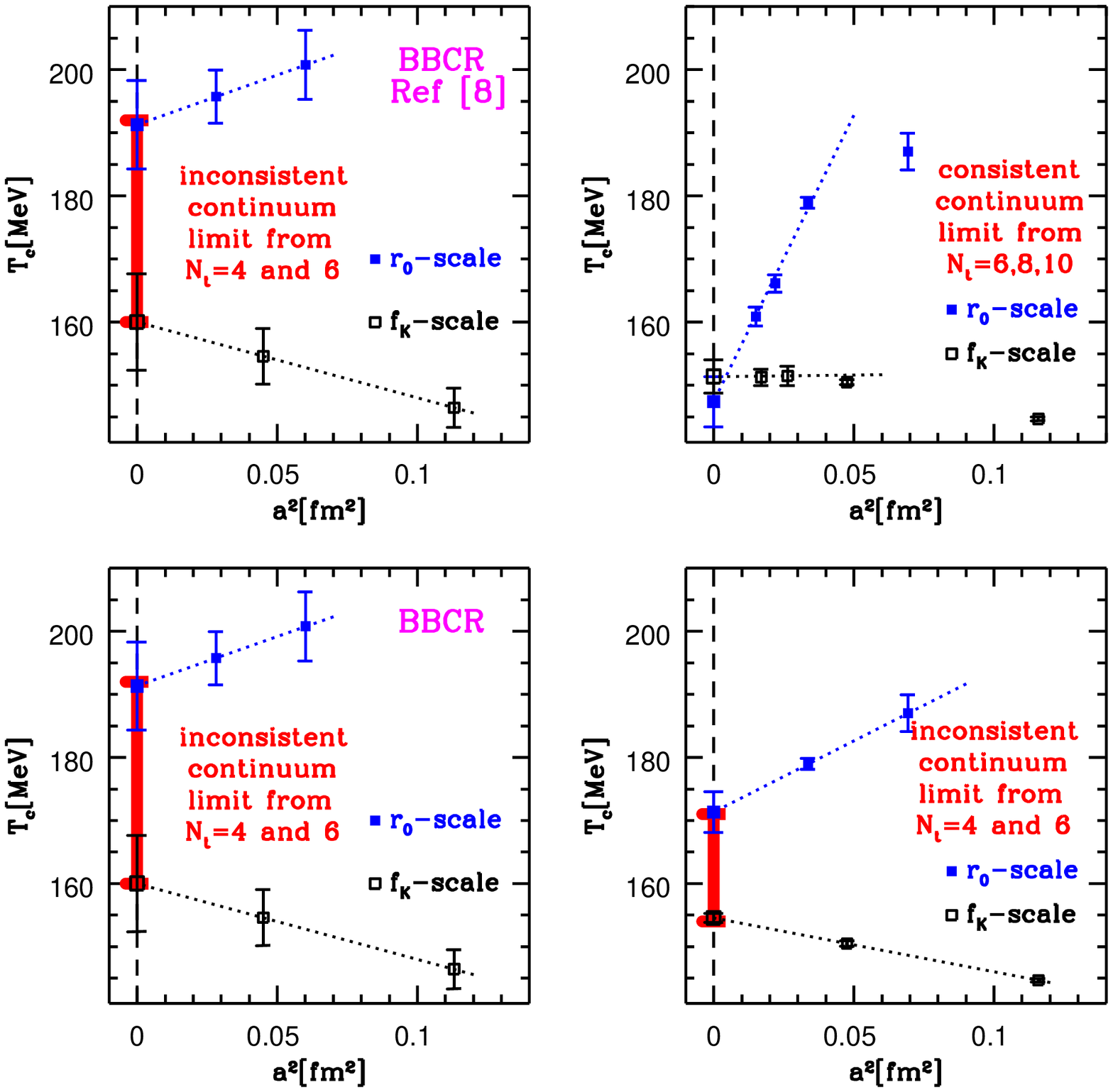}\hspace*{0.5cm}\includegraphics*[height=6.0cm,bb=300 435 592 690]{Figs/note3b.eps}
\end{center}\caption{\label{note3}
Continuum extrapolations based on $N_t$=4 and 6 (left panel: inconsistent
continuum limit) and using  $N_t$=6,8 and 10 (right panel: consistent
continuum limit).  
}
\end{figure}

Note that the renormalized chiral susceptibility used above to define
$T_c$ was normalized by $T^4$. Due to the broadness of the peak
a normalization by $T^2$ (which is applied by the other collaborations)
would increase $T_c$ by about 10~MeV.
This means that the
Wuppertal-Budapest result on the chiral susceptibility is consistent
with the MILC result. There is however a significant inconsistency with
the RBC-Bielefeld result. What are the differences between the two analyses
and how do they contribute to the 40~MeV discrepancy? 
The most important contributions to the discrepancy are shown by 
Figure \ref{tc_diff}. The first difference is the different normalization
of the chiral susceptibility mentioned before. This may account for $\approx 10$~MeV
difference. The overall
errors can be responsible for another 10~MeV. The origin of the remaining
20~MeV is somewhat more complicated. One possible explanation can be
summarized as follows. 
In Ref. \cite{Cheng:2006qk} only $N_t$=4 and 6 were used, which 
correspond to
lattice spacings a=0.3 and 0.2~fm, or $a^{-1}$=700MeV and 1GeV. 
These lattices are
quite coarse and it seems to be obvious, that no unambiguous scale can be
determined for these lattice spacings. 
The overall scale in Ref. \cite{Cheng:2006qk} was set by $r_0$ and no
cross-check was done by any other quantity independent of the static
potential (e.g. $f_k$). This choice might lead to an ambiguity for the
transition temperature, which is 
illustrated for the Wuppertal-Budapest 
data on Figure \ref{note3}.
Using only $N_t$=4 and 6 the continuum extrapolated 
transition temperatures are quite different
if one took $r_0$ or $f_K$ to determine the overall scale. This inconsistency
indicates, that these lattice spacing are not yet in the scaling region
(similar ambiguity is obtained by using the p4 action of \cite{Cheng:2006qk}).  
Having $N_t$=4,6,8 and 10 results 
this ambiguity disappears (as usual $N_t$=4 is off), 
these lattice spacings are already in the scaling region 
(at least within the present accuracy). 

The ambiguity related to the inconsistent continuum limit is 
unphysical, and it is resolved as we approach the continuum limit
(c.f. Figure \ref{note3}). The differences between the $T_c$ values for 
different observables are physical, it is a consequence of the crossover
nature of the QCD transition.

\section{Equation of state}

In the previous sections we discussed the nature of the QCD transition
and its characteristic scale. Now we extend the analysis to cover a 
larger temperature range~\cite{Aoki:2005vt}. In order to describe
the equilibrium properties of the quark-gluon plasma and/or the
hadronic phase one has to determine the equation of state. The equation
state describes the functional relationship between various thermodynamical
quantities. The most common way to start with is to calculate the pressure
as a function of the temperature. Using this function 
the temperature dependence of other quantities
can be determined (energy density, entropy density, speed of sound etc.), too.

Several recent papers discuss the equation of state. For the pure SU(3)
theory several lattice actions were 
used~\cite{Boyd:1996bx,Okamoto:1999hi,Namekawa:2001ih}.
In all of these cases the equation of state was given upto about $4T_c$.
There are few percent differences between the various results, however,
these differences can be traced back to the scale setting problem. Note,
that defining a scale in physical units is in principle
impossible for the  
pure SU(3) theory, experimentally measurable quantities should be compared
with results obtained in full QCD. Thus, for the pure SU(3) case only 
dimensionless combinations (e.g. ratios) can be considered as predictions
(dimensionless combinations can be obtained also in full QCD).
It is worth mentioning that until recently it was technically impossible
to calculate the equation of state to much higher temperatures ($T\gg T_c$).
In a recent work this old problem was solved\cite{Endrodi:2007tq}.

There are several full QCD results for the equation of state, though
none of them can be considered as full result. Unimproved dynamical
results in the staggered formalism were
published~\cite{Blum:1994zf,Bernard:1996cs}. Another important result
used ${\cal O}(a)$ improved Wilson fermions. As for the transition temperature,
also for the equation of state improved staggered fermions provide the 
fastest way to approach the physical quark mass and continuum limits.
The most important results  are obtained by the p4fat3 
action (see e.g. \cite{Karsch:2000ps,Cheng:2007jq}). Other improved
staggered results can be found for the ASQTAD action \cite{Bernard:2006nj}
and for the stout-smeared action~\cite{Aoki:2005vt}. It is illustrative to 
summarize the uncertainties of \cite{Karsch:2000ps} (many of them 
were cured in their --and other's-- later publications). This sort of summary
nicely shows how uncertainties are eliminated through computational and
technical progress.

(1) It is of particular importance to use physical quark masses both at 
T=0 and T>0. Until now the only published work which used 
physical quark masses are the one with stout-smeared 
improvement~\cite{Aoki:2005vt}. In earlier works~\cite{Karsch:2000ps}
pion masses of e.g. 600~MeV were used. Since the physical pion mass is
smaller than the transition temperature, it is obviously important
to use pion contributions with the proper Boltzmann weights.                

(2) In order to approach the continuum limit one has to use small  
enough lattice spacing. At least $N_t$=6 and 8 is needed (as we discussed      
earlier e.g. $N_t$=4 can not be used to set the scale reliably).     

(3) In the staggered formalism one has instead of three degenerate 
pseudo-Goldstone bosons ($\pi^+$, $\pi^-$ and $\pi^0$) only one. The others
are separated from this single one by a gap, which can be as large as 
several hundred MeV. The size of the gap depends on the choice of the action
and on the lattice spacing. As we have seen the stout-smeared improvement
is the best choice to reduce this taste symmetry violation. 

(4) In several studies~\cite{Karsch:2000ps,Bernard:2006nj} an inexact
Monte-Carlo technique was used, the so-called R-algorithm. Recently, an
exact algorithm appeared on the market, which allows to perform 2+1
flavour staggered simulations (RHMC algorithm).  The first large scale
analysis, which
used an exact algorithm for staggered thermodynamics was Ref. 
\cite{Aoki:2005vt}, which was then followed by \cite{Cheng:2007jq}.

(5) For a long time all staggered analyses used the non-LCP approach.
In this approximation there is a serious mismatch between the pion masses.
E.g. if one cooled down the analyzed systems at $0.8T_c$ and at $3.2T_c$ 
to vanishing temperature, the pion mass would be twice as large for the 
second system. This is clearly un-physical. The first work with the 
the proper line of constant physics 
was Ref.~\cite{Aoki:2005vt}
(using the heavy quark potential to set the relative scales),
which was then followed by \cite{Bernard:2006nj,Cheng:2007jq}. 

For large homogeneous systems the pressure is proportional to the 
free-energy density, which is the logarithm of the partition function $Z$.
\be
p=\frac{T}{V}\ln Z.
\ee
On a space-time lattice one determines the dimensionless $pa^4$ combination.
\be
pa^4=\frac{1}{N_tN_s^3}\ln Z.
\ee
Since the free energy has divergent terms, when we approach the continuum   
limit, one has to renormalize. As it was done earlier, this renormalization
can be achieved by subtracting the T=0 term. To that end one has to
carry out simulations on  $T=0$ lattices. The partition function on
T=0 lattices will be denoted by $Z_0$. The size of this T=0 lattice is
$N_{s0}^3\cdot N_{t0}$. The renormalized pressure is usually normalized
by $T^4$ which leads to a dimensionless combination
\be
\frac{p_R}{T^4}=N_t^4\left[ \frac{1}{N_tN_s^3}\ln
 Z - \frac{1}{N_{t0}N_{s0}^3}\ln
Z_0 \right].
\ee
In the rest of this review we omit the index $R$, since we use only
renormalized quantities. This renormalization prescription automatically
fulfills the $p(T=0)=0$ condition. It is worth mentioning that 
for a fixed lattice spacing $a$ the
weight of the terms proportional to $1/a^2$ (thus the diverging term) is
much larger for the pressure than for the chiral susceptibility. It 
is particularly true for large temperatures. Thus we have to determine 
the difference between two almost equal numbers, which needs high
numerical accuracy. This is one of the most important reason, why only 
$N_t$=4 and 6 published results available for the pressure, whereas 
for the transition temperature there are $N_t$=4,6,8 and 10 published
results, too. Another reason for the different levels of results is    
related to the lattice spacings. For large temperatures even the $N_t$=4    
analyses need small lattice spacings and relatively large T=0 lattices. 
E.g. on $N_t=4$ lattices at $T=2.5T_c$ one needs the same $T=0$ lattices
as for the $T_c$ determination on $N_t=10$ lattices. Quite recently, a 
new method appeared, which eliminates this difficulty and provides a 
renormalization by using T>0 lattice simulations\cite{Endrodi:2007tq}.

As usual for a fixed $N_t$ we tune the temperature by changing the 
gauge coupling $\beta$. In order to avoid any non-physical mismatch
we keep the system along the LCP. Thus, determining $\ln Z$ and 
$\ln Z_0$ along the proper LCP-defined ($\beta,am_q$) line gives
us the pressure (for simplicity $m_q$ denotes both the light and the
strange quark masses). We discussed the simulation algorithms based on
importance sampling in Chapter \ref{sect_alg}. Unfortunately, these
algorithms are not able to directly provide $Z$ or $\ln Z$, only       
derivatives of the partition functions can be determined. Therefore, 
the most straightforward technique is the integral 
method~\cite{Engels:1990vr}. The pressure is obtained as an integral 
of its derivatives along a line in the multi-dimensional $(\beta, am_q)$
space. 
\bea
\frac{p}{T^4}=
N_t^4\int^{(\beta,am_q)}_{(\beta_0,am_{q0})}
d (\beta,am_q)\left[
\frac{1}{N_tN_s^3}
\left(\begin{array}{c}
{\partial \ln Z}/{\partial \beta} \\
{\partial \ln Z}/{\partial (am_{q})}
\end{array} \right )-\right.\nonumber\\
\left.\frac{1}{N_{t0}N_{s0}^3}
\left(\begin{array}{c}
{\partial \ln Z_0}/{\partial \beta} \\
{\partial \ln Z_0}/{\partial (am_q)}
\end{array} \right )
\right].
\eea
Since the integrand is the gradient of the pressure, the value of the 
integral is independent of the integration path. Nevertheless, it is 
useful to integrate along the line of constant physics. In this case 
the endpoints of the integration paths will be just on the LCP, which we
need. As we will see later a slightly modified path is even 
more appropriate (in order to 
carry out the chiral extrapolations at $T=0$).

The lower end of the integration path should be chosen to ensure zero 
pressure. This goal can be reached by using temperatures well below the 
$T_c$. 
It is straightforward to calculate the 
derivatives, they are just the expectation values of the 
various terms of the staggered fermion
and gauge actions (\ref{hatas_stag},\ref{hatas_sym}). 
\begin{align}
\frac{\partial \ln Z}{\partial \beta}=\left< -S_g/\beta \right> &&
\frac{\partial \ln Z}{\partial (am_{ud})}=\left<\bar{\psi}\psi_{ud} \right> &&
\frac{\partial \ln Z}{\partial (am_{s})}=\left<\bar{\psi}\psi_{s} \right>.
\end{align}
The pressure can be written as 
\bea
\frac{p}{T^4}&=&
N_t^4\int^{(\beta,am_{ud},am_s)}_{(\beta_0,am_{ud0},am_{s0})}
d (\beta,am_{ud},am_s)\times \nonumber\\
&&\left[
\frac{1}{N_tN_s^3}
\left(\begin{array}{c}
\langle{\rm -S_g/\beta}\rangle \\
\langle\bar{\psi}\psi_{ud}\rangle \\
\langle\bar{\psi}\psi_{s}\rangle
\end{array} \right) 
-
\frac{1}{N_{t0}N_{s0}^3}
\left(\begin{array}{c}
\langle{\rm -S_g/\beta}\rangle_0 \\
\langle\bar{\psi}\psi_{ud}\rangle_0 \\
\langle\bar{\psi}\psi_{s}\rangle_0
\end{array} \right)\right].
\eea
Here $\left<\dots \right>_0$ denotes the expectation values calculated
on $T=0$ lattices.

The integral method was originally introduced for pure gauge theories.
Since these theories --at least in their simplest formulations-- have only
one parameter $\beta$ the pressure can be given by an integral over $\beta$.
Earlier staggered works used the same strategy, which --as we pointed out
already-- does not correspond the physical LCP. The proper solution
is to use the line of constant physics and avoid any mismatch of the 
spectrum.

The above formulas give the pressure as a function of the gauge coupling 
$\beta$. Clearly, one needs $p$ as a function of the temperature. 
To that end we need the  
$\beta$ dependence of the lattice spacing $a$. This can be the dependence
in absolute units (MeV) or in relative units ($T/T_c$). The relative  
units are somewhat easier to determine, e.g. one can calculate the static
potential for each $\beta$ and compare them directly (or compare some 
characteristic points of them $r_0$ or $r_1$). In order to give the lattice
spacing in physical units one has to insert the physical value of $r_0$ or
$r_1$ (unfortunately, --as it was discussed earlier-- they are not very precisely known). 

The energy density ($\epsilon$), the entropy density ($s$) and the speed 
of sound can be derived using the pressure and various thermodynamic 
relations:
\begin{align}\label{esc}
\epsilon = T(\partial p/\partial T)-p, && s = (\epsilon + p) T, && c_s^2
=\frac{dp}{d\epsilon}.
\end{align}
The derivatives of $p$ can be calculated numerically.

There is another popular method to determine the energy density. The energy density can be 
written as 
$\epsilon (T)={T^2}/{V}{\partial{\log Z}}/{\partial {T}}$. Using this form
and the relationship between the temperature, volume and the lattice spacing one can easily
show that 
\be
\frac{\epsilon-3p}{T^4}=-\frac{N_t^3}{N_s^3}a \frac{d (\log Z)}{d a}.
\ee
Thus, the expression $\epsilon-3p$ can be directly determined by using the total
derivative with respect to the lattice spacing. There are different names for this
quantity, Sometimes it is called ``interaction measure'' (at very high temperatures its 
value tends to zero, reflecting the non-interactive feature of the system), or 
``trace anomaly''. The above total derivative can be written as a            
derivative with respect to $\beta$ and the quark masses (one uses the chain rule). 
Renormalization is carried out analogously as in the case of the pressure. 
Adding three times the pressure to the trace anomaly gives the energy density.

The energy density can be also obtained from the pressure (\ref{esc}). This choice is
particularly useful, if one uses larger than physical quark masses at $T=0$ and uses
chiral perturbation theory for the extrapolation to the physical value. The form of
the chiral extrapolation is not known for all relevant quantities. For the 
chiral condensate $\left< \bar{\psi}\psi\right>$, which is needed for the pressure,
the leading form is linear in the quark masses. The precise extrapolation form for the 
for the gauge action or for the trace anomaly is not known. Thus, in order to 
determine the pressure one integrates along an LCP defined by a larger quark mass, 
after which the integration path is at fixed $\beta$. Along this last path
the integrand is the chiral condensate, for which chiral perturbation theory predicts 
the functional form. Using this technique one can avoid uncertainties in the 
chiral extrapolation. 

\begin{figure}\begin{center}
\includegraphics[width=8.5cm,angle=0]{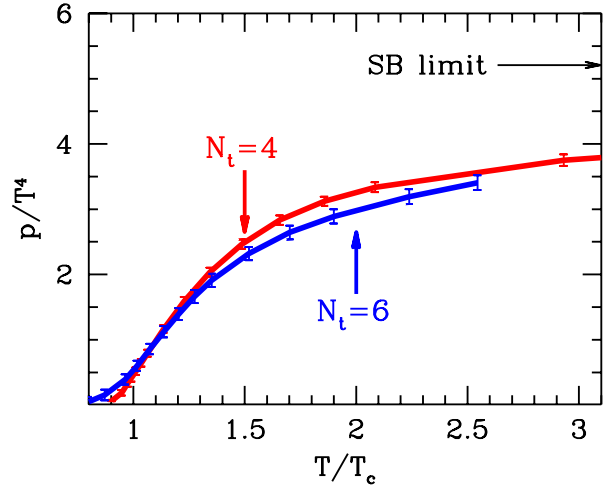}
\end{center}\caption{\label{eos_WB}
The pressure determined by the Wuppertal-Budapest group~\cite{Aoki:2005vt}.
}
\end{figure}

\begin{figure}\begin{center}
\includegraphics[width=8cm,angle=0]{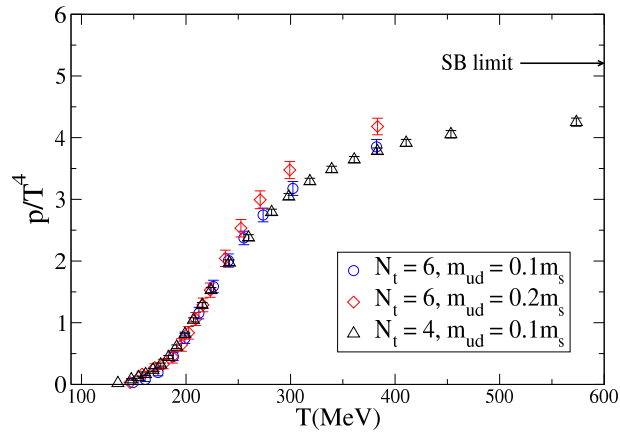}
\end{center}\caption{\label{eos_milc}
The pressure determined by the MILC collaboration~\cite{Bernard:2006nj}.
}
\end{figure}

\begin{figure}\begin{center}
\includegraphics[width=8.5cm,angle=0]{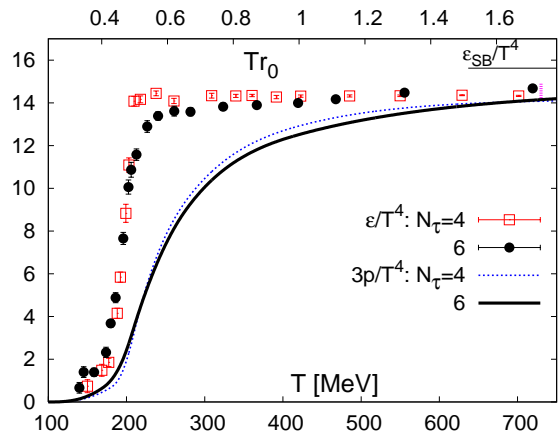}
\end{center}\caption{\label{eos_bie}
Pressure and energy density determined by the RBC-Bielefeld collaboration~\cite{Cheng:2007jq}.
}
\end{figure}

All three collaborations have results on the equation of state. The 
Wuppertal-Budapest group used physical quark masses and $N_t=$4,6 
lattices~\cite{Aoki:2005vt}.
The result is shown on Figure~\ref{eos_WB}. The MILC and RBC-Bielefeld 
collaborations used somewhat higher quark masses. Their results
are shown on Figures~\ref{eos_milc} and~\ref{eos_bie}. The 
RBC-Bielefeld collaboration also has a few points on $N_t=8$ lattices.

\newpage
\section{Note added in proof}

Since the submission of this review both the Wupperal-Budapest
and the HotQCD collaborations improved on their data. 
The Wuppertal-Budapest group used physical
quark masses also for the $T=0$ simulations and took even finer lattice 
spacings ($N_t=12$ and $N_t=16$ at one point) at finite 
temperature~\cite{Aoki:2009sc}. The $T_c$ values
remained consistent with the previous results.
The HotQCD collaboration uses
two lattice actions (asqtad and p4) and they extended their 
simulations to $N_t=8$ lattices~\cite{Bazavov:2009zn}. They also determined
the equation of state on $N_t=8$ lattices. 
The results of the two groups remained inconsistent.

A possible reason for the discrepancy could be the uncertainty coming
from the scale setting. Different quantities can be used to set 
the lattice spacing and the results should not depend on this choice.
It is important to check that the $T=0$ simulations, which are
needed for scale setting, are in the scaling regime. For staggered fermions,
due to the taste symmetry violation at finite lattice spacings, this is
particularly important. The Wuppertal-Budapest group has carried out such 
an analysis.

\begin{figure}[b]
\centerline{\includegraphics*[width=14cm,bb=32 405 580 702]{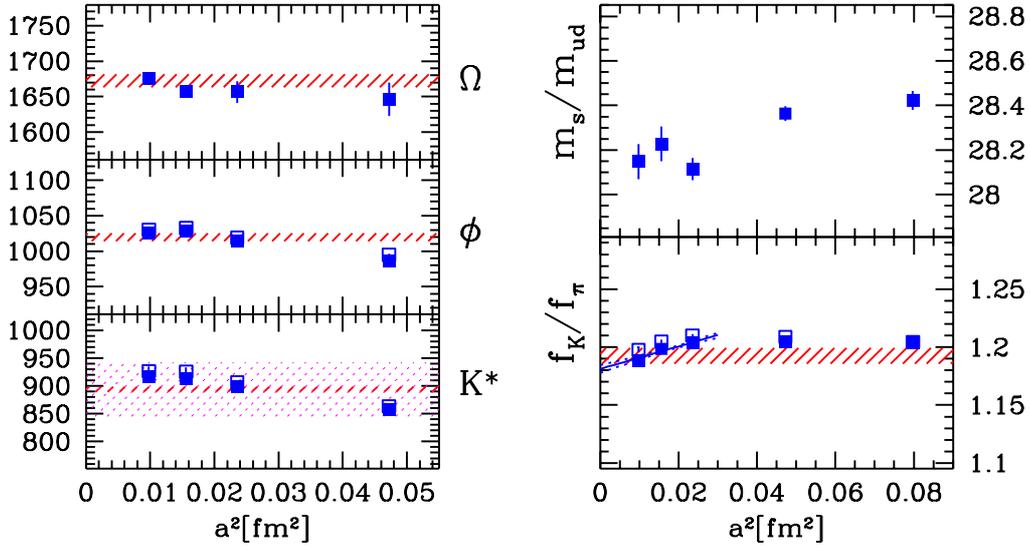}}
\caption{\label{fi:spect}
Left panel: masses of $\Omega$ baryon, $\phi(1020)$ meson and $K^*(892)$ meson in MeV 
on our four finest lattices as a
function of the lattice spacing squared. Right panel: quark mass ratio and $f_K/f_\pi$ for
all five ensembles. See text for a detailed explanation.
}
\end{figure}

Figure~\ref{fi:spect} shows the masses of the $\Omega$ 
baryon, $\phi(1020)$ meson and $K^*(892)$ meson as well as the ratio of
the quark masses and $f_K/f_\pi$ obtained from $T=0$ 
simulations of the Wuppertal-Budapest 
collaboration~\cite{Aoki:2009sc}. The agreement to the experimental values
indicates that the finite temperature results are independent of which
quantity ($\Omega$, $K^*$ or $\Phi$ mass, or the pion decay constant) 
is chosen for scale setting.

\begin{figure}
\centerline{
\includegraphics*[width=7.3cm,bb=0 430 300 704]{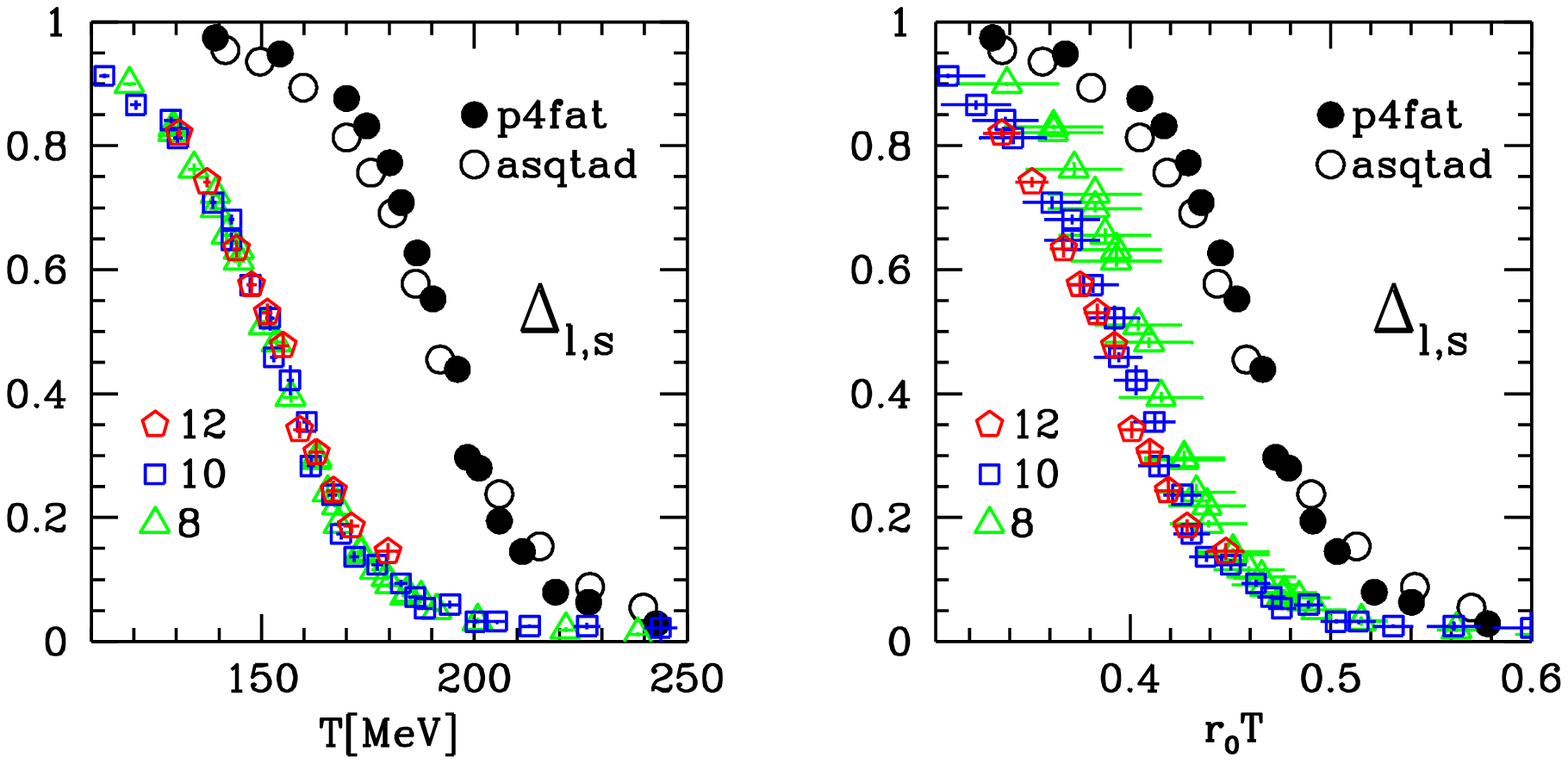}
\includegraphics*[width=7.3cm,bb=0 430 300 704]{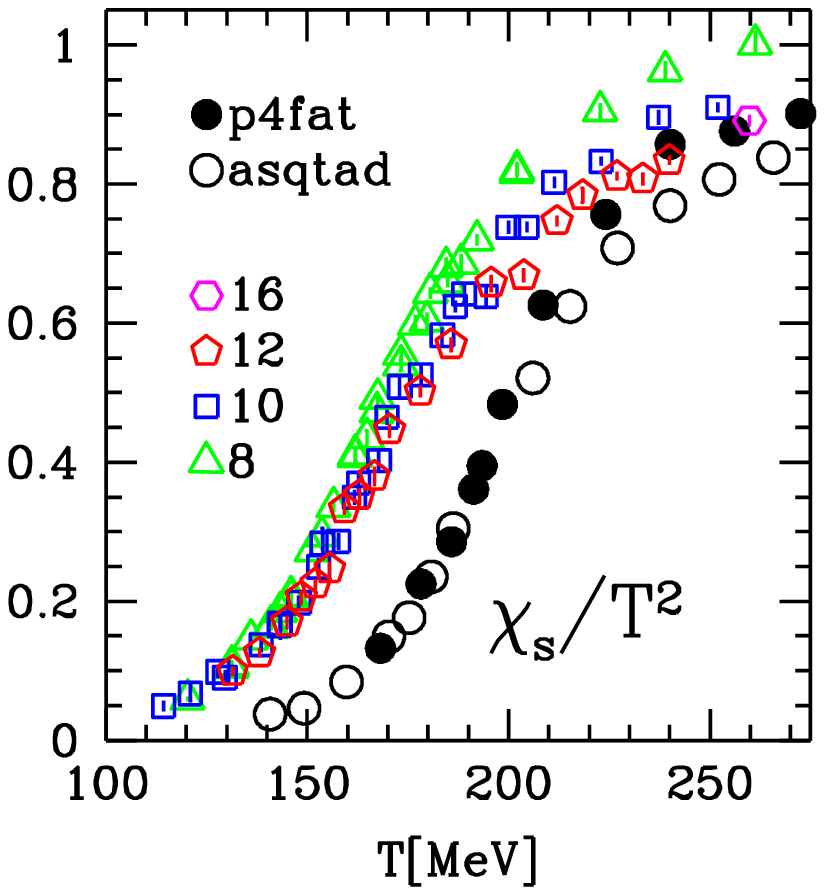}}
\caption{\label{fi:pbp_qns}
Left panel: renormalized chiral condensate as a function of the
temperature. The triangles, squares and pentagons correspond to
$N_t=6,8$ and 10 results of the Wuppertal-Budapest group, respectively.
The 'HotQCD' results are also shown by the closed and open circles. 
Right panel: strange quark number susceptibility.
}
\end{figure}

On Figure \ref{fi:pbp_qns}(left) the renormalized chiral condensate ($\Delta_{ls}$)
is shown as a function of the
temperature. We used the Wuppertal-Budapest data as well as 
the $N_t=8$ data of the 'hotQCD' collaboration from \cite{Karsch:2008fe}.
We can see a huge disagreement
between the curves in the transition regime. The shift between the curves
of the different groups is in the order of $35$ MeV.

The strange quark number susceptibility is
shown in Figure \ref{fi:pbp_qns}(right).
The $N_t=12$ data of the Wuppertal-Budapest group is shown with one additional
$N_t=16$ point at a
high temperature. The comparison with the results of the 'hotQCD' 
collaboration (see Reference
\cite{Karsch:2007dp}) brings us to a similar conclusion as for the 
chiral condensate. Around the transition point there
is an approximately $20$ MeV shift between the results of the two groups.

The most recent $T_c$ values published by the different groups are given in 
Table~\ref{ta:tc_results}. The latest results of the Wuppertal-Budapest group
are consistent with the ones
from 2006. The small difference comes from the fact that the experimental
value of $f_K$ has changed slightly since 2006. The discrepancy
between the Wuppertal-Budapest results and the 'HotQCD' ones 
is still present and has to be resolved by future
work.
\begin{table}[hb]
\begin{center}
\begin{tabular}{|c|c|c|c|c|c|c|}
\hline
 & $\Delta\chi_{\bar{\psi}\psi}/T^4$ & $\Delta\chi_{\bar{\psi}\psi}/T^2$ & 
$\Delta\chi_{\bar{\psi}\psi}$ & $\Delta_{l,s}$ & L &  $\chi_s$\\
\hline
Wuppertal-Budapest '09 & 146(2)(3) & 152(3)(3) & 157(3)(3) & 155(2)(3) & 170(4)(3) & 169(3)(3) \\
Wuppertal-Budapest '06 & 151(3)(3)  & - & - & - & 176(3)(4) & 175(2)(4)\\
RBCBC (ref.~\cite{Cheng:2006qk})         & - & 192(4)(7) & - & - & 192(4)(7) & -\\
\hline
\end{tabular}
\end{center}
\caption{
\label{ta:tc_results} 
Continuum extrapolated transition temperatures at the physical point 
for different observables and in different works. 
The first three columns give $T_c$ obtained from the chiral
susceptibility using different normalizations. The other three columns
give $T_c$ from the renormalized chiral condensate, renormalized 
Polyakov-loop and the strange quark number susceptibility.
}
\end{table}


\chapter{Finite chemical potential}\label{chap_mu}

In the last part of this review we discuss non-vanishing baryonic chemical potentials.
These studies are extremely difficult, since the infamous sign-problem spoils
any method based on importance sampling. Since the phase space is huge all 
lattice calculations use importance sampling. Thus for non-vanishing baryonic chemical 
potentials no direct simulations are possible. Until recently (till 2001) the most
advocated method was the Glasgow-method~\cite{Barbour:1997ej}, 
which was in principle theoretically correct, unfortunately only in the
infinite statistics limit. As it turned out even after producing several million      
configurations on tiny lattices the method did not work in practice. 
The reasons for this unlucky situation will be discussed later. There were 
several model studies, for systems other than QCD. Though these studies 
might give some insight to the real question (full QCD) they are of 
very limited practical use, therefore we do not discuss them any more.

In the following we show how one can introduce the baryonic chemical potential
in lattice field theory and illustrate the infamous sign problem. After that we
show the first method, which opened the way for quantitative predictions
in full QCD at non-vanishing baryonic chemical potentials (overlap improving 
multi-parameter reweighting). This method is still
one of the most accurate techniques, for many questions probably the best one we
know. Since then several other techniques were suggested, which we also discuss briefly.

Later we will discuss the potentials and goals of the forthcoming years. As we will
see, it is not yet possible to determine the continuum limit at 
non-vanishing baryonic chemical potentials. Thus, we will use non-improved
staggered {(\ref{hatas_stag})} and Wilson {(\ref{hatas_wilson_gauge})}
actions. Once the techniques are more established and more CPU power is available
than today, one should systematically analyze actions and decide which one
is the least CPU-demanding when we approach the continuum limit.

\section{Chemical potential on the lattice}
In continuum we use the grand canonical potential to treat non-zero 
chemical potential and use the corresponding $\mu N$ term (N is the
particle number). In the Euclidean lattice formulation the 
particle number is proportional to $\bar{\psi}\gamma_4\psi$.    
Thus, the most obvious solution for non-zero chemical potentials
would be to add a $\mu \sum_x \bar{\psi}\gamma_4\psi$
term to the action. It is easy to show that this choice leads to  
a quadratic divergence. Note, however, that a term  
of the form $\mu \sum_x \bar{\psi}\gamma_4\psi$ corresponds 
to a constant purely imaginary vector potential. Since we describe
gauge fields by link variables, it is straightforward to define     
non-vanishing chemical potentials also by link variables~\cite{Hasenfratz:1983ba}.
Based on these ideas it is clear, how to introduce $\mu$ on the lattice.
We multiply the forward timelinks $U_{x;4}$ by $e^{a\mu}$ and the     
the backward timelinks $U^\dagger_{x;4}$ by $e^{-a\mu}$, otherwise 
the form of the action remains the same. The staggered action  
{(\ref{hatas_stag})} reads: 
\bea\label{hatas_stag_mu}
S_{f,\rm{staggered}}(\mu)=\sum_x&\left[am\bar{\chi}\chi+
\frac{1}{2}\sum_{\nu=1\dots 3}\alpha_{x;\nu}
\left(\bar{\chi}_x U_{x;\nu}\chi_{x+a\hat{\nu}}-
\bar{\chi}_xU^\dagger_{x-a\hat{\nu};\nu}\chi_{x-a\hat{\nu}}
\right)+\right. \nonumber \\
&\alpha_{x;4}\left. \left(\bar{\chi}_x U_{x;4}e^{a\mu}\chi_{x+a\hat{4}}-
\bar{\chi}_xU^\dagger_{x-a\hat{4};4}e^{-a\mu}\chi_{x-a\hat{4}}
\right) \right].
\eea
For several quark fields (flavours) one has to introduce the 
chemical potentials for each flavours, they can be the same or   
they can be different. In the following we set the chemical 
potential of the strange quark to zero, whereeas the up and 
down quarks have the same chemical potentials. This choice is 
motivated by the physical conditions in heavy ion collisions
(the initial state has  zero strangeness, through the strong interactions
only strange-antistrange pairs can be produced --which does not change the
total strangeness, the only way to produce non-vanishing strangeness is 
through the weak interaction, which is subdominant).
Baryons with three light quarks have a baryonic chemical potential
$\mu_B$, which is three times the chemical potential of the light quarks.

For some questions one introduces the isospin chemical potential. To
that end the up and down quarks have opposite $\mu$ values. One can 
study systems with non-vanishing isospin chemical potentials by using
standard importance sampling methods, since the sign problem is not 
present in this case. In the rest of this review we want to deal with 
the sign problem and present various suggestions how to deal with it,
therefore we 
do not discuss the non-vanishing isospin chemical potential any longer.

\section{The sign problem}
In section \ref{sect_alg} we presented the available simulation 
algorithms, and discussed the necessary ingredients, particularly the 
positivity of the fermion determinant. At zero chemical potential this
positivity is garanteed by the $\gamma_5$ hermiticity of the fermion 
matrix. Unfortunately, at non-vanishing chemical potentials the $\gamma_5$
hermiticity is no longer fulfilled, the fermion determinant can take 
complex values. The partition function and the observables are real valued,
thus we can take the real part of the integrand $\Re \det M e^{-S_g}$.
The positivity of this quantity is, however, not garanteed, it can take
both positive and negative values. This is the so-called sign problem.

This feature (positive and negative signs in the integrand) has
two consequences. The more serious one is the impossibility to generate
configurations based on importance sampling (a function with negative
values can not be interpreted as a probability distribution). The 
other problem is related to the cancellation due to contributions of 
different signs. Even if we could generate the necessary configurations 
the sign of $\Re \det M e^{-S_g}$ for the individual configurations
oscillates, and there are large cancellations in the average, which
reduces the numerical accuracy.

\begin{figure}
\centerline{\includegraphics*[width=10cm,bb=10 230 590 640]{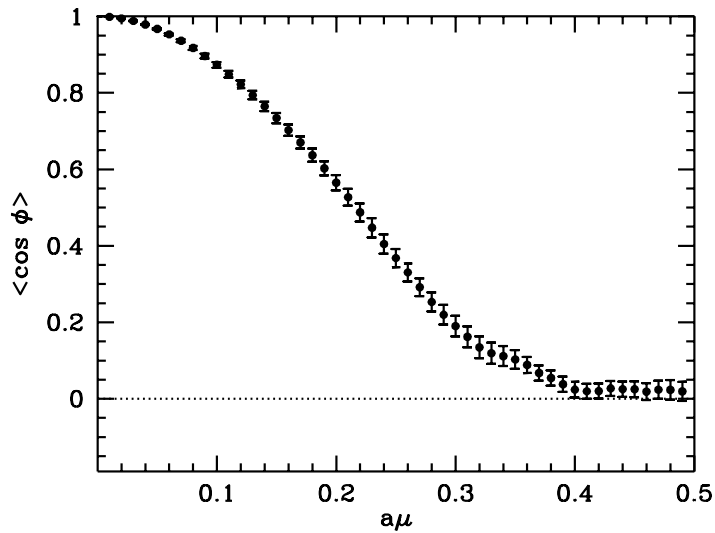}}
\caption{\small\label{fig_sign}
The expectation value of $\cos\phi$ as a function of $a\mu$, 
The phase of the fermion determinant is denoted by $\phi$. 
}   
\end{figure}
In order to illustrate the second problem let us write the determinant as
\be
\det M =\left| \det M \right| e^{i\phi}.
\ee
One can study the oscillation of the phase on a give ensemble. 
In order to do that we evaluate the determinant on each configuration
and calculate $\cos \phi$, which appears in the real part. The average 
of the $\cos \phi$ factors are shown as a function of the chemical potential
on Figure \ref{fig_sign}.
The configurations were obtained on a $8^3\cdot 4$ lattice at vanishing 
chemical potential at $\beta=5.1991$ and with
$am_{ud}=0.025$, $am_s=0.2$ quark masses. The gauge coupling was 
tuned to the transition point.                                 

As it can be seen for small chemical potentials  the expectation value
of $\cos \phi$ can be determined quite precisely, for $a\mu \gsi 0.4$
the phase oscillation is so strong, that the average of $\cos \phi$
is consistent with zero, the sign problem became quite serious.

\section{Multi-parameter reweighting}

A simple, but powerful generalization of the Glasgow method is the
overlap improving multi-parameter reweighting~\cite{Fodor:2001au}.
The partition function at finite $\mu$ can be rewritten as:
\begin{eqnarray}
\label{rew}
&Z = \int {\cal D}Ue^{-S_{g}(\beta,U)}[\det M(m,\mu,U)]^{N_f/4}= \nonumber\\
&\int {\cal D}U e^{-S_{g}(\beta_0,U)}[\det M(m_0,\mu=0,U)]^{N_f/4}\\
& \left\{e^{-S_{g}(\beta,U)+S_{g}(\beta_0,U)}
\left[\frac{\det M(m,\mu,U)}{\det M(m_0,\mu=0,U)}\right]^{N_f/4}\right\} \nonumber,
\end{eqnarray}
where the second line contains a positive definite action
which can be used to generate the configurations and the terms
in the curly bracket in the last line are taken into
account as an observable.
The expectation value of any observable can be then written in the form:
\be
<O>_{\beta,m,\mu}=\frac{\sum O(\beta,m,\mu) w(\beta,m,\mu)}{\sum w(\beta,m,\mu)}
\label{rew_O}
\ee
with $w(\beta,m,\mu)$ being the weights of the configurations defined by
the curly bracket of eqn. (\ref{rew}).

\begin{figure}[h!]
\centerline{\includegraphics*[width=10cm]{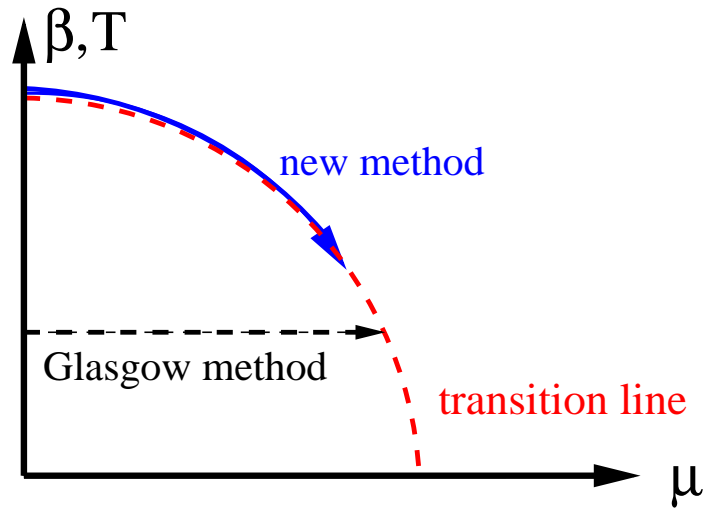}}
\caption{\label{method}
Comparison of the Glasgow method and the the new (multi-parameter 
reweighting) method. The Glasgow method collects an ensemble
deep in the hadronic phase, attaches weights to the individual
configurations and attempts to get information about the phase
diagram, thus informations about configurations on the other side 
of the phase line. There is no overlap between the original
typical configurations (hadronic phase) and the configurations 
of the new phase. This is the reason why the method fails. The
new technique (overlap improving multi-parameter 
reweighting) determines the 
phase line in a different way. First one tunes the system to 
the transition point at $\mu$=0. At this point the configuration
ensemble contains configurations from both phases. A 
simultaneous reweighting is done in $\beta$ (or in other
words in the temperature) and also in $\mu$. Since we are
looking for the phase line, thus for an equal ``mixture''
of the two phases, a careful change of the two parameters
keeps the system in this mixed phase. The overlap between 
$\mu$=0 and $\mu\neq$0 is much better, the phase line
can be determined. 
}
\end{figure}

The main difference from the Glasgow method is that reweighting is
done not only in $\mu$ but also in the other parameters of the
action (at least in $\beta$, but possibly also in $m$). This way
the overlap can be improved. If the starting point ($\beta_0, m_0, \mu_0=0$)
is selected to be at the $\mu=0$ transition point then a much better
overlap can be obtained with transition points at higher $\mu$.
A schematic figure shows the main differences between the
two techniques (see Figure \ref{method}).

Though all formulas are exact, nevertheless the practical applicability depends
on the fluctuation of the weights $w(U)$. In principle there might be two 
difficulties. The first one is the so-called overlap problem. 
In order to illustrate this question we study a reweighting, for which
both the simulation and the target parameters allows direct importance       
sampling based simulations, thus the determinants are for both ($\beta,m,\mu$)
sets positive, e.g. for $\mu=0$. 
\footnote{In such circumstances reweighting is clearly not needed,
however, it is provides us a useful illustration.} Since the weights are positive
real numbers, importance sampling is possible. Thus, there are some smaller subsets
of configurations, which are particularly important (these subsets are
selected by the importance sampling procedure). This is true for both parameter sets
($\beta,m,\mu$). In practice, the simulations results in some configurations
from these subsets. If the two subsets are disjoint (the typical
configurations of one of the subsets are quite different than the typical
configurations of the other subset) the so-called overlap problem appears. In this case
assigning new weights to the configurations does not help, since the most
important, typical configurations are simply missing. One of the most         
unconvenient features of the overlap problem is that it is very difficult to detect.
Let us look at an extreme example and assume that the two subsets are very 
different, e.g. they correspond to different phases of the physical system. 
For small ensembles usually no configurations can be obtained from the other phase.
The $w(U)$ will be of similar size and the result will have a artificially small
statistical error. If we increase the statistics a configuration, which is just
typical for the other phase, might appear. This single configuration receives
a large weight and it will change and even dominate the result. Clearly, as long as  
our ensemble is small and no such configuration is produced, one is not aware of
the overlap problem. This was exactly the most serious problem of the Glasgow 
method.

The other difficulty for the multy-parameter reweighting is the
sign-problem. The phase of the determinant appears in the weights $w$. For
large chemical potentials the sum of these complex weights can be even consistent
with zero (at least for small statistics). In these cases the expression
{(\ref{rew_O})} will be practically $0/0$. This feature is fortunately   
signalled by the jackknife analysis, since it uses different subsamples (the
cancellation in different subsemples are different, which influence the jackknife
error). A large statistical error is a clear sign of the sign-problem.

\subsection{Multi-parameter reweighting with Taylor-expansion}
There is a variant of the above multi-parameter reweighting technique,
which needs less computational power, particularly for large
lattices, though this method contains somewhat less informations of the
$\mu$ dependence. Let we discuss how it works.
The use of eqn. (\ref{rew}) requires the exact calculation
of determinants on each gauge configuration which is computationally
expensive. 
Instead of using the exact formula, one can make a Taylor expansion
for the determinant ratio in the weights~\cite{Allton:2002zi} (for simplicity
assuming no reweighting in the mass):
\be
\ln\left[\frac{\det M(\mu)}{\det M(0)}\right]^{N_f/4} =
\sum_{n=1}^{\infty} \frac{\mu^n}{n!}
\frac{\partial^n \ln \left[ \det M(0)\right]^{N_f/4}}{\partial \mu^n}
\equiv\sum_{n=1}^\infty
{R}_n\mu^n.
\ee
Taking only the first few terms of the expansion one gets
an approximate reweighting formula. The advantage of
this approximation is that the coefficients are
derivatives of the fermion determinant at $\mu=0$, which 
can be well approximated stochastically.
However, due to the termination of the series and the
errors introduced by the stochastic evaluation of the
coefficients we do not expect this method to work for 
as large $\mu$ values as the full technique.
Indeed, it has been shown in Ref.~\cite{deForcrand:2002pa} that
even for very small lattices (i.e. $4^4$) the
phase of the determinant is not reproduced 
by the Taylor expansion for $a\mu\ge 0.2$.

\subsection{Simulations at imaginary $\mu$}
The fermion determinant is positive definite if we use a purely imaginary 
chemical potential. So if the transition line $T_c(\mu)$ is an analytic
function then we can determine it for imaginary $\mu$ values and 
analytically continue back to real $\mu$-s~\cite{deForcrand:2002ci}. 
The analytic continuation
is in general impossible from just a finite number of points. However, 
taking a Taylor expansion in $\mu$ or $\mu/T$ one gets:
\be
\frac{T_c(\mu_B)-T_c(0)}{T}=a_2\left(\frac{\mu_B}{T}\right)^2+
a_4\left(\frac{\mu_B}{T}\right)^4+\dots
\ee
The coefficients $a_i$ can be determined from imaginary $\mu$ simulations.
One simply measures $T_c(\mu_I)$ for imaginary $\mu_I$-s and fits it
with a finite order polynomial in $\mu_I/T$.
Recently, a generalization of this method was proposed by using a more
general form of the action which still preserves the positivity
of the fermion determinant~\cite{Azcoiti:2004ri}.

Recently, instead of using the grand-canonical partition function
a canonical approach was also applied to study QCD at non-zero 
density~\cite{Kratochvila:2004wz,Alexandru:2005ix}.
This technique involves a Fourier integral for which the fermion
determinant at imaginary $\mu$ values is needed. The sign problem emerges
as fluctuations during the evaluation of this Fourier integral.

\subsection{Differences and similarities of the three techniques}

Although the three described methods seem different they are essentially
the same. The connection between exact reweighting and Taylor expansion
is obvious: the latter is an approximation of the former, using all 
non-vanishing
orders in the Taylor expansion gives exactly the same results as reweighting.

\begin{figure}[h!]
\centerline{\includegraphics*[width=10cm]{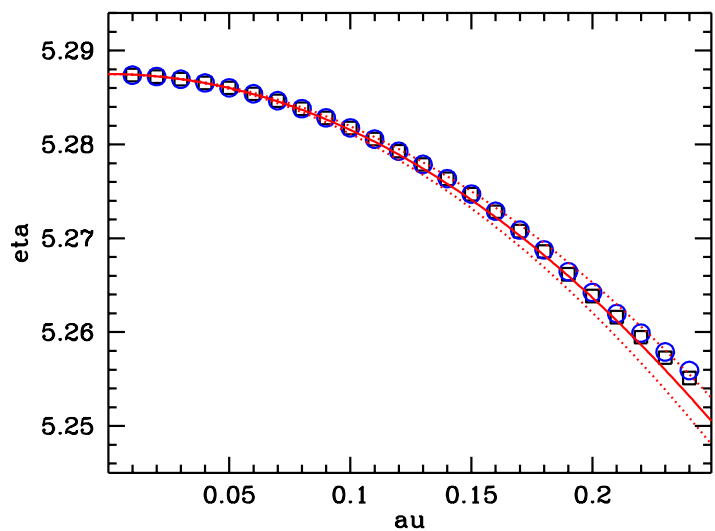}}
\caption{\label{fig_comp}
The $N_f=2$ phase diagram of Ref.~\cite{deForcrand:2002ci} obtained via
analytic continuation from imaginary $\mu$ (solid line; dotted lines show
the uncertainty) and the same system calculated by
exact multi-parameter reweighting (boxes) and Taylor expansion up to 
$\mu^4$ order (circles). 
There is a perfect agreement. To enhance
the differences the results were matched at $\mu=0$ points (note, that they agree within their uncertainties).
The errors are smaller than the symbol sizes.
}

\end{figure}

To see the connection between reweighting and analytic continuation is not so
straightforward. Since the phase diagram for imaginary $\mu$ is fitted
by a polynomial it yields the $\mu$ derivatives at $\mu=0$ (the closest point
to the real $\mu$ domain, since $\mu^2$ is the natural variable).
In this sense it should give the same results as the Taylor expansion method
in the same order. Thus, for moderate $\mu$ values the imaginary
$\mu$ method should also agree with 
exact reweighting. The agreement is demonstrated on Figure~\ref{fig_comp}.
In order to avoid difficulties when comparing different discretizations,
different quark masses, different choices to transform lattice
data into physical units and exact/non-exact Monte-Carlo generators we
applied the three methods using identical circumstances. We took
the same phase diagram as in Ref.~\cite{deForcrand:2002ci} 
and used their determination for the curvature of the phase line.
Their result is perfectly reproduced (upto four digits) by
multi-parameter reweighting with full determinants and also by 
the Taylor technique. As the chemical potential gets larger the
results start to deviate. This fact is an obvious consequence of the
higher order $\mu$ terms, which are missing both from the imaginary
chemical potential method and from the Taylor expansion technique.

As we mentioned previously in the case of staggered fermions a fractional
power of the fermion determinant is taken in order to have less than
four flavors. For $\mu >0$ this leads to an additional difficulty. The 
fourth root of a complex number cannot be taken unambiguously. There
are several ways to circumvent this problem. It has been shown in 
Ref.~\cite{Golterman:2006rw} that near the continuum limit these unambiguities 
dissappear and a unique fourth root can be defined. It has also been argued, 
however, that current lattices are not yet close enough to the continuum in 
this sense. The procedure the authors of Ref.~\cite{Golterman:2006rw} propose does
not work on todays lattices. An alternative method to choose among the Riemann
leaves is given in Ref.~\cite{Fodor:2001pe} which assumes analyticity of the 
fourth root along the real $\mu$ axis. Close to the continuum where the
procedure of Ref.~\cite{Golterman:2006rw} can be applied the two methods choose the same roots, thus they agree.
Since both Taylor expansion and analytic continuation from imaginary chemical
potentials implicitly assume analyticity they correspond to the same choice as that of Ref.~\cite{Fodor:2001pe}.

\subsection{Determining the phase diagram by Lee-Yang zeroes}

It is particularly convenient to determine the phase diagram by using the
method of Lee and Yang~\cite{Yang:1952be,Lee:1952ig}. The method is based on the
behaviour of the zeros of the partition function on the complex plane.
It can be effectively used, if the transition is strong enough (first order   
phase transition or a rapid crossover). \footnote{At $\mu=0$ our continuum extrapolated
analysis resulted in a weaker transition, therefore 
other methods were necessary.} 

Let us assume that the system ondergoes a first order phase tansition. In this
case at large volumes $V$ the two phases can coexist (in the vicinity of the 
transition: one of the phases is usually metastable). The partition function can be 
written as
\be\label{LY}
Z=e^{-\frac{V}{T} f_A}+e^{-\frac{V}{T} f_B},
\ee
where $f_A$ and $f_B$ are the free energy densities of the two phases.
At the transition point $T=T_c$  the two free energy densities
coincide $f_A=f_B$. Changing $\beta$ the system can be tuned away from $T_c$.
At the transition point we have $\beta=\beta_c$ and in its vicinity 
$T$ and $f_B$ can be Taylor expanded around $T_c$ and $f_A$ (the expansion
parameter is $\Delta\beta=\beta-\beta_c$ around $\beta_c$).
\begin{align}
T=T_c+c_1 \Delta\beta +{\cal O}(\Delta\beta^2) &&
f_B=f_A+c_2 \Delta\beta +{\cal O}(\Delta\beta^2)
\end{align}
with some  $c_1$ and $c_2$ coefficients.
Writing it back into {(\ref{LY})} the partition function $Z$ reads:
\be
Z=e^{-\frac{V}{T_c} f_A}\left(e^{aV\Delta\beta}+e^{bV\Delta\beta} \right),
\ee
where
\begin{align}
a=\frac{f_A c_1}{T_c^2} && b=a-\frac{c_2}{T_c}.
\end{align}
Rearranging the expression gives
\be
Z=2\exp\left({-\frac{V}{T_c} f_A + \frac{a+b}{2} V\Delta\beta}\right)
\cosh\left(\frac{a-b}{2}V\Delta\beta \right).
\ee
The first term can not be zero, however the second one vanishes for purely
imaginary $\Delta \beta$ with 
\be
\Im \Delta \beta =\frac{2}{(a-b)V}(k+\frac{1}{2})\pi,
\ee
where $k$ is an integer. Thus, for large enough volumes the partition function
Z($\beta$) has zeros on the complex plane. These are the Lee-Yang zeros.
The real part of these zeros are given by
$\Re \beta =\beta_c$ and $\Im \beta \propto 1/V$.
Thus, locating the Lee-Yang zeros the real part can be interpreted as the transition
point, whereas the 1/V scaling of the imaginary parts indicate a first 
order phase transition. In the $V\to\infty$ limit the imaginary parts tend to 
zero, which generates the singularity of the free energy at some real $\beta_c$.
For a crossover there are no singularities in the infinite volume limit, therefore
the zeroes do not approach the real axis. In numerical studies one usually
uses the first zero at $k=0$, since it is the closest one to the real axis,
along which the simulations are carried out.

The determination of the Lee-Yang zeros can be done by reweighting. 
We determine
$Z/Z_0$ for complex $\beta$ values in the vicinity of the 
simulation point $\beta_0$. To that end one has to add the weights $w$.
At $\mu=0$ this is particularly easy. The weights, for plaquette gauge
action, can be written as
\be
w(U)=e^{-S_g(\beta)+S_g(\beta_0)}=e^{(\beta-\beta_0)P}
\ee
Thus, measuring the averages of the plaquette variable $P_i$ the Lee-Yang zeroes 
are obtained by solving 
\be
\sum_i e^{(\beta-\beta_0)P_i} = 0
\ee
on the complex $\beta$ plane. The real part of the solution gives the transition
point. At non-vanishing chemical potentials the procedure is somewhat
more involved, but can be carried out, too. In this case one has to solve the  
following equation:
\be
\sum_i \left(\frac{\det M(\mu)}{\det M(\mu=0)} \right)_i e^{(\beta-\beta_0)P_i} = 0,
\ee
thus we  have to calculate the ratios of the determinants $\det M(\mu)/\det M(\mu=0)$ 
on each configuration.

\section {Results for the phase diagram}
In the following we review lattice results for the phase line separating
different phases and the critical point.

\subsection{Phase line}
All the discussed methods were used to give the phase line. The results
are in agreement although different regularizations and quite 
coarse lattices were used. Up to now all results were obtained for one set
of lattice spacings, on $N_t=4$ lattices.
(Let us emphasize, that different
discretizations should agree only at vanishing lattice spacings,
thus in the continuum limit. At non-vanishing lattice spacings
one usually has different results for different
lattice actions.) 

Using multi-parameter reweighting, the phase diagram was determined for
4 and 2+1 flavors of staggered 
fermions~\cite{Fodor:2001au,Fodor:2001pe,Fodor:2004nz}. 
For the physically interesting 
latter case both semi-realistic and realistic quark masses were used. The
phase diagram using physical quark masses is shown on Figure~\ref{pd_chir}
which will be discussed later in more detail.

\begin{figure}[h!]
\centerline{\includegraphics*[width=6cm]{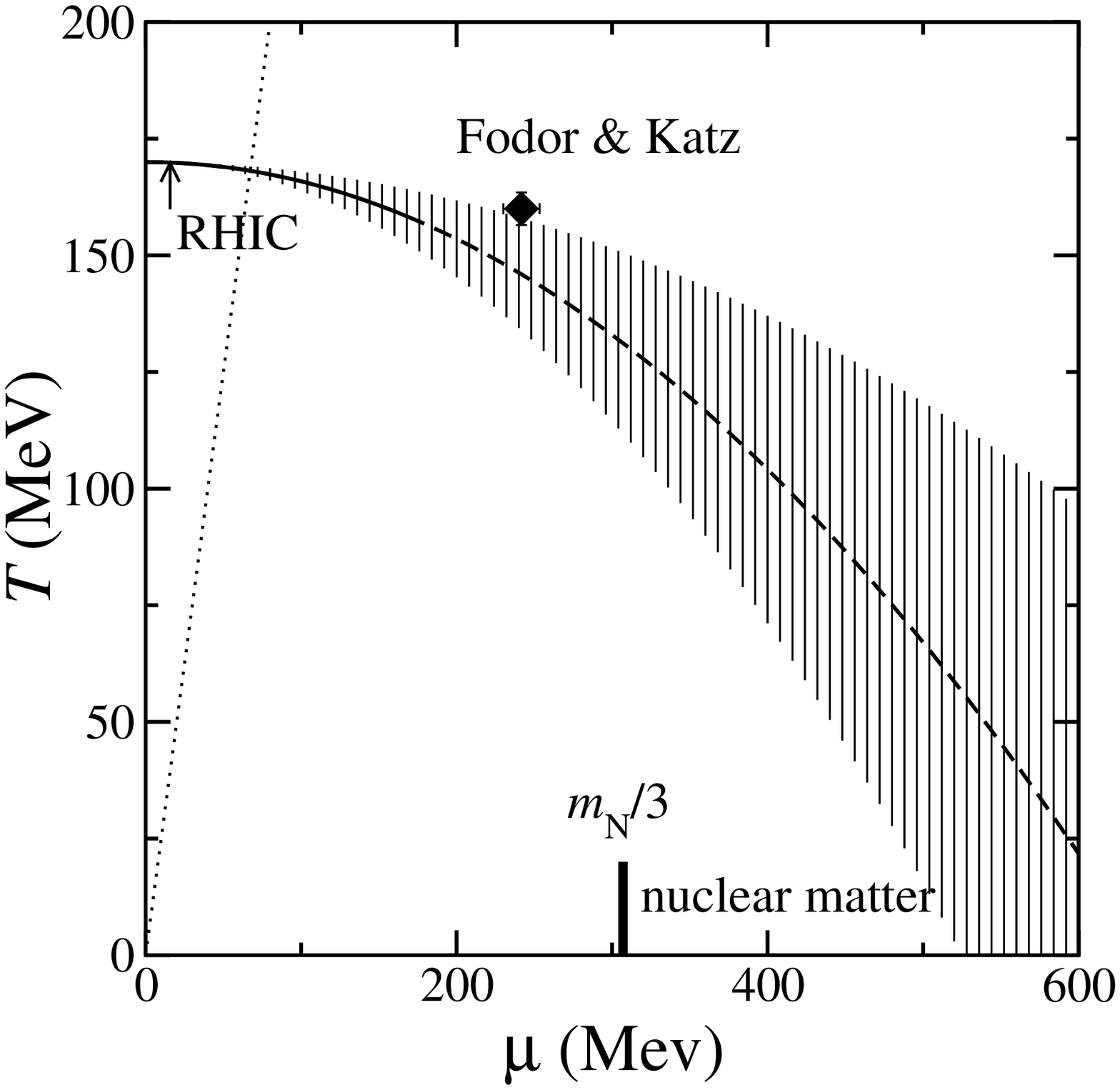}\hspace*{0.5cm}
\includegraphics*[width=8cm]{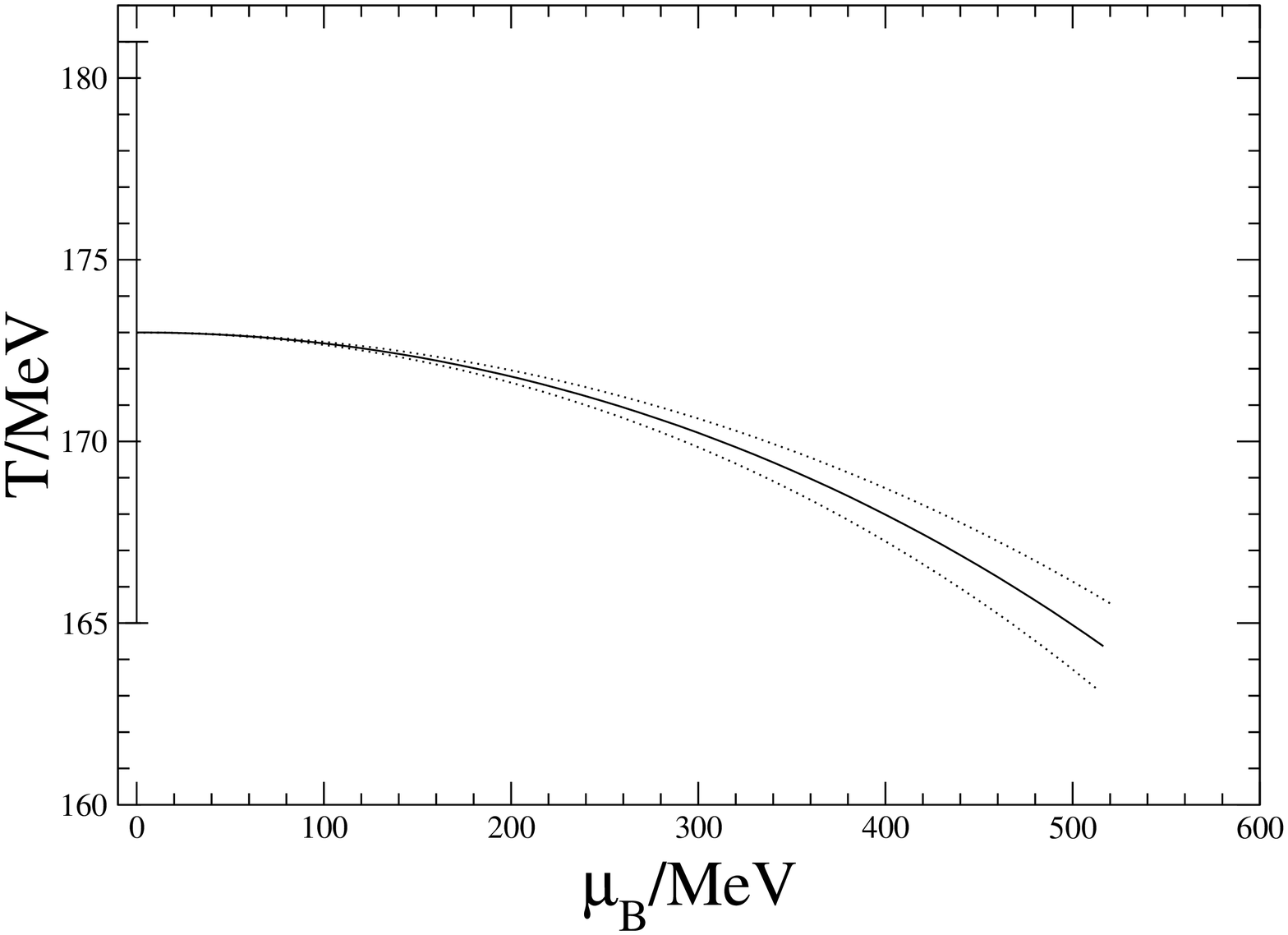}}
\caption{\label{fig_pd_2}
{\it Left:} The phase diagram  obtained from the Taylor expansion method
using two flavors of p4 improved fermions with a pion mass of 
$\approx$750~MeV (figure from Ref.~\cite{Allton:2002zi}).
{\it Right:} The phase diagram via analytic continuation from 
imaginary $\mu$  using two flavors of standard staggered fermions
and a pion mass of $\approx$230~MeV~\cite{deForcrand:2002ci}.
}
\end{figure}

The phase diagram obtained via Taylor expansion~\cite{Allton:2002zi} is 
shown on the left panel of Figure~\ref{fig_pd_2}. Two flavors of p4 improved
staggered fermions were used in this analysis. The critical point of 
Ref.~\cite{Fodor:2001pe} is also shown as a comparison. Note that although
different lattice actions were used at a finite lattice spacing there is a 
good agreement.

The right panel of Figure~\ref{fig_pd_2} shows the phase diagram obtained
by analytic continuation from imaginary $\mu$. The same method was also 
applied to four flavors of staggered fermions in Ref.~\cite{D'Elia:2002gd}.
Consistent results were found with a 
generalization of the method which made it possible to reach
somewhat larger values of $\mu$~\cite{Azcoiti:2005tv}.

In the case of multi-parameter reweighting the absolute temperature scale was
determined by a $T=0$ spectrum determination while in the case of the other
methods only perturbative $\beta$ functions were applied.

\begin{figure}[h!]
\centerline{\includegraphics*[width=10cm]{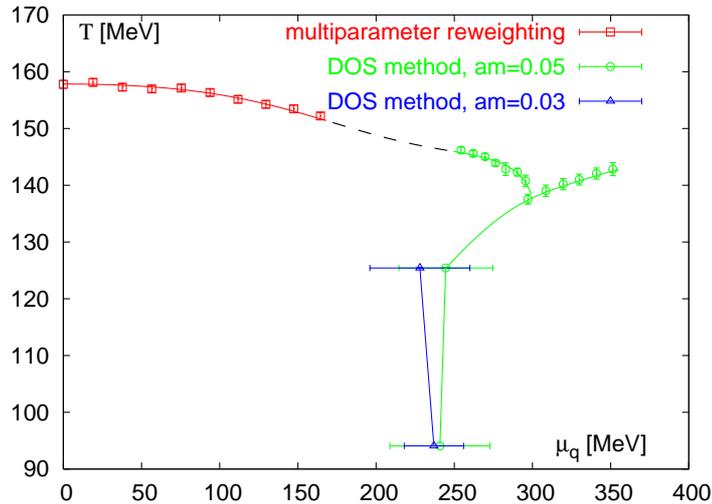}}
\caption{\label{dos}
The phase diagram of four flavour staggered QCD obtained with the density
of states method on $N_t=4$ lattices. A triple point was found at around
300~MeV quark chemical potential~\cite{Fodor:2007vv}.
}
\end{figure}

The latest result on the phase line comes from a combination
of multi-parameter reweighting and the density of states 
method~\cite{Fodor:2007vv}. The
phase diagram of four flavor staggered QCD 
was determined up to three times larger chemical potentials
than with previous methods. A triple point was found around 900~MeV baryonic
(300~MeV quark) chemical potenticals (see Figure~\ref{dos}).

\subsection{The critical point}
One of the most important features of the phase diagram is the 
possible critical point
separating a crossover region from a first-order phase transition regime.
If such a point exists, its location is an unambiguous prediction of QCD.

Since real phase transitions only occur at infinite volume a determination of
the order of the transition and thus locating the critical point is only
possible via a finite size scaling analysis. At a single, finite volume 
everything is analytic, no real phase transitions 
--and thus no critical point--  exist.

There are different possible strategies to locate the critical point.
One can use Lee-Yang zeroes, Binder-cumulants or the convergence radius
of the free energy density. These techniques will be discussed below.
They can be applied directly, by determining the appropriate observables
at finite $\mu$ using one of the methods described before. Another possibility
is to start from a non-physical point (using small quark masses) where the
critical point is located at zero or purely imaginary $\mu$ values and then
determine the quark mass dependence of the critical point and extrapolate to
the physical quark masses. The extrapolation, as usual, 
introduces errors, which are difficult to control.

At finite volumes the transition between the hadronic and quark-gluon phases
is always continuous, the free energy density is analytic for all real values
of the parameters of the action. Nevertheless, the partition function has 
zeroes even for finite volumes at complex values of the parameters. For
a first-order phase transition these zeroes approach the real axis when the
volume is increased -- thus generating the singularity of the free energy for 
real parameter values. A detailed analysis shows that the imaginary
part of these Lee-Yang zeroes scales as $1/V$ for large volumes.
For a crossover the Lee-Yang zeroes do not approach the real axis when
the volume is increased.
Therefore inspecting the volume dependence of the imaginary parts of 
the Lee-Yang zeroes one can distinguish a first-order transition from an 
analytic crossover.

Binder cumulants can also be used to locate critical points. In the 
infinite volume limit they converge to 1 in case of first order phase
transitions and specific values (determined by the universality class)
for second order phase transitions. For details see e.g. 
Ref.~\cite{Karsch:2001nf}
where the critical point of three flavor QCD at $\mu=0$ was determined
using this technique.

The convergence radius of the Taylor expansion of the free energy gives
the distance from the expansion point to the nearest singularity. If
all expansion coefficients are positive then the singularity is at a
real value of the expansion parameter which can than only be the critical 
point. As discussed before, this can only happen at infinite volume. 
The expansion coefficients have to be extrapolated to infinite volume, 
one has to be ensured that they are all positive and then the convergence
radius can be calculated from them. Since the last two steps would require
the knowledge of all coefficients (especially the positivity condition), 
this method gives a
lower limit on the location of the critical point.

A necessary (but not satisfactory) condition of the existence of the critical 
point is a crossover at $\mu=0$. We have seen in the previous sections that 
using staggered fermions, this is indeed the case.

\begin{figure}[h!]
\centerline{\includegraphics*[width=12cm]{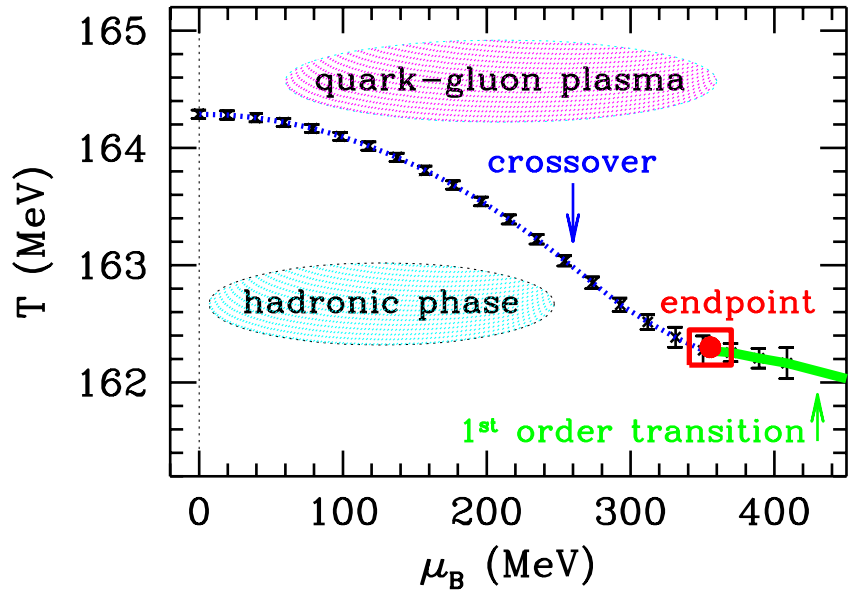}}
\caption{\label{pd_chir}
The phase diagram obtained with multi-parameter reweighting using
2+1 flavors of standard staggered fermions corresponding to the
physical pion mass. The dotted part of the transition line is the crossover
region while the solid line is of first order. 
The box shows the critical endpoint separating them.
}
\end{figure}

The multi-parameter reweighting combined with the Lee-Yang-zero analysis was
used to locate the critical point. The first study was done with 
semi-realistic quark masses corresponding to a pion mass of 
$\approx$230~MeV~\cite{Fodor:2001pe}.
The critical point was found at 
$T_E=160 \pm 3.5$~MeV and $\mu_E= 725 \pm 35$~MeV. 
The whole study was repeated using larger volumes and physical quark masses in
Ref.~\cite{Fodor:2004nz}. The results can be seen on Figure~\ref{pd_chir}.
The critical point is located at $T_E=162 \pm 2$~MeV and 
$\mu_E= 360 \pm 40$~MeV. One can see that the critical point moved to 
a smaller value of $\mu$ as the quark masses were decreased. This is in 
complete agreement with expectations. It is important to emphasize again 
that both of these results were obtained for one set of lattice spacings,
the continuum extrapolation is still missing.

The Taylor-expansion technique was used to determine the mass dependence
of the critical point as discussed above. Starting from the three-flavor
critical point where the phase transition is of second order at $\mu=0$,
the derivative $d\mu_E/dm$ was determined. A linear extrapolation to larger
quark masses using only this first derivative gave $\mu_E \approx 420$~MeV
for the location of the critical point for physical quark 
masses~\cite{Ejiri:2003dc}. Although
an extrapolation to very distant quark masses was done which introduces 
unknown and possibly large errors, this value is in  agreement with 
the value obtained directly via multi-parameter reweighting.

Another application of the Taylor-expansion method was done in 
Ref.~\cite{Gavai:2004sd} using two flavors of staggered fermions. 
The convergence radius of the series was estimated
using the first few coefficients. The authors found 
$T_E \approx 0.95 T_c$ and $\mu_E \approx 1.1 T_E$ which is significantly 
smaller than the multi-parameter reweighting result. We should not forget, 
however, that --as discussed before-- this result can only be considered as
a lower limit on the location of the critical point.

For small enough quark masses the critical point can be located at a purely
imaginary $\mu$. Approaching the point where the critical point reaches 
$\mu=0$ one can determine the derivative $d\mu_E^2/dm$ for negative
values of $\mu_E^2$. This analysis was carried out in 
Refs.~\cite{deForcrand:2003hx,deForcrand:2006pv}. For  negative
values of $\mu^2$ the critical quark mass $m_c$ was located and the
derivative $dm_c/d\mu^2$ was determined (which is just the inverse of the
above quoted derivative).

In Ref.~\cite{deForcrand:2003hx} $dm_c/d\mu^2$ was found rather small which
by a rough, linear extrapolation would suggest a much larger value
of $\mu_E$ for physical quark masses than found e.g. by multi-parameter 
reweighting.
More interestingly, when a similar analysis was done using an exact 
simulation algorithm instead of the previously applied approximate R 
algorithm, $dm_c/d\mu^2$ was found to be negative (but consistent with zero on
the two-$\sigma$ level)~\cite{deForcrand:2006pv}. Further calculations
increased the significanse of this result greatly~\cite{deForcrand:2008vr}.
However, both calculations were done on coarse, $N_t=4$ lattices only. 
Conventionally, a positive value is expected for the derivative
which was also observed
with multi-parameter reweighting (larger quark masses lead to a larger
value of $\mu_E$). 
Effective model calculations also support the positive
sign (for a recent study, see e.g.~\cite{Kovacs:2006ym}).
Future lattice studies on finer lattices, and eventually a continuum 
extrapolation will give the final answer.

\section{Equation of state at $\mu>0$}

Besides the transition line and the critical point one can also determine the
equation of state above and slightly below the transition line. The pressure
can be calculated similarly to the $\mu=0$ case using the integral method.
The only difference is that now we also have to include the 
chemical potential as an integration variable:
\bea
\frac{p}{T^4}=
N_t^4\int^{(\beta,am_q,a\mu)}_{(\beta_0,am_{q0},a\mu_0)}
d (\beta,am_q,a\mu)\left[
\frac{1}{N_tN_s^3}
\left(\begin{array}{c}
{\partial \ln Z}/{\partial \beta} \\
{\partial \ln Z}/{\partial (am_{q})} \\
{\partial \ln Z}/{\partial (a\mu)} \\
\end{array} \right ) \right.-\nonumber \\
\left. 
\frac{1}{N_{t0}N_{s0}^3}
\left(\begin{array}{c}
{\partial \ln Z_0}/{\partial \beta} \\
{\partial \ln Z_0}/{\partial (am_q)}\\
{\partial \ln Z_0}/{\partial (a\mu)}
\end{array} \right )
\right].
\eea
This expression is analogous to the one applied at $\mu=0$. The partial 
derivatives of $\ln Z$ correspond to the same observables as before. The 
only new one is ${\partial \ln Z_0}/{\partial (a\mu)}$ which is proportional
to the $n_q$ quark number density. To carry out the integration
one has to calculate the expectation values of these observables at non-vanishing
$\mu$. This can be achieved by any of the previously discussed methods.
In the following we discuss shortly the case of reweighting.

\begin{figure}
\centerline{\includegraphics*[width=8cm]{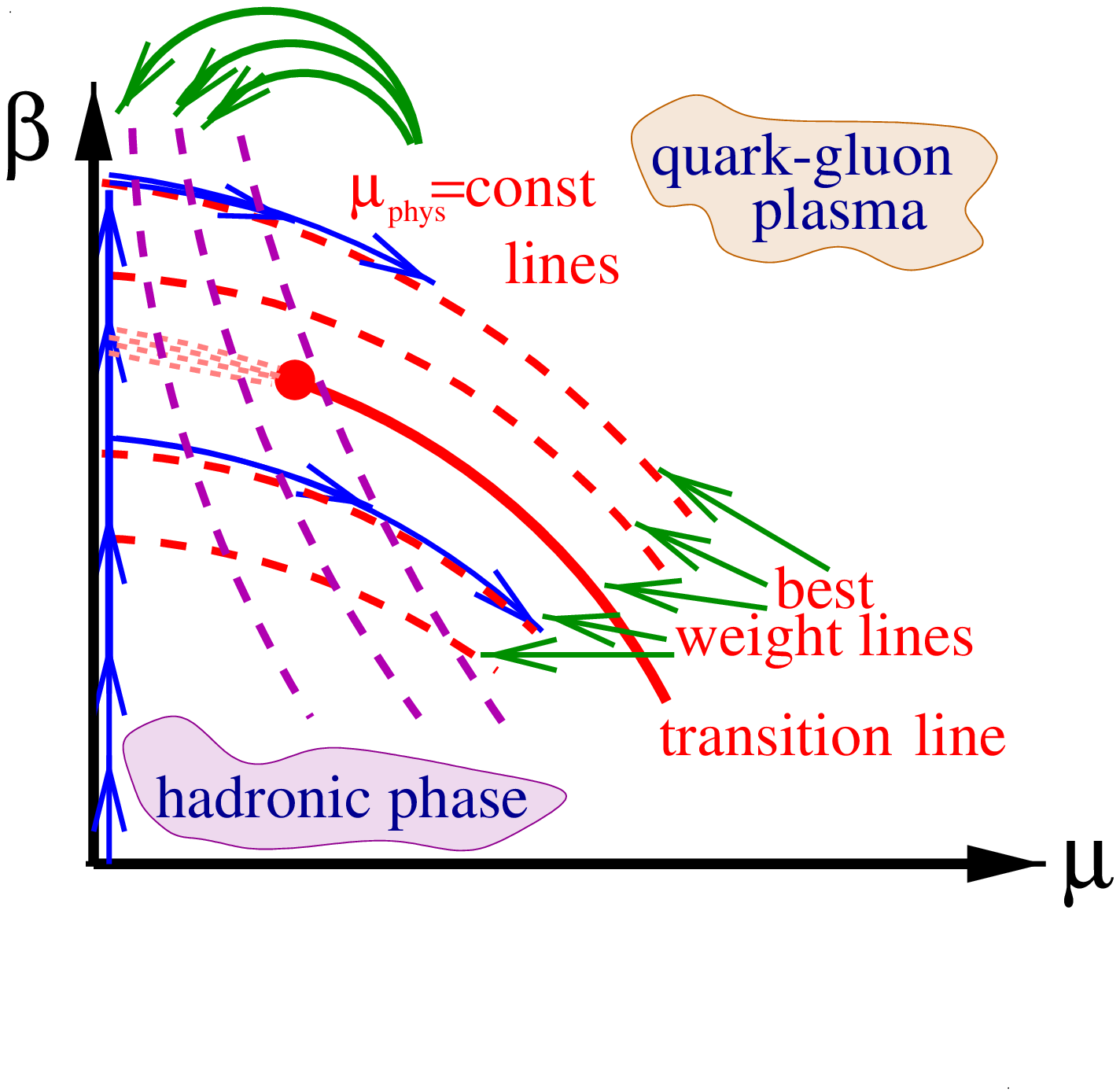}}
\caption{\small\label{fig_method_i}
The contours used in the integral method 
to evaluate the pressure on the $\beta$--$\mu$ plane (blue lines). 
First, we integrate from $\mu=0$ up to some $\beta_0$ 
along the LCP. Then a best weight line is followed to the 
target $\beta, \mu$ values (red dashed lines).
}
\end{figure}

Any observable can be evaluated at finite $\mu$ by reweighting (with or 
without Taylor expansion) using eqns. (\ref{rew}) and (\ref{rew_O}). 
In order to maximize the overlap between the configurations
generated at $\mu=0$ and the target ensemble, one has to choose
the starting point of the reweighting ($\beta_0$ at $\mu=0$) properly.
One possibility is to minimize the spread of the weights of the
configurations appearing in eqn. (\ref{rew_O}). This leads to
the best weight lines, which are illustrated on Figure~\ref{fig_method_i}.
The pressure can be calculated by following the lines indicated on the
figure.

\begin{figure}
\centerline{\includegraphics*[width=9cm,bb=10 270 590 695]{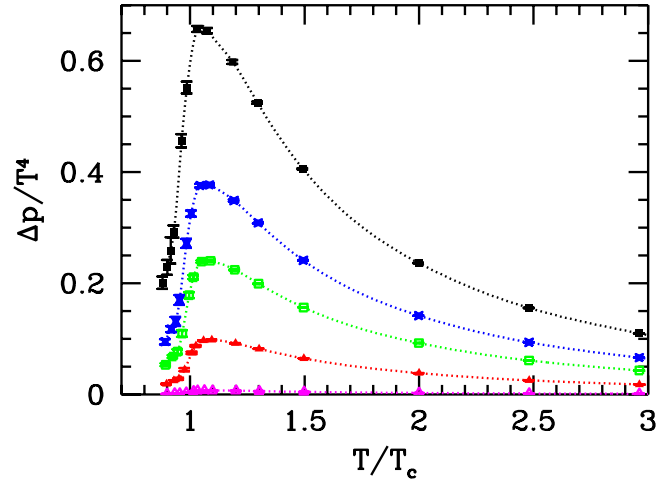}}
\caption{\small\label{fig_dp_mu}
The $\Delta p$ pressure difference for
several baryonic chemical potentials. The curves (from bottom to top) 
correspond to
$\mu_B=$100, 210, 330, 410 and 530~MeV.
}
\end{figure}

\begin{figure}
\centerline{\includegraphics*[width=9cm,bb=10 270 590 695]{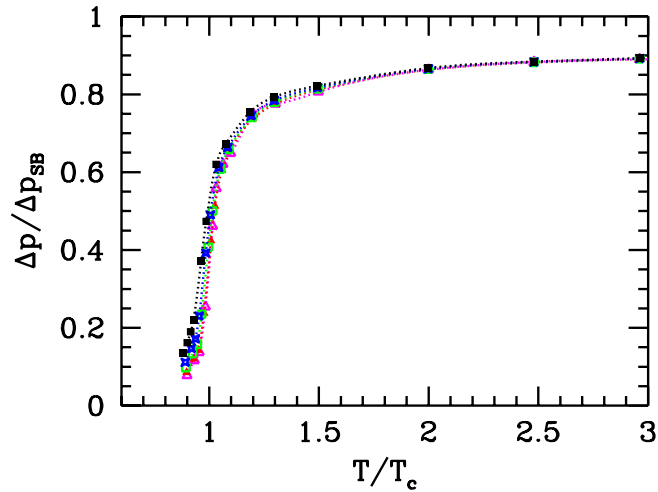}}
\caption{\small\label{fig_dpsb_mu}
The $\Delta p$ pressure difference normalized by the Stefan-Boltzmann
limit for different baryonic chemical potentials.
The curves correspond to
$\mu_B=$100 (purple), 210 (red), 330 (green), 410 (blue) and 530~MeV (black).
They all seem to show a universal $\mu$ independent behavior.
}
\end{figure}

Since the pressure at small chemical potentials differs only slightly
from the $\mu=0$ pressure, it is useful to define the difference as:
$\Delta p=p(\mu)-p(\mu=0)$. Figure~\ref{fig_dp_mu} shows the pressure
for five chemical potentials obtained by multi-parameter 
reweighting~\cite{Fodor:2002km,Csikor:2004ik}. For this analysis
standard staggered fermions were used on $N_t=4$ lattices. 
It is interesting that normalizing the shown curves by the Stephan-Boltzmann
value at the given $\mu$-s, one gets an almost universal behaviour 
(see Figure~\ref{fig_dpsb_mu}).

\begin{figure}
\centerline{\includegraphics*[width=7cm]{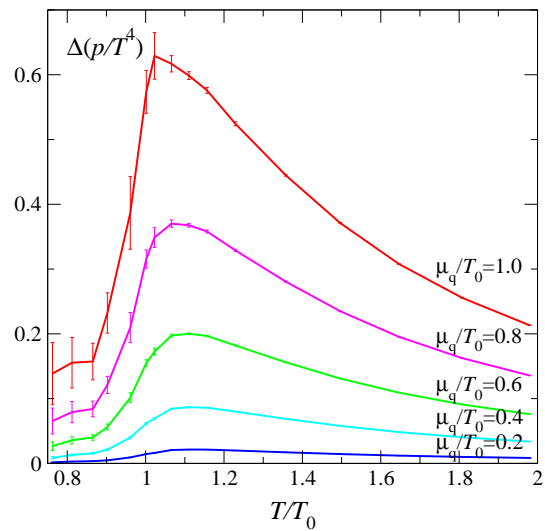}}
\caption{\small\label{eos_mu_bie}
The $\Delta p$ pressure difference calculated by the Bielefeld-Swansea
collaboration~\cite{Allton:2003vx} for five different $\mu$ values. 
The quark chemical potential, normalized by the transition temperature $T_0$
ranges from 0.2 to 1.
}
\end{figure}

The pressure difference has also been determined by the Bielefeld-Swansea
collaboration~\cite{Allton:2003vx,Allton:2005gk}. Their result is shown on 
Figure~\ref{eos_mu_bie}. It was obtained with p4 staggered 
fermions on $N_t=4$ lattices~\cite{Allton:2003vx}. Since no continuum limit was taken in either
case, the results do not have to be in completely consistent. Nevertheless
they seem to show a nice qualitative agreement. 

\chapter{Conclusions, outlook}\label{chap_sum}

A lattice result can be considered as a full result if two conditions are
fulfilled. The first condition is related to the quark masses.
We need results for physical
quark masses, or in other words $m_\pi$$\approx$135~MeV and 
$m_K$$\approx$500~MeV. 
Controlled extrapolation in the quark masses 
around the transition temperature are not easy. Note, however, that 
today computers are quite often capable to deal with physical
or almost physical quark masses. 
The second condition is the continuum extrapolation.
It can be reached only by measuring quantities
at non-vanishing lattice spacings and then extrapolating to
vanishing lattice spacings. Luckily enough the choice of the 
action tells us what sort of functional form of the lattice
spacings we expect for the deviation from the continuum limit result.
If this asymptotic behavior is already present, we are
in the so-called  scaling region.
E.g. for small enough lattice spacings results obtained
by the standard Wilson action deviate from the continuum result
by a linear term in the lattice spacing; for the staggered
formalism this dependence is quadratic in the lattice spacings.
Clearly, one should have an evidence that the results are already in
this scaling region, which is described by the  
asymptotic lattice spacing dependence. For this check results
at several lattice spacings are needed. Note, that thermodynamic
studies are carried out on lattices, which have smaller temporal 
than spatial extensions. Typically one uses $N_t$=4,6,8 and 10, which
--as a rule of thumb-- 
correspond to lattice spacings: $\approx$0.3, 0.2, 0.15 and
0.12~fm, respectively (particularly at small $N_t$ values,
the lattice spacing in physical unit is quite ambiguous, different
physical quantities give different results, this ambiguity
disappears when we approach the continuum limit). 

Let us summarize what is known about our specific question,
about the phase diagram of QCD. 
In some cases the result can be considered as a full one (at least 
using one specific formalism e.g. staggered one). In other
cases one can estimate that the full result can be obtained in
a year or two. There are however questions, which need much more
time to clarify, particularly the controlled continuum
limit is a difficult task.

a.) At vanishing chemical chemical potential the nature of the 
T$\neq$0 QCD transition is an
analytic crossover \cite{Aoki:2006we}. The result was obtained 
with physical quark masses in the staggered formalism. This
result can be considered as the full one. (As for any
result of such type and huge complexity at least one independent
analysis of the same depths is required to exclude any
mistakes.) There are two,
though unlikely possible uncertainties
of this finding. One of them is a question, what happens if
2+1 flavor staggered QCD happens to be {\it not} in the QCD universality
class.  Though we do not have any theoretical proof for this universality
class question, there is no sign for such a problematic scenario. 
Staggered lattice results for the whole 
spectrum and decay rates are in complete agreement with the experiments.
Nevertheless it would be important to repeat the calculation with other
fermion formulations (e.g. with Wilson fermions). This can be done
with computer resources which are about an order of magnitude
larger than the presently available ones.
Since the rapid crossover
is a remnant of the chiral transition of the massless theory, it
would be very interesting to study the question what happens in the
chiral limit. This question needs the same symmetry on the lattice
as in the continuum theory. The best choice is the overlap fermion.
Calculations with overlap fermions are usually  two orders of
magnitude more CPU-demanding than calculations with Wilson 
fermions.
The other inconvenient possibility is related to the continuum 
extrapolation. It is possible --though quite unlikely-- that after
observing a consistent and finite continuum limit for the chiral 
susceptibility a completely different
(phase transition-like, therefore divergent) continuum limit 
appears at
even smaller lattice spacings. Note, that the transition
turned out to be weaker and weaker as one decreased the lattice 
spacings. Thus, the above scenario --real phase transition in the
continuum limit-- would mean a completely
opposite lattice spacing dependence as it was observed. This is
the main reason, why one considers this possibility quite unlikely.
In order to go to even smaller
lattice spacings (e.g. $N_t$=12 or 14) 
approximately 1--2 orders of magnitude more CPU capacity
is needed. 

b.) We know the starting point of the phase diagram, namely the
transition temperature of the crossover at vanishing chemical
potential. Since the transition is an analytic one, there 
is no unique transition temperature. Different observables
lead to different transition temperatures. According to
Ref. \cite{Aoki:2006br} the width of the transition can be as large as
$\approx$40~MeV. Thus, transition temperatures, depending
on the definition, can be typically between 
$\approx$150 and 190~MeV. The actual values are still debated.
E.g. for the peak in the $\chi/T^2$ distribution ($\chi$ is the 
unrenormalized chiral susceptibility) Ref. \cite{Cheng:2006qk} used two
different lattice spacings, namely $N_t$=4 and 6, 
and obtained 192~MeV. For the peak in the
$\chi_r/T^4$ distribution  ($\chi_r$ is the
renormalized chiral susceptibility) Ref. \cite{Aoki:2006br} used 
also finer lattices with four
different lattice spacings, namely $N_t$=4,6,8 and 10, 
and obtained 151~MeV (note, that for $\chi_r/T^2$
the obtained value is about 10~MeV higher). It is worth
mentioning that for Ref. \cite{Aoki:2006br} the transition
temperature is independent of the choice of the overall scale,
whereas Ref. \cite{Cheng:2006qk} is still in the
scaling violation region (for their large lattice spacings
the ratios of physical quantities are several sigma away from their
experimental values). Since the available
CPU-capacity is enough to carry out independent lattice simulations on
$N_t$=8 and perhaps even on $N_t$=10 lattices, this controversy will 
be resolved in a year or two. The two possible but unlikely
uncertainties, mentioned in the previous paragraph, are relevant
also for the transition temperature. Therefore, one should 
determine the $\mu$=0 transition temperature using other 
formulations of lattice QCD (e.g. Wilson fermions or chiral
fermions), and double check the results with even smaller
lattice spacings.

c.) The curvature of the phase diagram at vanishing chemical
potential is known at a$\approx$0.3~fm lattice spacings. 
Results were obtained by standard and p4 improved lattice actions
for 2, 2+1 and 4 
flavors~\cite{Allton:2002zi,deForcrand:2002ci,D'Elia:2002gd,Fodor:2004nz}.
Though different choices of the QCD action should not necessarily
give the same result at this rather large lattice spacing, 
results are in good agreement. If one takes the same action
different techniques (multi-parameter reweighting with full
determinant, Taylor expansion, analytic continuation) give 
the same result upto several digits. This nice agreement
shows that the available methods are consistent.
Clearly, the major drawback
of these findings is the lack of the continuum extrapolation.
Similarly to the determination of the crossover temperature 
the continuum extrapolation might change the  a$\approx$0.3~fm
results quite significantly. The available computer
resources allow the determination of the curvature in the
continuum limit in a year or two. It is important to emphasize 
again, that the staggered formalism has an unclarified 
uncertainty, therefore the whole calculation should be 
repeated in the Wilson formalism, too. Such a Wilson analysis
is about an order of magnitude more CPU-demanding than the 
staggered one.

d.) There are several results for the existence and/or
location of the critical point on the
temperature versus chemical potential plane. All these
results were obtained at quite large lattice spacings 
a$\approx$0.3~fm. We discussed in detail the difficulties
and the problematic features of the available methods.
Let us point out a more general difficulty, which is
related to the continuum extrapolation.
As one determines the nature of the transition at vanishing
chemical potentials, it turns out that the transition
gets weaker and weaker for smaller and smaller lattice spacings.
This feature suggests, that 
the critical point, if it exists, might be at 
larger chemical potential in the continuum limit than on
$N_t$=4 lattices.
Unfortunately, for large chemical potentials the available
methods are less reliable, which is particularly true for the
staggered formalism (see Ref.~\cite{Golterman:2006rw} for a discussion
on the staggered eigenvalue quartets, which suggests to use quite small
lattice spacings). Searching for features at relatively large
chemical potentials and at small lattice spacings is a very difficult
and particularly CPU-demanding task. It is unlikely that
the available methods with the present computer-resources
can give a continuum  extrapolated staggered result in a few years. The 
available methods are all applicable for Wilson fermions, too.
On the one hand Wilson fermions do not have problems related to
the rooting of the determinant (c.f. \cite{Golterman:2006rw}), on the
other hand the full diagonalization of the Wilson matrix 
is about two orders of magnitude more CPU-consuming. Furthermore,
we do not have much experience how Wilson thermodynamics approaches
the continuum limit, therefore it is hard to say what temporal
extensions are needed to approach the continuum limit. The overlap
formalism has all the symmetries of the theory even at non-vanishing
lattice spacings, which is an advantage when we look for critical
behavior. Though the available methods are applicable also for overlap
fermions, the CPU-costs would be very large. To summarize:
the presently available resources do not allow to extrapolate
into the continuum limit. 
Results on the critical point at one or two non-vanishing lattice
spacings can not be considered as full results\footnote{Note, that 
the authors of this review emphasized this limitation, namely 
the lack of the continuum extrapolation even in the
abstract of their endpoint analysis \cite{Fodor:2004nz}}.

e.) There is one exploratory lattice result on the triple point of 
QCD \cite{Fodor:2007vv}. The lattice spacing is quite large,
the volumes are small and four flavor is applied to avoid the 
rooting problem. This density of states method reached 
approximately three times
larger chemical potentials than other methods in the literature.
The CPU-costs (for this factor of three) were about two orders
of magnitude larger than for the other methods. The method
works, but it is clear that due to limited resources
the continuum limit statement on the triple point can not be given
in the near future.

\bibliography{pd_review}

\providecommand{\href}[2]{#2}\begingroup\raggedright\begin{thebibliography}{10}

\bibitem{Kuti:1980gh}
J.~Kuti, J.~Polonyi, and K.~Szlachanyi, {\it {Monte Carlo Study of SU(2) Gauge
  Theory at Finite Temperature}},  {\em Phys. Lett.} {\bf B98} (1981) 199.

\bibitem{McLerran:1980pk}
L.~D. McLerran and B.~Svetitsky, {\it {A Monte Carlo Study of SU(2) Yang-Mills
  Theory at Finite Temperature}},  {\em Phys. Lett.} {\bf B98} (1981) 195.

\bibitem{Celik:1983wz}
T.~Celik, J.~Engels, and H.~Satz, {\it {The Order of the Deconfinement
  Transition in SU(3) Yang- Mills Theory}},  {\em Phys. Lett.} {\bf B125}
  (1983) 411.

\bibitem{Kogut:1982rt}
J.~B. Kogut {\em et~al.}, {\it {Deconfinement and Chiral Symmetry Restoration
  at Finite Temperatures in SU(2) and SU(3) Gauge Theories}},  {\em Phys. Rev.
  Lett.} {\bf 50} (1983) 393.

\bibitem{Gottlieb:1985ug}
S.~A. Gottlieb {\em et~al.}, {\it {The Deconfining Phase Transition and the
  Continuum Limit of Lattice Quantum Chromodynamics}},  {\em Phys. Rev. Lett.}
  {\bf 55} (1985) 1958.

\bibitem{Brown:1988qe}
F.~R. Brown, N.~H. Christ, Y.~F. Deng, M.~S. Gao, and T.~J. Woch, {\it {Nature
  of the Deconfining Phase Transition in SU(3) Lattice Gauge Theory}},  {\em
  Phys. Rev. Lett.} {\bf 61} (1988) 2058.

\bibitem{Fukugita:1989yb}
M.~Fukugita, M.~Okawa, and A.~Ukawa, {\it {ORDER OF THE DECONFINING PHASE
  TRANSITION IN SU(3) LATTICE GAUGE THEORY}},  {\em Phys. Rev. Lett.} {\bf 63}
  (1989) 1768.

\bibitem{Pisarski:1983ms}
R.~D. Pisarski and F.~Wilczek, {\it {Remarks on the Chiral Phase Transition in
  Chromodynamics}},  {\em Phys. Rev.} {\bf D29} (1984) 338--341.

\bibitem{Klevansky:1992qe}
S.~P. Klevansky, {\it {The Nambu-Jona-Lasinio model of quantum
  chromodynamics}},  {\em Rev. Mod. Phys.} {\bf 64} (1992) 649--708.

\bibitem{Montvay:1994cy}
I.~Montvay and G.~Munster, {\it {Quantum fields on a lattice}}, . Cambridge,
  UK: Univ. Pr. (1994) 491 p. (Cambridge monographs on mathematical physics).

\bibitem{Nielsen:1980rz}
H.~B. Nielsen and M.~Ninomiya, {\it {Absence of Neutrinos on a Lattice. 1.
  Proof by Homotopy Theory}},  {\em Nucl. Phys.} {\bf B185} (1981) 20.

\bibitem{Nielsen:1981xu}
H.~B. Nielsen and M.~Ninomiya, {\it {Absence of Neutrinos on a Lattice. 2.
  Intuitive Topological Proof}},  {\em Nucl. Phys.} {\bf B193} (1981) 173.

\bibitem{Neuberger:1997fp}
H.~Neuberger, {\it {Exactly massless quarks on the lattice}},  {\em Phys.
  Lett.} {\bf B417} (1998) 141--144,
  [\href{http://xxx.lanl.gov/abs/hep-lat/9707022}{{\tt hep-lat/9707022}}].

\bibitem{hep-lat/9802011}
M.~Luscher, {\it {Exact chiral symmetry on the lattice and the Ginsparg- Wilson
  relation}},  {\em Phys. Lett.} {\bf B428} (1998) 342--345,
  [\href{http://xxx.lanl.gov/abs/hep-lat/9802011}{{\tt hep-lat/9802011}}].

\bibitem{Fodor:2003bh}
Z.~Fodor, S.~D. Katz, and K.~K. Szabo, {\it {Dynamical overlap fermions,
  results with hybrid Monte- Carlo algorithm}},  {\em JHEP} {\bf 08} (2004)
  003, [\href{http://xxx.lanl.gov/abs/hep-lat/0311010}{{\tt hep-lat/0311010}}].

\bibitem{Cundy:2004pz}
N.~Cundy, {\it {Dynmical overlap}}, . Prepared for 58th Scottish Universities
  Summer School in Physics (SUSSP58): A NATO Advanced Study Institute and EU
  Hadron Physics 13 Summer Institute, St. Andrews, Scotland, 22-29 Aug 2004.

\bibitem{DeGrand:2004nq}
T.~A. DeGrand and S.~Schaefer, {\it {Physics issues in simulations with
  dynamical overlap fermions}},  {\em Phys. Rev.} {\bf D71} (2005) 034507,
  [\href{http://xxx.lanl.gov/abs/hep-lat/0412005}{{\tt hep-lat/0412005}}].

\bibitem{Egri:2005cx}
G.~I. Egri, Z.~Fodor, S.~D. Katz, and K.~K. Szabo, {\it {Topology with
  dynamical overlap fermions}},  {\em JHEP} {\bf 01} (2006) 049,
  [\href{http://xxx.lanl.gov/abs/hep-lat/0510117}{{\tt hep-lat/0510117}}].

\bibitem{Lang:2005jz}
C.~B. Lang, P.~Majumdar, and W.~Ortner, {\it {QCD with two dynamical flavors of
  chirally improved quarks}},  {\em Phys. Rev.} {\bf D73} (2006) 034507,
  [\href{http://xxx.lanl.gov/abs/hep-lat/0512014}{{\tt hep-lat/0512014}}].

\bibitem{Fukaya:2006vs}
{\bf JLQCD} Collaboration, H.~Fukaya {\em et~al.}, {\it {Lattice gauge action
  suppressing near-zero modes of H(W)}},  {\em Phys. Rev.} {\bf D74} (2006)
  094505, [\href{http://xxx.lanl.gov/abs/hep-lat/0607020}{{\tt
  hep-lat/0607020}}].

\bibitem{Bernard:2006ee}
C.~Bernard, M.~Golterman, and Y.~Shamir, {\it {Observations on staggered
  fermions at non-zero lattice spacing}},  {\em Phys. Rev.} {\bf D73} (2006)
  114511, [\href{http://xxx.lanl.gov/abs/hep-lat/0604017}{{\tt
  hep-lat/0604017}}].

\bibitem{Hasenfratz:1993sp}
P.~Hasenfratz and F.~Niedermayer, {\it {Perfect lattice action for
  asymptotically free theories}},  {\em Nucl. Phys.} {\bf B414} (1994)
  785--814, [\href{http://xxx.lanl.gov/abs/hep-lat/9308004}{{\tt
  hep-lat/9308004}}].

\bibitem{DeGrand:1995ji}
T.~A. DeGrand, A.~Hasenfratz, P.~Hasenfratz, and F.~Niedermayer, {\it {The
  Classically perfect fixed point action for SU(3) gauge theory}},  {\em Nucl.
  Phys.} {\bf B454} (1995) 587--614,
  [\href{http://xxx.lanl.gov/abs/hep-lat/9506030}{{\tt hep-lat/9506030}}].

\bibitem{Polonyi:1983tm}
J.~Polonyi and H.~W. Wyld, {\it {Microcanonical Simulation of Fermionic
  Systems}},  {\em Phys. Rev. Lett.} {\bf 51} (1983) 2257.

\bibitem{Scalettar:1986uy}
R.~T. Scalettar, D.~J. Scalapino, and R.~L. Sugar, {\it {NEW ALGORITHM FOR THE
  NUMERICAL SIMULATION OF FERMIONS}},  {\em Phys. Rev.} {\bf B34} (1986)
  7911--7917.

\bibitem{Duane:1987de}
S.~Duane, A.~D. Kennedy, B.~J. Pendleton, and D.~Roweth, {\it {Hybrid Monte
  Carlo}},  {\em Phys. Lett.} {\bf B195} (1987) 216--222.

\bibitem{Gottlieb:1987mq}
S.~A. Gottlieb, W.~Liu, D.~Toussaint, R.~L. Renken, and R.~L. Sugar, {\it
  {Hybrid Molecular Dynamics Algorithms for the Numerical Simulation of Quantum
  Chromodynamics}},  {\em Phys. Rev.} {\bf D35} (1987) 2531--2542.

\bibitem{Clark:2006wp}
M.~A. Clark and A.~D. Kennedy, {\it {Accelerating Staggered Fermion Dynamics
  with the Rational Hybrid Monte Carlo (RHMC) Algorithm}},  {\em Phys. Rev.}
  {\bf D75} (2007) 011502, [\href{http://xxx.lanl.gov/abs/hep-lat/0610047}{{\tt
  hep-lat/0610047}}].

\bibitem{Heller:1999xz}
U.~M. Heller, F.~Karsch, and B.~Sturm, {\it {Improved staggered fermion actions
  for QCD thermodynamics}},  {\em Phys. Rev.} {\bf D60} (1999) 114502,
  [\href{http://xxx.lanl.gov/abs/hep-lat/9901010}{{\tt hep-lat/9901010}}].

\bibitem{Morningstar:2003gk}
C.~Morningstar and M.~J. Peardon, {\it {Analytic smearing of SU(3) link
  variables in lattice QCD}},  {\em Phys. Rev.} {\bf D69} (2004) 054501,
  [\href{http://xxx.lanl.gov/abs/hep-lat/0311018}{{\tt hep-lat/0311018}}].

\bibitem{Aoki:2005vt}
Y.~Aoki, Z.~Fodor, S.~D. Katz, and K.~K. Szabo, {\it {The equation of state in
  lattice QCD: With physical quark masses towards the continuum limit}},  {\em
  JHEP} {\bf 01} (2006) 089,
  [\href{http://xxx.lanl.gov/abs/hep-lat/0510084}{{\tt hep-lat/0510084}}].

\bibitem{Bernard:2005mf}
C.~Bernard {\em et~al.}, {\it {The equation of state for QCD with 2+1 flavors
  of quarks}},  {\em PoS} {\bf LAT2005} (2006) 156,
  [\href{http://xxx.lanl.gov/abs/hep-lat/0509053}{{\tt hep-lat/0509053}}].

\bibitem{Cheng:2006qk}
M.~Cheng {\em et~al.}, {\it {The transition temperature in QCD}},  {\em Phys.
  Rev.} {\bf D74} (2006) 054507,
  [\href{http://xxx.lanl.gov/abs/hep-lat/0608013}{{\tt hep-lat/0608013}}].

\bibitem{Aoki:2006we}
Y.~Aoki, G.~Endrodi, Z.~Fodor, S.~D. Katz, and K.~K. Szabo, {\it {The order of
  the quantum chromodynamics transition predicted by the standard model of
  particle physics}},  {\em Nature} {\bf 443} (2006) 675--678,
  [\href{http://xxx.lanl.gov/abs/hep-lat/0611014}{{\tt hep-lat/0611014}}].

\bibitem{Sommer:1993ce}
R.~Sommer, {\it {A New way to set the energy scale in lattice gauge theories
  and its applications to the static force and alpha-s in SU(2) Yang-Mills
  theory}},  {\em Nucl. Phys.} {\bf B411} (1994) 839--854,
  [\href{http://xxx.lanl.gov/abs/hep-lat/9310022}{{\tt hep-lat/9310022}}].

\bibitem{Aubin:2004wf}
C.~Aubin {\em et~al.}, {\it {Light hadrons with improved staggered quarks:
  Approaching the continuum limit}},  {\em Phys. Rev.} {\bf D70} (2004) 094505,
  [\href{http://xxx.lanl.gov/abs/hep-lat/0402030}{{\tt hep-lat/0402030}}].

\bibitem{Gray:2005ur}
A.~Gray {\em et~al.}, {\it {The Upsilon spectrum and m(b) from full lattice
  QCD}},  {\em Phys. Rev.} {\bf D72} (2005) 094507,
  [\href{http://xxx.lanl.gov/abs/hep-lat/0507013}{{\tt hep-lat/0507013}}].

\bibitem{Alexandrou:2008tn}
{\bf European Twisted Mass} Collaboration, C.~Alexandrou {\em et~al.}, {\it
  {Light baryon masses with dynamical twisted mass fermions}},  {\em Phys.
  Rev.} {\bf D78} (2008) 014509, [\href{http://xxx.lanl.gov/abs/0803.3190}{{\tt
  0803.3190}}].

\bibitem{Gockeler:2005rv}
M.~Gockeler {\em et~al.}, {\it {A determination of the Lambda parameter from
  full lattice QCD}},  {\em Phys. Rev.} {\bf D73} (2006) 014513,
  [\href{http://xxx.lanl.gov/abs/hep-ph/0502212}{{\tt hep-ph/0502212}}].

\bibitem{Aoki:2008sm}
{\bf PACS-CS} Collaboration, S.~Aoki {\em et~al.}, {\it {2+1 Flavor Lattice QCD
  toward the Physical Point}},  \href{http://xxx.lanl.gov/abs/0807.1661}{{\tt
  0807.1661}}.

\bibitem{Amsler:2008zz}
{\bf Particle Data Group} Collaboration, C.~Amsler {\em et~al.}, {\it {Review
  of particle physics}},  {\em Phys. Lett.} {\bf B667} (2008) 1.

\bibitem{Schwarz:2003du}
D.~J. Schwarz, {\it {The first second of the universe}},  {\em Annalen Phys.}
  {\bf 12} (2003) 220--270,
  [\href{http://xxx.lanl.gov/abs/astro-ph/0303574}{{\tt astro-ph/0303574}}].

\bibitem{Witten:1984rs}
E.~Witten, {\it {Cosmic Separation of Phases}},  {\em Phys. Rev.} {\bf D30}
  (1984) 272--285.

\bibitem{Applegate:1985qt}
J.~H. Applegate and C.~J. Hogan, {\it {Relics of Cosmic Quark Condensation}},
  {\em Phys. Rev.} {\bf D31} (1985) 3037--3045.

\bibitem{Halasz:1998qr}
A.~M. Halasz, A.~D. Jackson, R.~E. Shrock, M.~A. Stephanov, and J.~J.~M.
  Verbaarschot, {\it {On the phase diagram of {QCD}}},  {\em Phys. Rev.} {\bf
  D58} (1998) 096007, [\href{http://xxx.lanl.gov/abs/hep-ph/9804290}{{\tt
  hep-ph/9804290}}].

\bibitem{Berges:1998rc}
J.~Berges and K.~Rajagopal, {\it {Color superconductivity and chiral symmetry
  restoration at nonzero baryon density and temperature}},  {\em Nucl. Phys.}
  {\bf B538} (1999) 215--232,
  [\href{http://xxx.lanl.gov/abs/hep-ph/9804233}{{\tt hep-ph/9804233}}].

\bibitem{Schaefer:2004en}
B.-J. Schaefer and J.~Wambach, {\it {The phase diagram of the quark meson
  model}},  {\em Nucl. Phys.} {\bf A757} (2005) 479--492,
  [\href{http://xxx.lanl.gov/abs/nucl-th/0403039}{{\tt nucl-th/0403039}}].

\bibitem{Herpay:2005yr}
T.~Herpay, A.~Patkos, Z.~Szep, and P.~Szepfalusy, {\it {Mapping the boundary of
  the first order finite temperature restoration of chiral symmetry in the
  (m(pi) - m(K))-plane with a linear sigma model}},  {\em Phys. Rev.} {\bf D71}
  (2005) 125017, [\href{http://xxx.lanl.gov/abs/hep-ph/0504167}{{\tt
  hep-ph/0504167}}].

\bibitem{Karsch:2003va}
F.~Karsch {\em et~al.}, {\it {Where is the chiral critical point in 3-flavor
  QCD?}},  {\em Nucl. Phys. Proc. Suppl.} {\bf 129} (2004) 614--616,
  [\href{http://xxx.lanl.gov/abs/hep-lat/0309116}{{\tt hep-lat/0309116}}].

\bibitem{deForcrand:2007rq}
P.~de~Forcrand, S.~Kim, and O.~Philipsen, {\it {A QCD chiral critical point at
  small chemical potential: is it there or not?}},  {\em PoS} {\bf LAT2007}
  (2007) 178, [\href{http://xxx.lanl.gov/abs/0711.0262}{{\tt 0711.0262}}].

\bibitem{Endrodi:2007gc}
G.~Endrodi, Z.~Fodor, S.~D. Katz, and K.~K. Szabo, {\it {The nature of the
  finite temperature QCD transition as a function of the quark masses}},  {\em
  PoS} {\bf LAT2007} (2007) 182, [\href{http://xxx.lanl.gov/abs/0710.0998}{{\tt
  0710.0998}}].

\bibitem{Bernard:2004je}
{\bf MILC} Collaboration, C.~Bernard {\em et~al.}, {\it {QCD thermodynamics
  with three flavors of improved staggered quarks}},  {\em Phys. Rev.} {\bf
  D71} (2005) 034504, [\href{http://xxx.lanl.gov/abs/hep-lat/0405029}{{\tt
  hep-lat/0405029}}].

\bibitem{Kaczmarek:2002mc}
O.~Kaczmarek, F.~Karsch, P.~Petreczky, and F.~Zantow, {\it {Heavy quark
  anti-quark free energy and the renormalized Polyakov loop}},  {\em Phys.
  Lett.} {\bf B543} (2002) 41--47,
  [\href{http://xxx.lanl.gov/abs/hep-lat/0207002}{{\tt hep-lat/0207002}}].

\bibitem{Fodor:2004ft}
Z.~Fodor, S.~D. Katz, K.~K. Szabo, and A.~I. Toth, {\it {Grand canonical
  potential for a static quark anti-quark pair at mu not equal 0}},  {\em Nucl.
  Phys. Proc. Suppl.} {\bf 140} (2005) 508--510,
  [\href{http://xxx.lanl.gov/abs/hep-lat/0410032}{{\tt hep-lat/0410032}}].

\bibitem{Aoki:2006br}
Y.~Aoki, Z.~Fodor, S.~D. Katz, and K.~K. Szabo, {\it {The QCD transition
  temperature: Results with physical masses in the continuum limit}},  {\em
  Phys. Lett.} {\bf B643} (2006) 46--54,
  [\href{http://xxx.lanl.gov/abs/hep-lat/0609068}{{\tt hep-lat/0609068}}].

\bibitem{Boyd:1996bx}
G.~Boyd {\em et~al.}, {\it {Thermodynamics of SU(3) Lattice Gauge Theory}},
  {\em Nucl. Phys.} {\bf B469} (1996) 419--444,
  [\href{http://xxx.lanl.gov/abs/hep-lat/9602007}{{\tt hep-lat/9602007}}].

\bibitem{Okamoto:1999hi}
{\bf CP-PACS} Collaboration, M.~Okamoto {\em et~al.}, {\it {Equation of state
  for pure SU(3) gauge theory with renormalization group improved action}},
  {\em Phys. Rev.} {\bf D60} (1999) 094510,
  [\href{http://xxx.lanl.gov/abs/hep-lat/9905005}{{\tt hep-lat/9905005}}].

\bibitem{Namekawa:2001ih}
{\bf CP-PACS} Collaboration, Y.~Namekawa {\em et~al.}, {\it {Thermodynamics of
  SU(3) gauge theory on anisotropic lattices}},  {\em Phys. Rev.} {\bf D64}
  (2001) 074507, [\href{http://xxx.lanl.gov/abs/hep-lat/0105012}{{\tt
  hep-lat/0105012}}].

\bibitem{Endrodi:2007tq}
G.~Endrodi, Z.~Fodor, S.~D. Katz, and K.~K. Szabo, {\it {The equation of state
  at high temperatures from lattice QCD}},  {\em PoS} {\bf LAT2007} (2007) 228,
  [\href{http://xxx.lanl.gov/abs/0710.4197}{{\tt 0710.4197}}].

\bibitem{Blum:1994zf}
T.~Blum, L.~Karkkainen, D.~Toussaint, and S.~A. Gottlieb, {\it {The beta
  function and equation of state for QCD with two flavors of quarks}},  {\em
  Phys. Rev.} {\bf D51} (1995) 5153--5164,
  [\href{http://xxx.lanl.gov/abs/hep-lat/9410014}{{\tt hep-lat/9410014}}].

\bibitem{Bernard:1996cs}
{\bf MILC} Collaboration, C.~W. Bernard {\em et~al.}, {\it {The equation of
  state for two flavor QCD at N(t) = 6}},  {\em Phys. Rev.} {\bf D55} (1997)
  6861--6869, [\href{http://xxx.lanl.gov/abs/hep-lat/9612025}{{\tt
  hep-lat/9612025}}].

\bibitem{Karsch:2000ps}
F.~Karsch, E.~Laermann, and A.~Peikert, {\it {The pressure in 2, 2+1 and 3
  flavour QCD}},  {\em Phys. Lett.} {\bf B478} (2000) 447--455,
  [\href{http://xxx.lanl.gov/abs/hep-lat/0002003}{{\tt hep-lat/0002003}}].

\bibitem{Cheng:2007jq}
M.~Cheng {\em et~al.}, {\it {The QCD Equation of State with almost Physical
  Quark Masses}},  {\em Phys. Rev.} {\bf D77} (2008) 014511,
  [\href{http://xxx.lanl.gov/abs/0710.0354}{{\tt 0710.0354}}].

\bibitem{Bernard:2006nj}
C.~Bernard {\em et~al.}, {\it {QCD equation of state with 2+1 flavors of
  improved staggered quarks}},  {\em Phys. Rev.} {\bf D75} (2007) 094505,
  [\href{http://xxx.lanl.gov/abs/hep-lat/0611031}{{\tt hep-lat/0611031}}].

\bibitem{Engels:1990vr}
J.~Engels, J.~Fingberg, F.~Karsch, D.~Miller, and M.~Weber, {\it
  {Nonperturbative thermodynamics of SU(N) gauge theories}},  {\em Phys. Lett.}
  {\bf B252} (1990) 625--630.

\bibitem{Aoki:2009sc}
Y.~Aoki {\em et~al.}, {\it {The QCD transition temperature: results with
  physical masses in the continuum limit II}},  {\em JHEP} {\bf 06} (2009) 088,
  [\href{http://xxx.lanl.gov/abs/0903.4155}{{\tt 0903.4155}}].

\bibitem{Bazavov:2009zn}
A.~Bazavov {\em et~al.}, {\it {Equation of state and QCD transition at finite
  temperature}},  {\em Phys. Rev.} {\bf D80} (2009) 014504,
  [\href{http://xxx.lanl.gov/abs/0903.4379}{{\tt 0903.4379}}].

\bibitem{Karsch:2008fe}
{\bf RBC} Collaboration, F.~Karsch, {\it {Equation of state and more from
  lattice regularized QCD}},  {\em J. Phys.} {\bf G35} (2008) 104096,
  [\href{http://xxx.lanl.gov/abs/0804.4148}{{\tt 0804.4148}}].

\bibitem{Karsch:2007dp}
F.~Karsch, {\it {Recent lattice results on finite temperature and density QCD,
  part I}},  {\em PoS} {\bf CPOD07} (2007) 026,
  [\href{http://xxx.lanl.gov/abs/0711.0656}{{\tt 0711.0656}}].

\bibitem{Barbour:1997ej}
I.~M. Barbour, S.~E. Morrison, E.~G. Klepfish, J.~B. Kogut, and M.-P. Lombardo,
  {\it {Results on finite density QCD}},  {\em Nucl. Phys. Proc. Suppl.} {\bf
  60A} (1998) 220--234, [\href{http://xxx.lanl.gov/abs/hep-lat/9705042}{{\tt
  hep-lat/9705042}}].

\bibitem{Hasenfratz:1983ba}
P.~Hasenfratz and F.~Karsch, {\it {Chemical Potential on the Lattice}},  {\em
  Phys. Lett.} {\bf B125} (1983) 308.

\bibitem{Fodor:2001au}
Z.~Fodor and S.~D. Katz, {\it {A new method to study lattice QCD at finite
  temperature and chemical potential}},  {\em Phys. Lett.} {\bf B534} (2002)
  87--92, [\href{http://xxx.lanl.gov/abs/hep-lat/0104001}{{\tt
  hep-lat/0104001}}].

\bibitem{Allton:2002zi}
C.~R. Allton {\em et~al.}, {\it {The QCD thermal phase transition in the
  presence of a small chemical potential}},  {\em Phys. Rev.} {\bf D66} (2002)
  074507, [\href{http://xxx.lanl.gov/abs/hep-lat/0204010}{{\tt
  hep-lat/0204010}}].

\bibitem{deForcrand:2002pa}
P.~de~Forcrand, S.~Kim, and T.~Takaishi, {\it Qcd simulations at small chemical
  potential},  {\em Nucl. Phys. Proc. Suppl.} {\bf 119} (2003) 541--543,
  [\href{http://xxx.lanl.gov/abs/hep-lat/0209126}{{\tt hep-lat/0209126}}].

\bibitem{deForcrand:2002ci}
P.~de~Forcrand and O.~Philipsen, {\it The qcd phase diagram for small densities
  from imaginary chemical potential},  {\em Nucl. Phys.} {\bf B642} (2002)
  290--306, [\href{http://xxx.lanl.gov/abs/hep-lat/0205016}{{\tt
  hep-lat/0205016}}].

\bibitem{Azcoiti:2004ri}
V.~Azcoiti, G.~Di~Carlo, A.~Galante, and V.~Laliena, {\it {Finite density QCD:
  A new approach}},  {\em JHEP} {\bf 12} (2004) 010,
  [\href{http://xxx.lanl.gov/abs/hep-lat/0409157}{{\tt hep-lat/0409157}}].

\bibitem{Kratochvila:2004wz}
S.~Kratochvila and P.~de~Forcrand, {\it {QCD at small baryon number}},  {\em
  Nucl. Phys. Proc. Suppl.} {\bf 140} (2005) 514--516,
  [\href{http://xxx.lanl.gov/abs/hep-lat/0409072}{{\tt hep-lat/0409072}}].

\bibitem{Alexandru:2005ix}
A.~Alexandru, M.~Faber, I.~Horvath, and K.-F. Liu, {\it {Lattice QCD at finite
  density via a new canonical approach}},  {\em Phys. Rev.} {\bf D72} (2005)
  114513, [\href{http://xxx.lanl.gov/abs/hep-lat/0507020}{{\tt
  hep-lat/0507020}}].

\bibitem{Golterman:2006rw}
M.~Golterman, Y.~Shamir, and B.~Svetitsky, {\it {Breakdown of staggered
  fermions at nonzero chemical potential}},  {\em Phys. Rev.} {\bf D74} (2006)
  071501, [\href{http://xxx.lanl.gov/abs/hep-lat/0602026}{{\tt
  hep-lat/0602026}}].

\bibitem{Fodor:2001pe}
Z.~Fodor and S.~D. Katz, {\it {Lattice determination of the critical point of
  QCD at finite T and mu}},  {\em JHEP} {\bf 03} (2002) 014,
  [\href{http://xxx.lanl.gov/abs/hep-lat/0106002}{{\tt hep-lat/0106002}}].

\bibitem{Yang:1952be}
C.-N. Yang and T.~D. Lee, {\it {Statistical theory of equations of state and
  phase transitions. I: Theory of condensation}},  {\em Phys. Rev.} {\bf 87}
  (1952) 404--409.

\bibitem{Lee:1952ig}
T.~D. Lee and C.-N. Yang, {\it {Statistical theory of equations of state and
  phase transitions. II: Lattice gas and Ising model}},  {\em Phys. Rev.} {\bf
  87} (1952) 410--419.

\bibitem{Fodor:2004nz}
Z.~Fodor and S.~D. Katz, {\it {Critical point of QCD at finite T and mu,
  lattice results for physical quark masses}},  {\em JHEP} {\bf 04} (2004) 050,
  [\href{http://xxx.lanl.gov/abs/hep-lat/0402006}{{\tt hep-lat/0402006}}].

\bibitem{D'Elia:2002gd}
M.~D'Elia and M.-P. Lombardo, {\it Finite density qcd via imaginary chemical
  potential},  {\em Phys. Rev.} {\bf D67} (2003) 014505,
  [\href{http://xxx.lanl.gov/abs/hep-lat/0209146}{{\tt hep-lat/0209146}}].

\bibitem{Azcoiti:2005tv}
V.~Azcoiti, G.~Di~Carlo, A.~Galante, and V.~Laliena, {\it {Phase diagram of QCD
  with four quark flavors at finite temperature and baryon density}},  {\em
  Nucl. Phys.} {\bf B723} (2005) 77--90,
  [\href{http://xxx.lanl.gov/abs/hep-lat/0503010}{{\tt hep-lat/0503010}}].

\bibitem{Fodor:2007vv}
Z.~Fodor, S.~D. Katz, and C.~Schmidt, {\it {The density of states method at
  non-zero chemical potential}},  {\em JHEP} {\bf 03} (2007) 121,
  [\href{http://xxx.lanl.gov/abs/hep-lat/0701022}{{\tt hep-lat/0701022}}].

\bibitem{Karsch:2001nf}
F.~Karsch, E.~Laermann, and C.~Schmidt, {\it {The chiral critical point in
  3-flavor QCD}},  {\em Phys. Lett.} {\bf B520} (2001) 41--49,
  [\href{http://xxx.lanl.gov/abs/hep-lat/0107020}{{\tt hep-lat/0107020}}].

\bibitem{Ejiri:2003dc}
S.~Ejiri {\em et~al.}, {\it {Study of QCD thermodynamics at finite density by
  Taylor expansion}},  {\em Prog. Theor. Phys. Suppl.} {\bf 153} (2004)
  118--126, [\href{http://xxx.lanl.gov/abs/hep-lat/0312006}{{\tt
  hep-lat/0312006}}].

\bibitem{Gavai:2004sd}
R.~V. Gavai and S.~Gupta, {\it {The critical end point of QCD}},  {\em Phys.
  Rev.} {\bf D71} (2005) 114014,
  [\href{http://xxx.lanl.gov/abs/hep-lat/0412035}{{\tt hep-lat/0412035}}].

\bibitem{deForcrand:2003hx}
P.~de~Forcrand and O.~Philipsen, {\it The qcd phase diagram for three
  degenerate flavors and small baryon density},  {\em Nucl. Phys.} {\bf B673}
  (2003) 170, [\href{http://xxx.lanl.gov/abs/hep-lat/030702}{{\tt
  hep-lat/030702}}].

\bibitem{deForcrand:2006pv}
P.~de~Forcrand and O.~Philipsen, {\it {The chiral critical line of N(f) = 2+1
  QCD at zero and non-zero baryon density}},  {\em JHEP} {\bf 01} (2007) 077,
  [\href{http://xxx.lanl.gov/abs/hep-lat/0607017}{{\tt hep-lat/0607017}}].

\bibitem{deForcrand:2008vr}
P.~de~Forcrand and O.~Philipsen, {\it {The chiral critical point of Nf=3 QCD at
  finite density to the order $(\mu/T)^4$}},
  \href{http://xxx.lanl.gov/abs/0808.1096}{{\tt 0808.1096}}.

\bibitem{Kovacs:2006ym}
P.~Kovacs and Z.~Szep, {\it {The critical surface of the SU(3)L x SU(3)R chiral
  quark model at non-zero baryon density}},  {\em Phys. Rev.} {\bf D75} (2007)
  025015, [\href{http://xxx.lanl.gov/abs/hep-ph/0611208}{{\tt
  hep-ph/0611208}}].

\bibitem{Fodor:2002km}
Z.~Fodor, S.~D. Katz, and K.~K. Szabo, {\it {The QCD equation of state at
  nonzero densities: Lattice result}},  {\em Phys. Lett.} {\bf B568} (2003)
  73--77, [\href{http://xxx.lanl.gov/abs/hep-lat/0208078}{{\tt
  hep-lat/0208078}}].

\bibitem{Csikor:2004ik}
F.~Csikor {\em et~al.}, {\it {Equation of state at finite temperature and
  chemical potential, lattice QCD results}},  {\em JHEP} {\bf 05} (2004) 046,
  [\href{http://xxx.lanl.gov/abs/hep-lat/0401016}{{\tt hep-lat/0401016}}].

\bibitem{Allton:2003vx}
C.~R. Allton {\em et~al.}, {\it {The equation of state for two flavor QCD at
  non-zero chemical potential}},  {\em Phys. Rev.} {\bf D68} (2003) 014507,
  [\href{http://xxx.lanl.gov/abs/hep-lat/0305007}{{\tt hep-lat/0305007}}].

\bibitem{Allton:2005gk}
C.~R. Allton {\em et~al.}, {\it {Thermodynamics of two flavor QCD to sixth
  order in quark chemical potential}},  {\em Phys. Rev.} {\bf D71} (2005)
  054508, [\href{http://xxx.lanl.gov/abs/hep-lat/0501030}{{\tt
  hep-lat/0501030}}].

\end{thebibliography}\endgroup

\end{document}